\documentclass[aps, two column, prb, superscriptaddress]{revtex4-2}
\usepackage{amsfonts}
\usepackage{amsmath}
\usepackage{subfigure}
\usepackage{graphicx, mathtools}
\usepackage[normalem]{ulem}
\usepackage{epsfig}
\usepackage{bm}
\usepackage{braket}
\usepackage{natbib}
\usepackage{enumitem}
\usepackage{longtable}
\usepackage{color}
\usepackage{xfrac}
\usepackage[UKenglish]{babel}
\usepackage[colorlinks=true,linkcolor=blue,urlcolor=blue,citecolor=blue,anchorcolor=blue]{hyperref}
\makeatletter
\usepackage{tikz}
\usetikzlibrary{decorations.markings}
\tikzset{->-/.style={decoration={markings, mark=at position #1 with {\arrow{>}}}, postaction={decorate}}}
\usepackage{circuitikz}[american voltages]
\makeatother

\begin{document}

\title{A Tale of Two Quantum Compass Models}

\author{Soumya Sur}\email{soumyasur@imsc.res.in}

\author{M. S. Laad}\email{mslaad@imsc.res.in}
\affiliation{Institute of Mathematical Sciences, Taramani, Chennai 600113, India}
\affiliation{Homi Bhabha National Institute Training School Complex,
Anushakti Nagar, Mumbai 400085, India}
\author{Arya Subramonian}\email{arya.subramonian@physics.uu.se}
\affiliation{Materials Theory Division, Department of Physics and Astronomy, Uppsala University, Sweden}
\author{S. R. Hassan}\email{shassan@imsc.res.in}
\affiliation{Institute of Mathematical Sciences, Taramani, Chennai 600113, India}
\affiliation{Homi Bhabha National Institute Training School Complex,
Anushakti Nagar, Mumbai 400085, India}

\date{\rm\today}

\begin{abstract}
We investigate two variants of quantum compass models (QCMs). The first, an orbital-only honeycomb QCM, is shown to exhibit a quantum phase transition (QPT) from a $XX$- to $ZZ$-ordered phase in the $3d$-Ising universality class, in accord with earlier studies. In a fractionalized parton construction, this describes a ``superfluid-Mott insulator'' transition between a higher-order topological superfluid and the toric code, the latter described as a $p$-wave resonating valence bond state of the partons. The second variant, the spinless fermion QCM on a square lattice, is of interest in context of cold-atom lattices with higher-angular momentum states on each atom.  We explore finite-temperature orbital order-disorder transitions in the itinerant and localized limits using complementary methods. In the itinerant limit, we uncover an intricate temperature ($T$)-dependent dimensional crossover from a high-$T$ quasi-$1d$ insulator-like state, via an incoherent bad-metal-like state at intermediate $T$, to a $2d$ symmetry-broken insulator at low $T$, well below the ``orbital'' ordering scale. Finally, we discuss how engineering specific, tunable and realistic perturbations in both these variants can act as a playground for simulating a variety of exotic QPTs between topologically ordered and trivial phases. In the cold-atom context, we propose a novel way to engineer a possible realisation of the exotic exciton Bose liquid phase at a QPT between a Bose superfluid and a charge density wave insulator. We argue that advances in design of Josephson junction arrays and manipulating cold-atom lattices offer the hope of simulating such novel phases of matter in the foreseeable future.
\end{abstract}
\maketitle

\section{Introduction}

\indent Frustration in quantum matter is increasingly recognized as a tangible ingredient for forming atypical quantum-ordered (QO) phases \cite{Wen}, \cite{savary} of matter, whether it derives from lattice geometry or competing exchange interactions between multiple degrees of freedom of the system. Generically, it induces macroscopic degeneracy of quasi-classical (Landau like) ordered states, suppressing the conventional ``solid''-like ordering tendencies. A subset of these emergent QO states possessing a gap to the excitations are instead characterized by various topological invariants like genus-dependent ground state degeneracy \cite{Wen2} or topological entanglement entropy \cite{Levin-wen, Kitaev-preskill}. Elementary excitations typically carry fractional charges by virtue of the intricate long-range entangled structure of the QO ground state. Best exemplified by the exactly soluble Kitaev honeycomb model \cite{Kitaev}, intense activity is focused on novel transition metal oxides (TMO) with strong spin-orbit interactions \cite{Yogesh, Moessner, Moessner2} as well as artificially constructed systems like Josephson Junction (JJ) arrays and ultra-cold atoms. In the later avatar, tantalizing links to numerous quantum computing (QC) architectures and error-correcting codes \cite{Nori, Sau, Bacon} hold out the promise for future QC technology.\\
\indent In a somewhat associated vein, models of interacting Majorana fermions have recently been explored in search for unconevntional ordered phases and associated quantum critical phenomena \cite{Affleck, Affleck2, Kamiya}. In condensed matter systems, Majorana fermions are shown (at least theoretically) to appear as low (or zero) energy excitations at the interfaces between topologically distinct phases \cite{Alicea}. When the Majorana zero modes (MZMs) are arranged on some lattice, intra- and inter-site correlations between them can lead to non-trivial consequences. For example, In a network of JJ coupled mesoscopic superconducting islands hosting quantum nanowires, the ratio of inter-island hybridization to local intra-island interaction can be explicitly controlled by tuning the condesate's charging energy and the JJ assisted single-particle tunneling (see \cite{Hassler} and references therein). Remarkably, such models of \textit{crossed} Majorana chains with a (strong) local interaction between four MZMs on each vertex have recently been invoked in the context of fracton topological order and fracton codes \cite{von Oppen}. They exhibit higher order topological states (HOTS) with corner ($d=2$) or hinge ($d=3$) MZMs, attesting to underlying excitations with restricted mobility along lines ($d=2$) or planes ($d=3$). Given such rich possibilities, it is obviously of great interest to investigate other realistic situations where these Majorana lattice models can arise as an emergent low-energy description.\\
\indent One such physically appealing platform consists of different lattice models of highly-frustrated quantum magnets which upon {\it partonization} give rise to interacting Majorana models \cite{Wen2}, \cite{Kitaev}, \cite{Fu}. Imposition of local constraints on the parton Hilbert space leads to emergent ($Z_{2}$) gauge fields which interact with the Majorana partons (matter fields). When the gauge interaction is weak or the theory is in some {\it deconfined} phase, fractionalized Majorana partons survive as long-lived gapped or gapless excitations. This happens exactly in the Kitaev's honeycomb model where the gauge fluxes have no dynamics. On the other hand, if the gauge coupling becomes sufficiently strong, Majorana particles will confine which leads to conventional (Landau-like) non-fractionalized phases. \\
\indent In this article, we investigate the honeycomb quantum compass model (HQCM) (see Eq.\eqref{eq1}) in light of the above parton-like description. This model bears a close similarity with the Kitaev's model except for some modifications (in the exchange easy-axis) on certain bonds. As we will see, this leads to a drastic change in its low-energy behavior. Interestingly, HQCM has been previously studied (\cite{Nussinov-Majorana}, called XXZ compass model there) using bond algebraic dualities. Here, we employ two different approaches; one based on a direct Jordan-Wigner (JW) transformation which maps HQCM to a Hubbard-like locally interacting fermion model with highly anisotropic ``compass-like'' $1d$ hopping and $p$-wave pairing \cite{Nussinov-vdbrink} (see Eq.\eqref{eq6}). This JW fermion model can be written further as an interacting Majorana fermion model (see Eq.\eqref{eq5}) that was previously studied \cite{von Oppen} in the context of fractons. Our second approach is based on a rotated, Kramers-Wannier duality applied to each $ZZ$ bonds of the honeycomb lattice. It maps the HQCM to a $2d$ transverse field Ising model (TFIM) with Ising bond disorder. The exact mapping to $2d$ TFIM already deteremines the underlying unversalitly class of both the HQCM and the JW-fermion compass-Hubbard model near their respective quantum critical points (QCP). These exact mappings have also other interesting consequences. The second-order quantum phase transition (QPT) between two Ising nematic ($XX$ and $ZZ$) phases of the HQCM, in the JW fermion description, can be interpreted as a QPT between a symmetry-protected topological superconductor and a $Z_{2}$ topologically ordered Mott insulator, lying in the $d=3$ classical Ising universality class.\\
\indent In terms of elementary excitations, the above transformations provide hints for the two different class of excitations present in HQCM: (i) two species of gapped, $1d$ fermionic \textit{spinons} which either moves along $x$ or $y$ direction of the dual square lattice constructed from the bond centers of the original honeycomb lattice. These appear as a novel consequence of low-dimensional symmetries present in the HQCM. (ii) Mapping to $2d$ TFIM further indicates that there exist two different kinds of $2d$ quasiparticles, which are dual analogs of ``droplet'' and ``transverse spin-flip'' excitations of the dual TFIM. To get a better understanding of these quasiparticles, we then investigate the JW fermionic version of HQCM by mapping it to two different $Z_{2}$ lattice gauge theories. In the first case, we incorporate a dimerized description of the $p$-wave Cooper pairs made from JW fermions. These dimers are labeled by Ising link variables and satisfy a local gauge constraint. By incorporating this restriction, we propose a low-energy $Z_{2}$ gauge theory of fluctuating Cooper pairs. In the superfluid phase, doubly occupied sites (or overlapping dimers) proliferate to maintain the ``phase coherence'' of the condensate. As we will see, proliferation of these \textit{doublons} leads to strong pairing of the $Z_{2}$ vortices, which are the topological excitations of the Cooper pair condensate. These paired vortices (or ``vortex-dipoles'') are analogous to the ``droplet'' excitations of the dual TFIM. When the QCP is reached, these paired vortices become gapless and proliferate in the insulating phase. Here, the \textit{doublons} are energetically expensive, which further results in the unbinding of $Z_{2}$ single-vortices, which obtain a gapped $2d$ band dispersion in the insulating phase. Fractionalization is inevitable as this insulator is a condensate of vortex-dipoles rather than single vortices \cite{Balents-Nayak}. \\
\indent To give our description additional rigor, we then re-express the fermionic compass-Hubbard model in the language of $Z_{2}$ slave-spin formalism \cite{Sigrist, Medici}. By incorporating the local constraints exactly in the partition function, we derive an effective low-energy $Z_{2}$ gauge theory coupled to fermionic spinons and Ising chargons, which are reminiscent of doubly occupied sites in the original microscopic model. When the chargon kinetic energy (or hopping) becomes large, these (bosonic) particles will condense, resulting in a superfluid. This is the \textit{Higgs} phase of the effective $Z_{2}$ gauge theory \cite{Fradkin-Shenker}. On the other hand, when chargon hopping is small (or there is a finite charge gap), chargons can be integrated out, resulting in a $Z_{2}$ gauge theory coupled to only $p$-wave paired (gapped) spinons. When the gauge coupling is weak, deconfinement occurs. This is exactly the $Z_{2}$ topologically ordered toric code phase \cite{Kitaev,Wen2}, the whoose existence has also been established using a strong-coupling (large $U/t$) perturbation theory. This deconfined toric code phase hosts gapped $Z_{2}$ vortex-like excitations (called \textit{visons}) which are reminiscent of the $Z_{2}$ single-vortex excitations of the Cooper pair condensate. \\
\indent Finally, we consider a simplified version of the above JW fermionized HQCM (Eq.\eqref{fqcm}) without the $p$-wave pairing term. This is the so-called fermionic QCM, and is of interest in it's own right. Specifically, such models are of interest in the context of cold-atom simulators \cite{Vincent, Nussinov-vdbrink} of fermionic models and quantum information. Generally, with a strong confining potential along $z$, one ends up with a planar optical lattice. In an optical lattice with two fermions in $s$ and $p$-states on each atom, the $s$-shell is inert, while the other fermion occupies either the $p_{x}$ or $p_{y}$ states (the $p_{z}$ state is rendered ineffective by the confinement potential).  Ignoring the inter-atomic $\pi$-type bonding between $p_{x,y}$ states, the $p_{x(y)}$-electrons hop only along $x(y)$ axes on the square lattice. Scattering in the $p$-wave channel can be made strong (and even change it's sign) by tuning into a Feshbach resonance. This leads to an intra-atomic Hubbard interaction.\\
\indent  For a different realization of the fermionic QCM, we go back to the Hamiltonian of HQCM (Eq.\eqref{eq1}), but now replace the $S^{x}_{i}S^{x}_{i+x}$ on the horizontal bonds of the equivalent brick-wall lattice by an XY-type couplings, $\sim (S^{x}_{i}S^{x}_{i+x} + S^{y}_{i}S^{y}_{i+x})$. Though it is hard to conceive of real $d$-band Mott insulators with such an orbital exchange, we notice that such $XY$ couplings along horizontal bonds of the honeycomb lattice, along with $ZZ$ couplings along vertical bonds, can be engineered in JJ arrays by using suitable directional coupling schemes between Josephson qubits \cite{Wendin}. It is easiest to simulate different inter-qubit spin-spin couplings using charge qubits. The $ZZ$-coupling along the vertical bonds of the brickwall lattice is achievable by capactively coupling two charge qubits. The circuit quantum Hamiltonian in this case is $H_{zz}=2e^{2}(C^{-1})_{i,i+z}n_{i}n_{i+z}$. This yields the coupling term $\sim J_{z}\sigma^{z}_{i}\sigma^{z}_{i+z}$. To achieve the $(XX+YY)$-coupling along the horizontal bonds, one can use a Josephson junction to couple the two charge qubits \cite{Siewert}. The Josephson coupling is off-diagonal in the charge basis, and leads to the effective model, $H_{XY}=\lambda(\sigma^{x}_{i}\sigma^{x}_{i+x}+\sigma^{y}_{i}\sigma^{y}_{i+x})$, with $\lambda$ related to the Josephson energy, $E_{J}$ of the coupler. Both of these situations lead to the {\it fermionic} ``orbital'' QCM on a square lattice {\it without} the $p$-wave Cooper pairing term.\\
\indent We then investigate its various low-$T$ phases (at half-filling) using (i) two-particle self-consistent (TPSC) approach \cite{Tremblay} in the weak-coupling regime and (ii) a strong-coupling ($t/U$) expansion valid in the large interaction ($U/t$) regime. In the absence of $p$-wave pairing, the above topological features (of the $p$-wave paired case) are totally replaced by conventional (Landau like) ordering phenomena. In the weak interaction limit, we find a $2d$ Ising orbital pseudo-spin density wave state at very low-$T$, due to Fermi surface nesting, and above some critical crossover temperature scale, the physics of almost decoupled $1d$ chains are observed in the single-particle self-energy. In the large-$U/t$ regime, we obtain a Mott insulating Ising antiferromagnet with subleading XY-type ring exchange interactions. We propose an experimentally realizable (in cold-atom systems) situation where the Ising interaction can be suppressed, which could lead to the realization of critical {\it bose liquid} and associated quantum Lifshitz criticality between two ordered phases. \\
\indent We conclude this section with an outline of the rest of the article. The HQCM is defined in the section \ref{hqcm-def}, where we analyze its symmetries and deduce its different dual descriptions. Based on these dual models, we argue about the nature of ground states and the QPT of HQCM in the same section. In this context, we develop a dimer-like description of the quantum fluctuating Cooper-pairs (made of JW fermions) to describe the above QPT in the subsection \ref{fluctuating}. The low-energy spinon-chargon $Z_{2}$ lattice gauge theory for the JW fermionized HQCM is then derived and subsequently analyzed in subsection \ref{spinon-chargon}. A qualitative discussion of the possible implications of the above results for general spin-orbital (Kugel-Khomskii) models is made in subsection \ref{t2g}. The fermionic QCM without $p$-wave pairing is defined in section \ref{fQCM} and then investigated using TPSC and strong-coupling perturbation theory in sections \ref{tpsc-section} and \ref{str-section} respectively. Finally, a summary of our conclusions is provided in section \ref{conc}.
\section{The Honeycomb quantum compass model}\label{hqcm-def}
%%%%%%%%%%%%%%%%%%%%%%%%%%%%%%%%%%%%%%%%%%%%%%%%%%%%%%%%%%%%%%%%%%%%%%%%%%%%%%%%%%%%%%%%%%%%%%%%%%%%%%%
\indent The HQCM is defined on a honeycomb (or topologically equivalent brick-wall) lattice (see Fig.\ref{fig1}) as
\begin{align}
H=J_{x}\sum_{i\in b}(S^{x}_{i,b}S^{x}_{i+\hat{x},w}+S^{x}_{i,b}S^{x}_{i-\hat{x},w})+J_{z}\sum_{i\in b}S^{z}_{i,b}S^{z}_{i+\hat{z},w}\label{eq1}
\end{align}
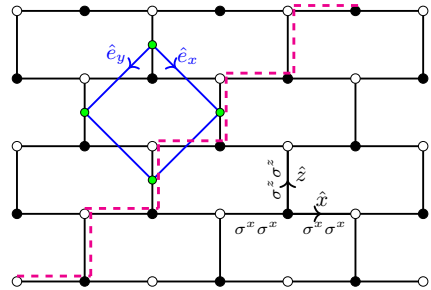
\begin{figure}
\scalebox{0.9}{
\begin{tikzpicture}
 \draw[thick] (0,0) --(6,0);
 \draw[thick] (0,1) --(6,1);
 \draw[thick] (0,2) --(6,2);
 \draw[thick] (0,3) --(6,3);
 \draw[thick] (0,4) --(6,4);
  \draw[thick] (0,1) --(0,2); 
  \draw[thick] (0,3) --(0,4);
 \draw[thick] (1,0) --(1,1);  \draw[thick] (1,2) --(1,3); 
  \draw[thick] (2,1) --(2,2);  \draw[thick] (2,3) --(2,4);
 \draw[thick] (3,0) --(3,1);  \draw[thick] (3,2) --(3,3); 
  \draw[thick] (4,1) --(4,2);  \draw[thick] (4,3) --(4,4);
 \draw[thick] (5,0) --(5,1);  \draw[thick] (5,2) --(5,3); 
  \draw[thick] (6,1) --(6,2);  \draw[thick] (6,3) --(6,4);
 
 %\draw[thick,dashed] (0,0) --(0,1); \draw[thick,dashed] (0,2) --(0,3); \draw[thick,dashed] (1,1) --(1,2); \draw[thick,dashed] (1,3) --(1,4); \draw[thick,dashed] (2,0) --(2,1); \draw[thick,dashed] (2,2) --(2,3); \draw[thick,dashed] (3,1) --(3,2); \draw[thick,dashed] (3,3) --(3,4); \draw[thick,dashed] (4,0) --(4,1); \draw[thick,dashed] (4,2) --(4,3); \draw[thick,dashed] (5,1) --(5,2); \draw[thick,dashed] (5,3) --(5,4); \draw[thick,dashed] (6,0) --(6,1); \draw[thick,dashed] (6,2) --(6,3);
 
 \draw[->-=0.5,thick,black] (4.0,1.0) --(5.0,1.0); \node [above right,black] at (4.3,1.0) {\small $\hat{x}$};
 \node [below right,black] at (4.1,1.0) {\scriptsize $\sigma^{x}\sigma^{x}$};
 \node [below right,black] at (3.1,1.0) {\scriptsize $\sigma^{x}\sigma^{x}$};
 \draw[->-=0.5,thick,black] (4.0,1.0) --(4.0,2.0);
 \node [right,black] at (4.0,1.6) {\small $\hat{z}$};
 \node [left,black,rotate=90] at (3.8,1.96) {\scriptsize $\sigma^{z}\sigma^{z}$};
 
\foreach \i in {0,2,4,6}
\foreach \j in {0,2,4}
 {
  \draw[fill=white] (\i,\j) circle [radius=0.07]; 
  }
\foreach \i in {1,3,5}
\foreach \j in {0,2,4}  
{
  \draw[fill=black] (\i,\j) circle [radius=0.07]; 
  }
\foreach \i in {0,2,4,6}
\foreach \j in {1,3}
 {
  \draw[fill=black] (\i,\j) circle [radius=0.07]; 
  }
\foreach \i in {1,3,5}
\foreach \j in {1,3}
 {
  \draw[fill=white] (\i,\j) circle [radius=0.07]; 
  }
\draw[thick,blue] (3.0,2.5) --(2.0,1.5); \draw[thick,blue] (1.0,2.5) --(2.0,1.5);
\draw[thick, dashed,magenta, line width=1.2] (0.0,0.08) --(1.09,0.08); \draw[thick, dashed,magenta, line width=1.2] (1.09,0.08) --(1.09,1.08); \draw[thick, dashed,magenta, line width=1.2] (1.0,1.08) --(2.09,1.08); \draw[thick, dashed,magenta, line width=1.2] (2.09,1.08) --(2.09,2.08); \draw[thick, dashed,magenta, line width=1.2] (2.09,2.08) --(3.09,2.08); \draw[thick, dashed,magenta, line width=1.2] (3.09,2.08) --(3.09,3.08); \draw[thick, dashed,magenta, line width=1.2] (3.09,3.08) --(4.09,3.08); \draw[thick, dashed,magenta, line width=1.2] (4.09,3.08) --(4.09,4.08); \draw[thick, dashed,magenta, line width=1.2] (4.09,4.08) --(5.09,4.08);

\draw[->-=0.35,thick,blue] (2.0,3.5) --(1.0,2.5); \draw[->-=0.35,thick,blue] (2.0,3.5) --(3.0,2.5);
\node [above left,blue] at (2.8,3.1) {$\hat{e}_{x}$}; \node [above right,blue] at (1.2,3.1) {$\hat{e}_{y}$};
%\node [above left] at (1,3) {$1$}; \node [above left] at (1,2) {$6$}; \node [below] at (2,3) {$2$}; \node [above] at (2,2) {$5$}; \node [above right] at (3,3) {$3$}; \node [below right] at (3,2) {$4$};
\draw[fill=green] (2.0,3.5) circle [radius=0.06]; \draw[fill=green] (1.0,2.5) circle [radius=0.06];
\draw[fill=green] (3.0,2.5) circle [radius=0.06]; \draw[fill=green] (2.0,1.5) circle [radius=0.06];

\end{tikzpicture}}
\caption{The HQCM on a honeycomb or brick-wall lattice, the $b$ and $w$ sub-lattices are shown by filled and empty circles. The dual square lattice is constructed from the mid-points of $zz$ links of the brick-wall lattice and the corresponding unit vectors are shown by blue lines connecting green circles.}\label{fig1}
\end{figure}
Here $b,\ w$ denote black and white sub-lattice indices of the brick-wall lattice. We choose $J_{x}, J_{z}$ to be positive real numbers (the $J_{x,z}<0$ case is related by local spin basis rotations). Apart from the global Ising spin-flip symmetry, this model has two different low-dimensional symmetries: (i) $F_{i}=\sigma^{x}_{i}\sigma^{x}_{i+\hat{z}}$ ($d=0$), and (ii) $\mathcal{S}_{c}=\prod_{i\in C}\sigma^{x}_{i}$ ($d=1$), where the operator string runs along the zig-zag paths, $C$ (one specific path is shown in Fig.\ref{fig1} by dashed, magenta lines). Our first step is to fermionize Eq.~\eqref{eq1} using the JW transformation,
\begin{align}
S^{+}_{i}=c^{\dag}_{i}\prod_{j<i}(1-2c^{\dag}_{j}c_{j})\ ,\ S^{z}_{i}=c^{\dag}_{i}c_{i}-\frac{1}{2}\label{eq2}
\end{align}
Here, the path of the JW operator string runs parallel to the horizontal ($x$) axis of the brick-wall lattice. This results in the following,
\begin{align}
H=&\frac{J_{x}}{4}\sum_{i\in b}\big[ (c_{i}-c_{i}^{\dag})_{b}(c_{i+\hat{x}}+c_{i+\hat{x}}^{\dag})_{w}+(c_{i-\hat{x}}-c_{i-\hat{x}}^{\dag})_{w}\nonumber\\
&\times(c_{i}+c_{i}^{\dag})_{b}\big]+J_{z}\sum_{i\in b}\big(n_{ib}-1/2\big)\big(n_{i+\hat{z},w}-1/2\big)\label{EqJW3}
\end{align}
\indent Introducing Majorana operators, $A^{w}_{i}=(c_{i}-c_{i}^{\dag})_{w}/2i$, $A^{b}_{i}=(c_{i}+c_{i}^{\dag})_{b}/2$, $B^{w}_{i}=(c_{i}+c_{i}^{\dag})_{w}/2$, and $B^{b}_{i}=(c_{i}-c_{i}^{\dag})_{b}/2i$, 
 so that $2iA^{b}_{i}B^{b}_{i}=(n_{ib}-1/2)$, $2iA^{w}_{i}B^{w}_{i}=(1/2-n_{iw} )$. So, we have
\begin{align}
H=iJ_{x}\sum_{i\in b}&(B^{b}_{i}B^{w}_{i+\hat{x}}+A^{w}_{i-\hat{x}}A^{b}_{i})\nonumber\\
&-4J_{z}\sum_{i\in b}(iA^{b}_{i}A^{w}_{i+\hat{z}})(iB^{b}_{i}B^{w}_{i+\hat{z}})\label{eq4}
\end{align}
As shown in Fig.\ref{fig1}, we now go over to a {\it dual} square lattice centered on $r$ (midpoint of a $zz$-bond, ($i,i+\hat{z}$)), whence
\begin{align}
H_{1}=iJ_{x}\sum_{r}(&B^{b}_{r}B^{w}_{r+\hat{e}_{x}}+A^{w}_{r-\hat{e}_{y}}A^{b}_{r})\nonumber\\
&-4J_{z}\sum_{r}(iA^{b}_{r}A^{w}_{r})(iB^{b}_{r}B^{w}_{r})\label{eq5}
\end{align}
\indent This is exactly the model introduced phenomenologically in Ref.\cite{von Oppen}, which we have shown to arise directly from a spin-$1/2$ Hamiltonian with short-range, two-body interactions. Next we combine $A,\ B$ Majorana operators at the ends of each $zz$-bonds into complex fermions as $d_{r}=(A^{b}_{r}+iA^{w}_{r})$, $f_{r}=(B^{w}_{r}+iB^{b}_{r})$, with $(d^{\dag}_{r}d_{r}-1/2)=2iA^{b}_{r}A^{w}_{r}$, $-2iB^{b}_{r}B^{w}_{r}=(d^{\dag}_{r}d_{r}-1/2)$, to derive a model of crossed, interacting $p$-wave Kitaev chains with on-site Hubbard interaction,
\begin{align}
H_{2}=&\frac{J_{x}}{4}\sum_{r}\big[(f_{r}-f^{\dag}_{r})(f_{r+\hat{e}_{x}}+f_{r+\hat{e}_{x}}^{\dag})+(d_{r}-d^{\dag}_{r})\nonumber\\
&\times(d_{r+\hat{e}_{y}}+d^{\dag}_{r+\hat{e}_{y}})\big]+J_{z}\sum_{r}(n_{r}^{d}-1/2)(n_{r}^{f}-1/2)\label{eq6}
\end{align} 
\indent The low dimensional symmetries of this model are: (i) total fermion number parity along each of the chains, i.e., $P_{x}=\prod_{r\in C_{x}} (1-2n^{f}_{r})$, and $P_{y}=\prod_{r\in C_{y}} (1-2n^{d}_{r})$, where $C_{x}$ ($C_{y}$) denote chains along $e_{x}$ ($e_{y}$) axes. These are basically fermionic analogs of the $d=1$ ($Z_{2}$) symmetries ($\mathcal{S}_{c}$) of the HQCM, defined on zig-zag paths of the original brick-wall lattice. (ii) The Hamiltonian Eq.\eqref{eq6} also commutes with the following local ($Z_{2}$) operators: $W_{\scriptsize\square}=\prod_{\langle ij\rangle \in \square} U_{ij}$, with $U_{r,r+\hat{e}_{y}}=iA^{w}_{r}A^{b}_{r+\hat{e}_{y}}$, and $U_{r,r+\hat{e}_{x}}=iB^{b}_{r}B^{w}_{r+\hat{e}_{x}}$ (indicated by the orange-colored Majorana links in Fig.\ref{fig2}). In the HQCM coordinates, $W_{\square}$ is the same as $F_{i}F_{i+2\hat{x}}$. As we will see, these ``gauge-like'' symmetries have important consequences for the low-energy properties of HQCM.\\
\indent We can also recast Eq.~\eqref{eq1} in the above {\it dual} square lattice indices as we did while going from Eq.~\eqref{eq4} to Eq.~\eqref{eq5},
\begin{align}
H=J_{x}\sum_{r}(S^{x}_{r,b}S^{x}_{r+\hat{e}_{x},w}+S^{x}_{r,b}S^{x}_{r+\hat{e}_{y},w})+J_{z}\sum_{r}S^{z}_{r,b}S^{z}_{r,w}\label{eq7}
\end{align}
\indent At this point, we use the Mattis-Nam trick (a two-site version of the Kramers-Wannier duality) (\cite{Mattis}) to obtain a revealing result:
\begin{align}
H_{QI}=J_{x}\sum_{r}\sum_{\alpha=x,y}(2V^{z}_{r+\hat{e}_{\alpha}})W^{x}_{r}W^{x}_{r+\hat{e}_{\alpha}}
+\frac{J_{z}}{2}\sum_{r}W^{z}_{r}\label{eq8}
\end{align}
\indent To get Eq.~\eqref{eq8}, we used $\big(S^{x}_{r,w}, S^{y}_{r,w}, S^{z}_{r,w}\big)=\big(2V^{z}_{r}W^{x}_{r},$ $ 2V^{y}_{r}W^{x}_{r}, -V^{x}_{r}\big)$, and $\big(S^{x}_{r,b}, S^{y}_{r,b}, S^{z}_{r,b}\big)=$ $\big(W^{x}_{r}, -2W^{y}_{r}V^{x}_{r}, -2W^{z}_{r}V^{x}_{r}\big)$, so that $V^{z}_{r}=2S^{x}_{r,b}S^{x}_{r,w}$. Obviously, $[V_{r}^{z}, H_{QI}]=[V^{z}_{r}, H]=0$, for all $r$. \\
\indent Instead of labeling the conserved Ising fields $V^{z}_{r}$ on the lattice sites ($r$), we can imagine them sitting on the lattice links ($r,r+\hat{e}_{\alpha}$). Eq.\eqref{eq8} then becomes a $2d$ TFIM with bond disorder. Via the standard $Z_{2}$ particle-vortex duality \cite{Fisher}, this can be transformed to a $2d$ quantum Ising gauge theory with static background charges $Q_{i}=\prod_{(ij)\in \square}(2V^{z}_{ij})$. In the language of Ising model, $Q_{i}$ measures whether an elementary square plaquette is frustrated or not. While the ($T=0$) ground states belong to a sector with all $Q_{i}=1$ (i.e., no frustration), at any finite $T$, a fraction of $Z_{2}$ ``fluxes'' will always be excited, which will act as an ``emergent disorder'' to the system. As a gauge choice (for $T=0$), we choose $V^{z}_{r,r+\hat{\alpha}}=+1/2$ at every link of the lattice. \\
\indent Different dual manifestations of HQCM, Eqs.~\eqref{eq6} and \eqref{eq8} offer various perspectives on the ground states and the nature of QPTs as a function of $\alpha=|J_{z}|/|J_{x}|$. Mapping to the TFIM (Eq.~\eqref{eq8}) exactly determines the nature, location (in the parameter space) and the universality class of the QPT of these dual models. From the well-known numerical results of TFIM \cite{Pfeuty}, we already know that there is a long-range magnetic order, $\lim_{|r|\rightarrow \infty}|\langle w^{x}_{i}w^{x}_{i+r}\rangle|\sim m_{0}^{2}$ for $\alpha<\alpha_{c}$, which is separated from a disordered paramagnetic phase with $|\langle w^{x}_{i}w^{x}_{i+r}\rangle|\sim e^{-|r|/\xi}$ (for $\alpha>\alpha_{c}$) by a QCP at $\alpha_{c}\approx 3.0$. Here, $m_{0}\sim |\langle w^{x}_{i}\rangle|$ is the average magnetization (along $x$-direction) in the symmetry-broken phase. In the language of HQCM, the broken-symmetry phase of TFIM corresponds to a phase where the nearest neighbor $xx$-correlations dominate: $\langle s^{x}_{r}s^{x}_{r+\hat{e}_{x}}\rangle\sim \langle w^{x}_{r}w^{x}_{r+\hat{e}_{x}}\rangle$. For $\alpha>\alpha_{c}$, $zz$-correlations dominate, i.e., $\langle s^{z}_{r,b}s^{z}_{r,w}\rangle\sim \langle w^{z}_{r}\rangle$. A continuous transition between these Ising nematic phases occurs exactly at $\alpha=\alpha_{c}$.\\
\indent In the fermionic language (Eqs.~\eqref{eq5},\eqref{eq6}), things look even more interesting. A glance at Eq.~\eqref{eq5} shows that it is identical to a phenomenological model \cite{von Oppen} introduced recently, whence {\it all} their results hold here. In particular, a continuous transition from a $p$-wave superconductor to the $2d$ toric code occurs: thanks to Eq.~\eqref{eq8}, this occurs at $\alpha_{c}$ and falls in the $d=3$ classical Ising universality class. Moreover, the $p$-wave superfluid is an example of higher (second-) order topological phase with unpaired MZMs at the high symmetry corners of a finite (but large) square lattice. These MZMs are protected by the lattice reflection and discrete rotation symmetries, which results in degenerate ground states \cite{Foot1}. Symmetry protection of the corner MZMs can be seen easily: each corner mode can be thought of as an intersection of two edges that map into each other under $\pi/2$ rotation and $B^{w}_{r}\leftrightarrow B^{b}_{r}$. If we apply an edge perturbation, say along the horizontal direction, that either closes or reopens the edge band gap, the chiral charge of the corner mode will change by $l$, an integer. But the presence of 4-fold rotation symmetry dictates that an identical change will occur along the other edge. Hence, the total change of the chiral charge at the MZM will be $2l$. Hence, the parity of the number of MZMs per corner must be invariant under strong edge perturbations; this is a proper $Z_{2}$ invariant of the HOTS found above.\\
\indent The (bulk) ground state of the superconducting phase (for $\alpha<\alpha_{c}$) is best approximated by a $p$-wave BCS condensate of Cooper pairs,
\begin{align}
\ket{\psi_{GS}}_{\alpha<\alpha_{c}}\approx \exp{\bigg(\sum_{r_{i}\neq r_{j}}\sum_{\gamma}^{d,f}g_{\gamma}(r_{i}-r_{j})\psi^{\dag}_{r_{i}\gamma}\psi^{\dag}_{r_{j}\gamma}\bigg)}\ket{0}
\end{align} 
where $\psi^{T}_{r,\gamma}=(d_{r},f_{r})$ denotes fermion spinor operator, and $g_{\gamma}(r-r')=\frac{1}{L^{2}}\sum_{k}(v^{\gamma}_{k}/u^{\gamma}_{k})e^{ik\cdot(r-r')}$ is the Cooper-pair wavefunction in the position space. The BCS coherence factors $u^{\gamma}_{k}$ and $v^{\gamma}_{k}$ get trivially renormalized from their non-interacting values due to the on-site Hubbard interaction. Adiabatic continuity arguments are applicable here since the Bogoliubov excitations are always gapped in the $p$-wave superfluid phase.\\
\indent When $J_{z}>>J_{x}$, doubly occupied sites are energetically expensive, which suppresses charge fluctuations in the pair condensate. Hence, phase coherence is completely destroyed, and the system becomes a Mott insulator. Nevertheless, preformed $p$-wave Cooper pairs (between nearest neighbor sites) still exist, as the pairing interaction is explicitly present in the Hamiltonian. A super-exchange like perturbative (in $J_{x}/J_{z}$) expansion yields, to the leading ($4^{\text{th}}$) order, the famed $Z_{2}$ toric (surface) code \cite{toric} (see Appendix \ref{C}, for the details of this calculation),
\begin{align}
H_{eff}\sim -\mathcal{O}\bigg(\frac{J^{4}_{x}}{J_{z}^{3}}\bigg)\sum_{r}\bigg[\prod_{r\in P_{a}}\sigma^{z}_{r}+\prod_{r\in P_{b}}\sigma^{x}_{r}\bigg]\label{Htc}
\end{align}   
\begin{figure}
\scalebox{0.8}{
\begin{tikzpicture}
 
 \draw [blue, thick, line width=2.0] (-0.5,-0.5) --(5.5,5.5);
 \draw [blue, thick, line width=2.0] (-0.5,1.5) --(3.5,5.5);
 \draw [blue, thick, line width=2.0] (-0.5,3.5) --(1.5,5.5);
 \draw [blue, thick, line width=2.0] (1.5,-0.5) --(5.5,3.5);
 \draw [blue, thick, line width=2.0] (3.5,-0.5) --(5.5,1.5);
 \draw [blue, thick, line width=2.0] (-.5,5.5) --(5.5,-.5);
 \draw [blue, thick, line width=2.0] (-.5,3.5) --(3.5,-0.5);
 \draw [blue, thick, line width=2.0] (-0.5,1.5) --(1.5,-0.5);
 \draw [blue, thick, line width=2.0] (1.5,5.5) --(5.5,1.5);
 \draw [blue, thick, line width=2.0] (3.5,5.5) --(5.5,3.5);

\draw [orange, thick, line width=2.0] (1.5,1.5) --(2.5,2.5);
\draw [orange, thick, line width=2.0] (1.5,1.5) --(2.5,0.5);
\draw [orange, thick, line width=2.0] (2.5,0.5) --(3.5,1.5);
\draw [orange, thick, line width=2.0] (2.5,2.5) --(3.5,1.5);

\foreach \i in {-0.5,1.5,3.5,5.5}
\foreach \j in {-0.5,1.5,3.5,5.5}
  {\draw [fill=white,thick] (\i,\j) circle [radius=0.25];
  \draw [fill=green, very thin] (\i-0.09,\j-0.09) circle [radius=0.04];
 \draw [fill=green, very thin] (\i+0.11,\j-0.09) circle [radius=0.04];
 \draw [fill=green, very thin] (\i-0.09,\j+0.09) circle [radius=0.04];
 \draw [fill=green, very thin] (\i+0.11,\j+0.09) circle [radius=0.04];}

%\foreach \i in {-0.5,1.5,3.5}
  %   {\draw [fill=white,thick] (\i,5.5) circle [radius=0.25];
 % \draw [fill=green, very thin] (\i-0.09,5.5-0.09) circle [radius=0.04];
% \draw [fill=green, very thin] (\i+0.11,5.5-0.09) circle [radius=0.04];
% \draw [fill=green, very thin] (\i-0.09,5.5+0.09) circle [radius=0.04];
 %\draw [fill=green, very thin] (\i+0.11,5.5+0.09) circle [radius=0.04];}
 
\foreach \i in {0.5,2.5,4.5}
\foreach \j in {0.5,2.5,4.5}
  {\draw [fill=white, thick] (\i,\j) circle [radius=0.25];
  \draw [fill=green, very thin] (\i-0.09,\j-0.09) circle [radius=0.04];
 \draw [fill=green, very thin] (\i+0.11,\j-0.09) circle [radius=0.04];
 \draw [fill=green, very thin] (\i-0.09,\j+0.09) circle [radius=0.04];
 \draw [fill=green, very thin] (\i+0.11,\j+0.09) circle [radius=0.04];}

\draw [fill=red, very thin] (-0.5-0.09,5.5+0.09) circle [radius=0.06];
\draw [fill=red, very thin] (5.5+0.09,-0.5-0.09) circle [radius=0.06];
\draw [fill=red, very thin] (5.5+0.09,5.5+0.09) circle [radius=0.06];
\draw [fill=red, very thin] (-0.5-0.09,-0.5-0.09) circle [radius=0.06];

\draw [fill=orange, very thin] (1.5+0.11,1.5-0.09) circle [radius=0.06];
\draw [fill=orange, very thin] (1.5+0.11,1.5+0.09) circle [radius=0.06];
\draw [fill=orange, very thin] (3.5-0.11,1.5-0.09) circle [radius=0.06];
\draw [fill=orange, very thin] (3.5-0.11,1.5+0.09) circle [radius=0.06];
\draw [fill=orange, very thin] (2.5-0.11,2.5-0.09) circle [radius=0.06];
\draw [fill=orange, very thin] (2.5+0.11,2.5-0.09) circle [radius=0.06];
\draw [fill=orange, very thin] (2.5-0.11,0.5+0.09) circle [radius=0.06];
\draw [fill=orange, very thin] (2.5+0.11,0.5+0.09) circle [radius=0.06];
%\draw [green, dashed, line width=1.6] (-0.5-0.09,5.5+0.09) --(5.5+0.11,5.5+0.09);
%\draw [orange, dashed, line width=1.6] (-0.5-0.09,-0.5-0.09) --(-0.5-0.09,5.5+0.09);
%\draw [fill=red, very thin] (-0.5-0.09,5.5+0.09) circle [radius=0.04];
 %%%%%%%%%%%%%%%%%%%%%%%%

 \node [thick] at (1.5,.5) {\Large $b$}; \node [thick] at (3.5,.5) {\Large $b$}; 
 \node [thick] at (.5,1.5) {\Large $a$}; \node [thick] at (2.5,1.5) {\Large $a$};
 \node [thick] at (4.5,1.5) {\Large $a$};
 \node [thick] at (.5,3.5) {\Large $a$}; \node [thick] at (2.5,3.5) {\Large $a$}; 
 \node [thick] at (4.5,3.5) {\Large $a$};
 \node [thick] at (1.5,2.5) {\Large $b$}; \node [thick] at (3.5,2.5) {\Large $b$}; 
 \node [thick] at (1.5,4.5) {\Large $b$}; \node [thick] at (3.5,4.5) {\Large $b$};
 
 \node [left] at (1.92,3.9) {\scriptsize $B^{b}$}; \node [left] at (1.34,3.57) {\scriptsize $A^{b}$};
 \node [left] at (2.3,3.36) {\scriptsize $A^{w}$}; \node [left] at (1.68,3.1) {\scriptsize $B^{w}$};
\draw [->-=.54, thick, black] (1.5+0.11,3.5+0.09) to [bend right] (2.5-0.09,4.5-0.09);
\draw [->-=.54, thick, black] (1.5+0.11,3.5-0.09) to [bend right] (2.5-0.09,2.5+0.09);

%\fill[white!40!white] (5.1,-0.5) rectangle (5.5,5.5);
\end{tikzpicture}}
\caption{A pictorial depiction of Hamiltonian Eq.\eqref{eq5} is shown, the circles/sites contain four Majorana fermions with the hopping directions shown by arrows. This is equivalent to a network where four MZMs on each island are coupled to MZMs on neighboring islands in the manner shown. The four corner MZMs that constitute the second order topological superconductor are indicated by red circles.}\label{fig2}
\end{figure}
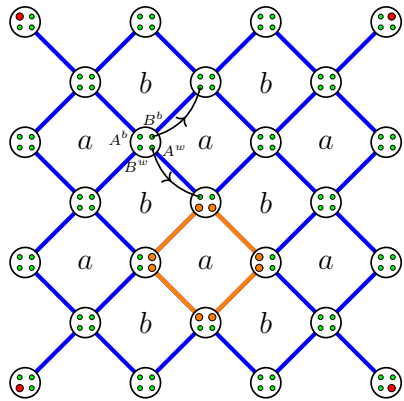
\indent The two different plaquette labels ($P_{a}$ and $P_{b}$) are used to differentiate topologically inequivalent $Z_{2}$ flux ($m$ particle) and $Z_{2}$ charge ($e$ particle) excitations of this model \cite{Kitaev}. These plaquettes are labeled by $a$ and $b$ in Fig.\ref{fig2}. Going beyond the $4$th order Hamiltonian is unnecessary; it turns out that the higher-order corrections do not make any qualitative changes to toric code physics as long as the charge gap is finite. The Eq.\eqref{Htc} is an exactly soluble model. Its ground state subspace corresponds to states with net zero charge and flux. The excitations are Ising anyons, which acquire a non-trivial ($Z_{2}$) Aharonov-Bohm phase (upon exchange) due to long-range statistical interaction between these particles.\\
\indent Slave particle or {\it parton} description also gives exact results for this Hamiltonian \cite{Wen2}. It turns out that the slave-fermion ``mean-field'' theory has the following structure,
\begin{align}
H_{m}=-J_{s}\sum_{i,\gamma}(\psi_{i\gamma}-\psi^{\dag}_{i\gamma})\sigma_{i,i+\gamma}(\psi_{i+\hat{\gamma},\gamma}+\psi^{\dag}_{i+\hat{\gamma},\gamma})\label{Hspinon}
\end{align}
where $J_{s}\sim \mathcal{O}(J^{4}_{x}/J^{3}_{z})$ is the effective spinon hopping in the large $\alpha$ regime. So, the spinons are minimally coupled to $Z_{2}$ gauge fields ($\sigma_{ij}$), the flux of which is determined by the plaquette interactions of Eq.\eqref{Htc}, i.e., $H_{TC}\sim -K\sum_{\square}\prod_{(ij)\in \square} \sigma_{ij}$. Therefore, in the $\alpha>>\alpha_{c}$ limit, the ground state is still described by a $p$-wave pair condensate of spinons, but with an additional Gutzwiller projection that strictly enforces the one fermion per site restriction,
\begin{align}
&\ket{\psi_{GS}}_{\alpha>>\alpha_{c}}\approx \nonumber\\
&\prod_{r}n_{r}(2-n_{r})\exp{\bigg(\sum_{r_{i}\neq r_{j}}\sum_{\gamma}^{d,f}\tilde{g}_{\gamma}(r_{i}-r_{j})\psi^{\dag}_{r_{i}\gamma}\psi^{\dag}_{r_{j}\gamma}\bigg)}\ket{0}\label{Gutzwiller}
\end{align}
 where $n_{r}=n^{d}_{r}+n^{f}_{r}$ measures the total fermion occupation at site $r$. The Cooper-pairs are extremely short-ranged, i.e., $\tilde{g}_{\gamma}\sim e^{-|r|/\xi_{\phi}}$ ($\xi_{\phi}$ is the phase-coherence length), and almost behave like hard-core dimers (see below).  This is nothing but a $p$-wave resonating valance bond (RVB) wavefunction \cite{RVB-PWA} that consists of all possible hard-core dimer coverings of the square lattice. Thus, the toric (surface) code phase is an RVB state in disguise. Therefore, we discover a second-order QPT between a higher-order topological superfluid (with highly overlapping Cooper pairs) and a Mott insulator (MI) with an RVB ground state (with hard-core Cooper pairs). It must be emphasized that this fermionic superfluid-MI QPT is associated with a competition between spinon pairing and localization in the partonized HQCM. The energy gap (to the excitations) vanishes at $\alpha_{c}$ like $\Delta(\alpha)\sim |\alpha_{c}-\alpha|^{\nu}\sim \xi^{-1}$, with $\xi$ the correlation length and $\nu=0.66$ \cite{Pfeuty}. Superfluid-MI transitions are vigorously studied in various contexts \cite{SIbook}. Our results imply that the QPT between Ising nematic ($xx$ and $zz$-ordered) phases in the HQCM is interpretable as a superfluid-MI transition of JW fermionic partons.\\
\indent We will now investigate what the dual models discussed above (Eqs.\eqref{eq6} and \eqref{eq8}) can tell us about the elementary excitations of HQCM. In the fermionic description, when $\alpha<\alpha_{c}$, the excitations are Bogoliubov fermions above the phase-coherent $p$-wave superfluid condensate. In the opposite limit ($\alpha>\alpha_{c}$), the excitations are ($Z_{2}$) charge-neutral fermionic spinons, which are Bogoliubov excitations of the $p$-wave RVB state \cite{Foot2}. Clearly, these gapped fermionic excitations have strictly one dimensional energy dispersion as a consequence of $1d$ fermion number parity conservation along the respective chains (see Eq.\eqref{eq6},\eqref{Hspinon}). On the other side, mapping to $2d$ TFIM indicates that HQCM must possess additional quasiparticles with $2d$ energy bands. These are apparently hidden when naively viewed from the perspective of Eq.\eqref{eq6}. The magnetically ordered phase of TFIM contains excitations having ``droplet''-like structures: A group of oppositely aligned spins inside the ordered background. The smallest-size droplet consists of one flipped spin, created by $w^{z}_{r}$ on the ordered ground state. These single droplet states are highly degenerate ($\sim N$) in the $J_{z}=0$ limit, but obtain a $2d$ dispersive energy band once $J_{z}$ is turned on. Perturbatively (in $\alpha$), the energy dispersion has the following tight-binding structure,
\begin{align}
\epsilon^{D}_{k}\approx J_{x}\bigg[8-\frac{\alpha^{2}}{4}(\cos{k}_{x}+\cos{k_{y}})+\mathcal{O}(\alpha^{4})\bigg] \label{droplet}
\end{align} 
Similarly, when $\alpha>>\alpha_{c}$, the first excited state of the TFIM is created by acting $w^{x}_{r}$ on the transverse-field polarized paramagnetic ground state. The ``transverse spin-flip'' quasiparticles are also mobile in $2d$ space. Perturbatively (in $\bar{\alpha}=2/\alpha$), their energy dispersion reads as follows,
\begin{align}
\epsilon^{t}_{k}\approx \frac{J_{z}}{2}\bigg[&\bigg(1+\frac{4}{9}\bar{\alpha}^{2}\bigg)-\bar{\alpha}(\cos{k}_{x}+\cos{k_{y}}) \nonumber\\
&-\frac{\bar{\alpha}^{2}}{2}(\cos{k}_{x}+\cos{k_{y}})^{2}+\cdot\cdot\cdot\bigg]
\end{align}  
These excitations are fully gapped in their respective ``disordered'' phases, but become gapless at the QCP and proliferate upon entering their respective ``ordered'' phases \cite{Foot3}. What do these $2d$ quasiparticles correspond to in the JW fermionized HQCM (Eq.\eqref{eq6})? \\
\indent To answer that, we follow here two complementary paths, the main objectives of which are to identify the relevant low-energy degrees of freedom (d.o.f.) in these phases and what type of symmetry constraints these emergent d.o.f. follow. In our first approach, the relevant dynamical d.o.f. are charge-$2e$ $p$-wave Cooper pairs, while in the second approach, these are (1) bosonic (Ising) {\it chargons}, (2) fermionic {\it spinons} and (3) emergent $Z_{2}$ gauge fields, which glue (1) and (2) back to the physical fermions. Both of these ``pictures'' ultimately lead to an effective $Z_{2}$ gauge theory description of the above superfluid-MI QPT, a detailed discussion of which will be presented in the upcoming sections (see \ref{fluctuating},\ref{spinon-chargon}).
%%%%%%%%%%%%%%%%%%%%%%%%%%%%%%%%%%%%%%%%%%%%%%%%%%%%%%%%%%%%%%%%%%%%%%%%%%%%%%%%%%%%%%%%%%%%%%%%%%%%%%
%%%%%%%%%%%%%%%%%%%%%%%%%%%%%%%%%%%%%%%%%%%%%%%%%%%%%%%%%%%%%%%%%%%%%%%%%%%%%%%%%%%%%%%%%%%%%%%%%%%%%%
\subsection{An effective low-energy theory of fluctuating $p$-wave Cooper pairs}\label{fluctuating}
Inspired by the short-range RVB theories of High-$T_{c}$ superconductivity \cite{Kivelson1, Kivelson2}, we conceptualize the $p$-wave Cooper pairs of the fermionized HQCM (Eq.\eqref{eq6}) as ``dimers" that are positioned along the nearest-neighbor links of the square lattice. This localized picture of the Cooper pairs (with short coherence length) is approximately valid when the phase coherence of the condensate is relatively weak near the QCP of the superfluid-MI QPT. As a result of the conservation of total fermion number parity along the $d$ and $f$ chains, $f$ ($d$) fermions can create coherent pairs \textit{only} in the $x$- ($y$-) directions. This results in a two-level local Hilbert space for the dimerized Cooper pair number (density) operator on every link of the square lattice. We label $\sigma^{x}_{i,i\pm\gamma}=\mp 1$ ($\gamma=x,y$) if the link $i\pm \gamma$ is occupied (or not occupied) by a charge $2e$ Cooper pair. Similarly, we define the canonically conjugate \textit{phase} operator, $\sigma_{ij}^{z}$ on the links of the (same) lattice.\\
\indent For $\alpha<\alpha_{c}$, charge fluctuations are dominant in the system (thus phase coherence of the condensate is maintained); so, both the horizontal ($xx$) and vertical ($yy$) dimers can overlap simultaneously at a given site. Thus, the dimers are not strictly hard-core in the superfluid phase. Therefore, the Cooper-pair number (or dimer) Hilbert space is subjected to the following local constraint,
\begin{align}
(-1)^{n_{i}}A^{s}_{i}=(-1)^{(n^{d}_{i}+n^{f}_{i})}\prod_{\gamma=\pm x,\pm y}\sigma^{x}_{i,i+\hat{\alpha}}= 1\label{ceq1}
\end{align}
Here, $n^{\gamma}_{i}=0,1$ ($\alpha=d,f$) denotes fermion occupancy at a given site. In this dimer-based description, it is not permissible for identical types of dimers (i.e., $xx$ or $yy$) to overlap at the same site as it violates Pauli's exclusion principle. To account for this, projection operators $P_{\gamma}\sim \prod_{i}(2+\sigma^{x}_{i-\hat{\gamma},i}+\sigma^{x}_{i,i+\hat{\gamma}})$ can be employed to eliminate unphysical states containing two $xx$ (or $yy$) dimers sharing a common lattice site. The dimer Hilbert space resembles that of $Z_2$ gauge theory when the condition in Eq.\eqref{ceq1} is satisfied. However, the BCS ground state is not an eigenstate of $A_s$ due to the large charge fluctuations of the condensate. Instead, it is an eigenstate of $(-1)^{n_i}A^{s}_{i}$ with eigenvalue unity. This leads to a modified $Z_{2}$ gauge invariance \cite{Tupitsyn}, and to account for this, \textit{chargon} degrees of freedom ($s^{x}_{i}$) are now introduced, which are Ising ``matter'' fields that track the number of doubly occupied sites (with two particles or holes),
\begin{align}
s^{x}_{i}\equiv (-1)^{(n^{d}_{i}+n^{f}_{i}-1)}
\end{align}
Therefore, the local constraint satisfied by the dimer Hilbert space is the following,
\begin{align}
s^{x}_{i}A_{i}^{s}\ket{\text{Phys}}=(-1)\ket{\text{Phys}} \label{ceq2}
\end{align}
\begin{figure}
\centering
\scalebox{0.8}{
\begin{tikzpicture}[scale=1.2]
\foreach \i in {0,...,4}  
  { \draw[thick, black] (\i,0) --(\i,4);}
\foreach \i in {0,...,4}  
  { \draw[thick, black] (0,\i) --(4,\i);}
%\draw[dashed, red, line width=1.2] (1.0,0.5) --(0.5,1.0);
%\draw[dashed, red, line width=1.2] (1.0,1.5) --(0.5,1.0);
%\draw[dashed, red, line width=1.2] (1.0,1.5) --(1.5,1.0);
%\draw[dashed, red, line width=1.2] (1.0,0.5) --(1.5,1.0);
\draw[green, line width=1.6] (0.5,1.0) --(1.5,1.0);
\draw[green, line width=1.6] (1.0,0.5) --(1.0,1.5);
\draw[orange, line width=1.4] (2.5,0.0) --(2.0,0.5);
\draw[orange, line width=1.4] (2.5,0.0) --(3.0,0.5);
\draw[orange, line width=1.4] (2.5,1.0) --(3.0,0.5);
\draw[orange, line width=1.4] (2.5,1.0) --(2.0,0.5);
\foreach \i in {0,...,4}
\foreach \j in {0,...,4}
 { \draw[fill=blue] (\j,\i) circle [radius=0.06];}
\node [right, black] at (2.3,0.5) {$B_{p}$};
\node [above right, black] at (1.0,1.0) {$A_{s}$};
\draw[-stealth, red,line width=0.7] (2.5,2.0) -- (2.5,3.0);
\draw[-stealth, red,line width=0.7] (2.5,3.0) -- (2.5,2.0);
\draw[-stealth, red,line width=0.7] (2.0,2.5) -- (3.0,2.5);
\draw[-stealth, red,line width=0.7] (3.0,2.5) -- (2.0,2.5);
\draw[dashed] (2,2.5) ellipse (1/6 and 4/6); \draw[dashed] (3,2.5) ellipse (1/6 and 4/6);
\draw (2.5,2) ellipse (4/6 and 1/6); \draw (2.5,3) ellipse (4/6 and 1/6);
%\foreach \i in {0,...,3}
%\foreach \j in {0,...,3}
% { \draw[fill=green] (\j+0.5,\i+0.5) circle [radius=0.06];}
\end{tikzpicture}}
\caption{An effective $Z_{2}$ gauge theory of fluctuating $p$-wave Cooper-pair dimers, defined on a square lattice: The dimers are represented by an Ising link variable (gauge fields), $\sigma^{x}_{ij}$ and the chargons (matter fields) are denoted by site-centered Ising variable, $s^{x}_{i}$.}\label{Cooper-pair}
\end{figure}
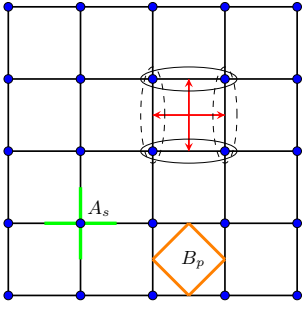
The Cooper-pair hopping, or the fluctuations in their number density, is primarily influenced by the $J_{x}$ term in the microscopic Hamiltonian (Eq.\eqref{eq6}). This term works to make the $Z_{2}$ \textit{phase} (of the condensate) more ``rigid''. Conversely, the $J_{z}$ term is diagonal in the fermion number basis, and it tracks whether a site is occupied by one or two ($xx$ and $yy$) dimers. In the superfluid phase, doubly occupied (electron or holes) sites or chargons are abundant. So, it is expected that the chargon creation operator ($s^{z}_{i}$) will obtain a \textit{long-range} order (under a fixed gauge), i.e., the chargons will condense ($\sim \langle s^{z}_{i}\rangle\neq 0$). On the other hand, in the insulator phase, a finite charge gap will lead to the absence of chargons at low energies.\\
\indent We now construct a minimal (leading-order) Hamiltonian in the $Z_{2}$ ``number-phase'' basis, which adheres to the above-mentioned gauge restriction. The Hamiltonian ($H_{eff}$) is given as follows, 
%-\sum_{ij}g'_{s}(i,j) s^{z}_{i}\bigg[\prod_{\langle mn\rangle\in (r_i,r_j)}\sigma^{z}_{mn}\bigg]s^{z}_{j}%
\begin{align}
&H_{eff}=H_{s}+H_{v}+H_{c}\label{heff-vortex}\\
&H_{s}=-\frac{1}{g_{s}}\sum_{i}s^{x}_{i}-g_{s}\sum_{\langle ij\rangle}s^{z}_{i}\sigma^{z}_{ij}s^{z}_{j}
+\cdot\cdot\cdot\cdot\label{h-s}\\
&H_{v}=-\lambda_{p}\sum_{\square}\prod_{\langle ij\rangle\in \square}\sigma^{z}_{ij}+\cdot\cdot\cdot\cdot\cdot\label{h-v}\\
&H_{c}=-\lambda_{c}\sum_{\langle ij\rangle}\sigma^{x}_{ij}+\lambda_{v}\sum_{r}\sum_{a=1,2}\sigma^{x}_{r}\sigma^{x}_{r+\hat{\gamma}_{a}}+\cdot\cdot\cdot\cdot\label{h-c}
\end{align}
Here, the product in $H_{v}$ is over the phase variables ($\sigma^{z}_{ij}$) on the elementary plaquettes (shown by the label $B_{p}$ in Fig.\ref{Cooper-pair}). The midpoint of the dimer links are labelled by $r$. The $\gamma_{1}$ ($\gamma_{2}$) denote unit vectors connecting the centers of any two neighboring dimer links (shown by red arrows in Fig.\ref{Cooper-pair}). All the coupling constants ($ \lambda_{c}$, $\lambda_{v}$, $\lambda_{p}$, and $g_{s}$) are positive real numbers. Finally, the physical Hilbert space also satisfies the local (gauge) constraint Eq.\eqref{ceq2}.\\

\indent We now examine the various parts of the Cooper-pair Hamiltonian (Eq.\eqref{heff-vortex}). The term proportional to $g_{s}$ represents the (leading order) interaction between the chargons and Cooper-pairs. Apart from it, rest of the terms in our effective Hamiltonian commute with $A_{s}$. The first term of Eq.\eqref{h-s} denotes the energy scale associated with creation of double occupancy (chargons) or the so-called ``charge gap'' (of the insulating phase). Deep inside the insulating phase (when $\alpha>>\alpha_{c}$), charge gap is large, i.e., $g_{s}\rightarrow 0$. In that case, doubly-occupied states (or chargons) becomes rare, i.e. $\langle s^{x}_{i}\rangle\approx 1$. Then, the low energy Hilbert space strictly follows $A^{s}_{i}=-1$ $\forall\ i$. On the other hand, when $g_{s}>>1$, the (bosonic) chargons are likely to condense, i.e., $\langle s^{z}_{i}\rangle\neq 0$; the effective ``phase-only'' Hamiltonian then becomes, $-g_{s}\langle s^{z}_{i}\rangle^{2}\sum_{\langle ij\rangle}\sigma^{z}_{ij}$. It gives rise to a finite {\it phase stiffness} to the Cooper-pair condensate. This is nothing but the $p$-wave superfluid phase. Chargon proliferation also leads to the confinement of the $Z_{2}$ vortex excitations which results in tightly bound $Z_{2}$ flux-pairs (see below). From symmetry grounds, we cannot rule out the possibility of having long-range (gauge invariant) interactions between chargons and phases (terms denoted by the ellipses in Eq. \eqref{h-s}),
\begin{align}
H'_{s}=-\sum_{ij}g'_{s}(i,j) s^{z}_{i}\bigg[\prod_{\langle mn\rangle\in (r_i,r_j)}\sigma^{z}_{mn}\bigg]s^{z}_{j}
\end{align}
To maintain the local gauge invariance, chargons must be coupled via a ``string'' of $\sigma^{z}_{ij}$ that ends at site $i$ and $j$. Using a perturbative series expansion, Fradkin and Shenker earlier found (away from the QCP) that the expectation values of these long-range interactions decay exponentially with the distance $|r_{i}-r_{j}|$ \cite{Fradkin-Shenker}, for any finite (but large) values of $\lambda_{p}$,
\begin{align}
\bigg\langle s^{z}_{i}\prod_{\langle mn\rangle\in (r_i,r_j)}\sigma^{z}_{mn} s^{z}_{j}\bigg\rangle \sim \exp{\big(-2r_{ij}e^{-8\lambda_{p}-2g_{s}}\big)}
\end{align}
Therefore, these higher-order correlations should only trivially renormalize the leading-order chargon hopping amplitude $g_{s}$. The plaquette interaction ($\sim \lambda_{p}$) also maintains the phase stiffness by attaching a finite energy cost of $\mathcal{O}(2\lambda_{p})$ to a single $Z_{2}$ vortex excitation ($\prod_{(ij)\in p}\sigma^{z}_{ij}=-1$) in the pair condensate; this term is analogous to the vortex {\it fugacity} and should be linked to the coupling strength $J_{x}$ of the microscopic theory. Here also, we neglect the higher-order (gauge invariant) interaction between plaquettes; such coupling parameters ($\lambda^{int}_{p}(\tilde{r}_{ij})$) are also expected to decay rapidly with distance and only renormalize the leading-order single-plaquette coupling $\lambda_{p}$.\\
\indent The coupling parameters, $\lambda_{c}$ and $\lambda_{v}$ are responsible for the fluctuations in the Cooper-pair phase ($\sigma^{z}_{ij}$), and they arise mainly from the onsite Coulomb interaction ($\sim J_{z}$) of the microscopic model. Indeed, $\lambda_{c}$ is equivalent to the $Z_{2}$-analog of Cooper-pair charging energy, while $\lambda_{v}$ represents the repulsive pair-interaction between any two dimers along specific links. To formally see this, consider the canonical expressions for the onsite charging energy and repulsive pair interaction: $U_{c}\sum_{r}(\delta n_{r})^{2}$ and $V_{c}\sum_{r,\gamma}(\delta n_{r}+\delta n_{r+\hat{\gamma}})^{2}$, where $\delta n_{r}=n_{r}-\langle n\rangle$. Now, the substitution $\delta n_{r}=(1-\sigma^{x}_{ij})/2$ and neglecting various unimportant constant factors, one gets $\lambda_{c}, \lambda_{v}$ terms as written in Eq.\eqref{h-c}. On general grounds, we can assume $\lambda_{c}$ to be always greater than $\lambda_{v}$.\\

\indent {\it Mott insulating phase :} If we consider $\lambda_{c}>>\lambda_{p}$ and $g_{s}\rightarrow 0$ (so, there is a large charge gap), the (classical) ground state subspace is highly degenerate, consists of all possible arrangements of hard-core dimers ($\sigma^{x}_{ij}=-1$) on the lattice that are consistent with the gauge constraint, $A^{s}_{i}=-1$. The plaquette interaction ($\lambda_{p}$) now acts as a dimer-flip (resonance) operation that tunnels between these degenerate states and lifts the macroscopic degeneracy. The underlying \textit{order by disorder} (OBD) mechanism is exactly equivalent to the one discussed in Ref.\cite{MSF} which treats the quantum dimer model as a limiting case (in the large $\lambda_{c}$ limit) of odd Ising gauge theory. For $\lambda_{c}>>\lambda_{v}$, the Ising gauge theory is always in the confining phase, and the ``odd''-Ising gauge constraint ($A^{s}_{i}=-1$) further results in translation symmetry breaking \cite{Jalabert, MS}. This translation-symmetry broken confined phase is nothing but a columnar valence bond solid (VBS) formed by $p$-wave Cooper pairs. The OBD mechanism prefers the columnar VBS, since this state allows for accessing maximum number of low-energy excitations (created by plaquette flip, $\lambda_{p}$) compared to other VB configurations. On the other hand, strong-coupling perturbation theory (in $1/\alpha=J_{x}/J_{z}$) finds a robust deconfined ($Z_{2}$ topologically ordered) toric code ground state (see Eq.\eqref{Htc}). Therefore, the effective $Z_{2}$ gauge theory of fluctuating Cooper pairs (Eq.\eqref{heff-vortex}) apperantly fails to produce a robust deconfied insulating phase. The impact of dimer-dimer repulsion ($\lambda_{v}$) is not considered yet. This term itself disfavor the columnar VBS state and with the aid of higher-order gauge symmetry-allowed (long-range) $\sigma^{x}\sigma^{x}$ coupling terms (between dimers), can significantly frustrate the confining columnar VBS order and lead to a deconfined liquid phase. In the quantum dimer model literature, $\lambda_{v}$ exactly translates to the ``potential'' term ($v$) that counts the number of {\it flippable} dimers. \\
\indent So far, we have neglected the interaction between fermions and Cooper-pairs, 
\begin{align}
H_{\sigma\text{-}f}=-\lambda_{f}\sum_{\langle ij\rangle, \gamma} (\psi^{\dag}_{i\gamma}\sigma^{z}_{ij}\psi^{\dag}_{j\gamma}+h.c.)
\end{align} 
which takes into account processes in which Cooper-pairs break into two fermions (Bogoliubov excitations) or vice-versa. Although, these are relatively high-energy processes that require an energy of $\mathcal{O}(2\Delta)$ ($\Delta$= Bogoliubov quasiparticle gap), these can lead to a renormalization of the coupling strength, $\lambda_{p}$, especially close to the QCP. Overall, it is likely that there might exist a finite region in the parameter space $(\lambda_{p}/\lambda_{c}, \lambda_{v}/\lambda_{c})$ where the disordered RVB liquid is stable (when $g_{s}\rightarrow 0$). However, explicitly determining this region is a non-trivial task. In order to provide stronger evidence for the existence of the deconfined insulating phase, we will follow a different track in the following section by constructing an effective spinon-chargon $Z_{2}$ gauge theory. This will be elaborated on in the upcoming sections.\\

\indent {\it Superfluid phase :} When $\lambda_{p}>>\lambda_{c}, \lambda_{v}$ (or $J_{x}>>J_{z}$), we are in the $p$-wave superfluid phase. Here, the density of $Z_{2}$ vortices is extremely low, since the vortex fugacity is large. Also, chargon condensation leads to a term, $\sim -h_{x}\sum_{\langle ij\rangle}\sigma^{z}_{ij}$. The finite $h_{x}\sim g_{s}\langle s^{z}_{i}\rangle^{2}$ creates a strong attractive potential which causes two single $Z_{2}$ vortices (of the Cooper-pair condensate) to firmly bind to each other. Fluctuations of the ``string'' (or its length) connecting any two single vortices are highly suppressed, as the confining potential grows rapidly ($\sim 2h_{x}l$) with the string length, $l$. On the other hand, these tightly bound vortex pairs or ``vortex-dipoles'' become almost free since there is no finite length $\sigma^{z}$ string attaching the vortex pairs. A single vortex-dipole obtains a $2d$ band dispersion similar to the single ``droplet'' quasi-particles of the dual Ising model (Eq.\eqref{droplet}). This can be easily verified from a perturbative expansion in $\beta_{c}=\lambda_{c}/\lambda_{p}$ and $\beta_{v}=\lambda_{v}/\lambda_{p}$,
\begin{align}
\epsilon_{k}^{2v}=\epsilon_{0}-2t(\cos{k_{x}}+\cos{k_{y}})-t'\cos{k_{x}}\cos{k_{y}}+\cdot\cdot\cdot
\end{align}
Here, $\epsilon_{0}\sim 4(\lambda_{p}+h_{x})$ is the threshold energy to create a vortex and $t,t'$ are functions of the coupling parameters, $\beta_{c}$ and $\beta_{v}$. To the leading order, $t\sim \mathcal{O}((\beta_{c})^{2})$ and $t'\sim \mathcal{O}((\beta_{c})^{4})$, since we assumed earlier that $\beta_{c}>\beta_{v}$ (or $\lambda_{c}>\lambda_{v}$).\\
\indent However, in the superfluid phase, two vortex-dipoles can decrease their overall energy even more by creating a ``vortex quadrupole'' made up of four $Z_{2}$ vortices located on adjacent plaquettes. Therefore, the physical characteristics of vortex dipoles are very similar to those of single droplet quasiparticles in the dual TFIM. Additionally, a single droplet excitation of the dual TFIM can be imagined as tightly bound $1d$ domain walls (either along $x$ or $y$ directions). In $2d$ TFIM, domain walls (or kinks) cannot separate due to a strong confining potential that scales linearly with separation, which confirms their resemblance to the single vortices. We can infer from these similarities that the energy gap of the vortex-dipoles and the binding energy of the vortex-quadrupole will both disappear precisely at the QCP, leading to the proliferation of free vortex-dipoles when $\alpha\gtrsim \alpha_{c}$. When the system just passes the QCP, off-diagonal long-range order of the chargons also disappears, causing the confining string that connects two single $Z_{2}$ vortices to become invisible. Consequently, single vortices become free. Although free single-vortex excitations are no longer directly available since the vortex-dipoles have already condensed, gapped $Z_{2}$ flux-like excitations (called \textit{visons}) remain as a residue of the $Z_{2}$-vortices within the insulating RVB phase. This will be elaborated further in the subsequent sections. \\

\indent {\it Correlation functions near criticality :} Before closing this section, we make some comments on the behavior of various correlation functions. To investigate the Cooper-pair correlation function in the superconducting phase, we consider the function, $G_{\phi}=\lim_{|r-r'|>>1}\langle \sigma^{z}_{r}\sigma^{z}_{r'}\rangle$; here the phase operators, $\sigma^{z}_{r}$ (defined on lattice links) are Cooper-pair (boson) creation (and annihilation) operators of the effective low energy theory (Eq.\eqref{ceq1}). This operator is not gauge invariant, so its expectation value has to be exactly zero by Elitzur's theorem. We can instead consider $\tilde{G}_{\phi}=\langle \tilde{\sigma}^{z}_{r}\tilde{\sigma}^{z}_{r'}\rangle$, where $\tilde{\sigma}^{z}_{ij}=s^{z}_{i}\sigma^{z}_{ij}s^{z}_{j}$. This is a gauge invariant object. When the chargons condense (in the superfluid phase), there will be a ``long-range order'' of Cooper pairs, which can be easily seen in the unitary gauge fixing, $s^{z}_{i}=1$ (The unitary gauge is defined by the gauge transformation: $s^{z}_{i}\rightarrow G_{i}s^{z}_{i}$ and $\sigma^{z}_{ij}\rightarrow G_{i}\sigma^{z}_{ij}G_{j}$, with $G_{i}=s^{z}_{i}$). So, under a fixed gauge, $\tilde{G}_{\phi}(r-r')\equiv G_{\phi}(r-r')\sim \text{constant}$, which implies phase coherence of the pair-condensate.\\
\indent Well inside the superfluid phase, $\phi\sim |\langle \tilde{\sigma}^{z}_{r}\rangle|\lesssim 1$ (because $\lambda_{v}$ and $g_{s}$ are finite and large). Virtually created vortex-pairs (by quantum fluctuations) will decrease the magnitude of ``phase'' order as we move towards the QCP, i.e., $\lim_{r_{ij}\rightarrow \infty}\langle \tilde{\sigma}^{z}_{i}\tilde{\sigma}^{z}_{j}\rangle \sim (\alpha_{c}-\alpha)^{\alpha_{z}}$. At the QCP, vortex-dipoles (single droplets of the dual TFIM) become gapless. In addition, various correlation functions (of local observables) become scale invariant. Therefore, the single boson correlation function (under the unitary gauge) behaves like a power law, $G_{\phi}(r)\sim 1/r^{\alpha_{b}}$. Inside the insulating phase (for $\alpha>\alpha_{c}$), all the  excitations are once again gapped (since the symmetry group here is $Z_{2}$). Thus, $G_{\phi}\sim e^{-|r-r'|/\xi_{\phi}}$ where $\xi_{\phi}$ is the Cooper-pair phase correlation length. \\
\indent The correlation function of $\sigma^{x}_{ij}$, that creates a $Z_{2}$ vortex dipole at neighboring plaquettes of the lattice (neighboring sites of the dual lattice), should also depict a power-law behavior at the QCP due to scale invariance, [The single-vortex creation operator is defined as $v_{\tilde{i}}\sim \prod_{(ij)\in C}\sigma^{x}_{ij}$, where the product is over lattice links that cut the path $C$ starting from a plaquette center (or dual lattice site) and ending at spatial infinity.]
\begin{align}
G_{2v}(r_{i}-r_{j})=&\langle v_{\tilde{i}}v_{\tilde{i}+\tilde{\alpha}}v_{\tilde{j}}v_{\tilde{j}+\tilde{\beta}}\rangle \nonumber\\
&\sim \langle \sigma^{x}_{i,i+\alpha} \sigma^{x}_{i,j+\beta}\rangle\sim |r_{i}-r_{j}|^{-\alpha_{2v}}
\end{align}
Here, $\bar{\alpha}, \bar{\beta}$ are (orthogonal) dual lattice unit vectors corresponding to $\alpha, \beta$ of the direct lattice and $r_i,\ r_j$ are the midpoints of the lattice links. Inside the superfluid phase ($\alpha<\alpha_{c}$), vortex dipoles are gapped; so, one expects, $G_{2v}(r)\sim e^{-|r|/\xi_{2v}}$. In the insulating phase ($\alpha>\alpha_{c}$), vortex dipoles condense, so $G_{2v}(r)$ becomes a finite constant ($\sim |\langle v_{r}v_{r+\hat{\alpha}}\rangle|^{2}$).\\

\indent Since the effective Hamiltonian (Eq.\eqref{heff-vortex}) is derived solely from symmetry arguments, the coupling constants were treated as phenomenological parameters whose exact relations to the microscopic couplings remain undetermined. However, the qualitative picture discovered here (mainly on the superconducting side) fits nicely to what was guessed earlier from the duality mappings and what will be shown in the next section. The present assertions will be further solidified there. We will derive an effective low-energy $Z_{2}$ gauge plus matter theory (almost exactly) from our microscopic model, Eq.\eqref{eq6}, using a $Z_{2}$ slave-spin representation for the JW fermions.
%%%%%%%%%%%%%%%%%%%%%%%%%%%%%%%%%%%%%%%%%%%%%%%%%%%%%%%%%%%%%%%%%%%%%%%%%%%%%%%%%%%%%%%%%%%%%
%%%%%%%%%%%%%%%%%%%%%%%%%%%%%%%%%%%%%%%%%%%%%%%%%%%%%%%%%%%%%%%%%%%%%%%%%%%%%%%%%%%%%%%%%%%%%
\subsection{JW fermionized HQCM as an effective chargon-spinon $Z_{2}$ gauge theory}\label{spinon-chargon}
An electron, as an elementary particle, carries both charge and spin. However, strongly correlated systems that are composed of a large number of these interacting electrons could allow for situations where the (collective) excitations carry only a fraction of the bare particle's quantum numbers. One such remarkable instance is when the motion of electrons is confined to one dimension. In this case, phase space constraints lead to only collective {\it soliton}-like excitations, which carry either the spin or charge of the bare electron \cite{Giamarchi}. In higher dimensions, the mechanism of quasiparticle fractionalization is different; it requires emergent (compact) gauge fields \cite{savary}. Depending on the parameters of the emergent gauge theory, the system can be in a phase where the fractionalized gauge charges survive as long-lived excitations. To build up its mathematical framework, one often represents the bare degrees of freedom (electron, spin) as a composite of various {\it partons}. We consider here one such description, in which the bare fermion operator ($c_{i\alpha}, c^{\dag}_{i\alpha}$) carrying both charge and spin is described as a composite of $Z_{2}$ slave-spin (carrying only the fermion's charge) and a spinon (carrying the fermion's spin and Fermi statistics) operator \cite{Sigrist, Medici}. The system is considered at half-filling ($\langle n_{i}\rangle=1$) with onsite Hubbard interaction. The local electronic Hilbert space has dimension four (in one orbital case). In the slave-spin language, empty and double-occupied sites are defined to have {\it charge} quantum number $N_{i}=1$ (with respect to the avg. filling, $\langle N\rangle=1$), and the other two singly-occupied states have $N_{i}=0$. Associated with these charge quantum numbers, we define a slave-spin state, defined as $N_{i}=(1-s^{x}_{i})/2$. The $Z_{2}$ charge neutral spinons are created (or destroyed) by $f^{\dag}_{i\alpha}$ ($f_{i\alpha}$) that obeys Fermi statistics. This leads to the following state correspondences between the two descriptions, $\ket{0}\rightarrow \ket{s^{x}=-1,n^{f}_{\sigma}=0}$, $c^{\dag}_{\sigma}\ket{0}\rightarrow \ket{s^{x}=+1,n^{f}_{\sigma}=1}$ (for $\sigma=\uparrow,\downarrow$), and $c^{\dag}_{\uparrow}c^{\dag}_{\downarrow}\ket{0}\rightarrow \ket{s^{x}=-1,n^{f}_{\uparrow}=1,n^{f}_{\downarrow}=1}$. So, the bare electron (creation) operator in this $Z_{2}$-slave-spin basis is defined as $c^{\dag}_{i\alpha}=s^{z}_{i}f^{\dag}_{i\alpha}$. This representation enlarges the local Hilbert space (its dimension becomes twice), so one needs to impose constraints (at every site) to rule out the unphysical states,
\begin{align}
(n^{c}_{i}-1)^{2}=\frac{1-s^{x}_{i}}{2}\ \ \ \text{where }n^{c}_{i}=\sum_{\alpha}c^{\dag}_{i\alpha}c_{i\alpha}
\end{align}
It relates the charge operator of the two representations. Also, $n^{c}_{i}=n^{f}_{i}=\sum_{\alpha}f^{\dag}_{i\alpha}f_{i\alpha}$. An equivalent description of the above constraint is given by the following condition,
\begin{align}
(-1)^{N_{i}+n^{f}_{i}}\ket{\text{Phys}}=(-1)\ket{\text{Phys}}\label{const-slave}
\end{align}
To strictly enforce it, we incorporate a projection operator, $P=\prod_{i}P_{i}$ in the partition function \cite{Senthil-Fisher}, 
\begin{align}
P_{i}=\frac{1}{2}[1+(-1)^{(N_{i}+n^{f}_{i}-1)}]=\frac{1}{2}\sum_{\sigma_{i}=\pm 1}e^{i\frac{\pi}{2}(1-\sigma_{i})(N_{i}+n^{f}_{i}-1)}
\end{align}
The JW fermionized HQCM Hamiltonian (Eq.\eqref{eq6}) now transforms to the following (Notice the change in notations, we now denote the JW fermions, $d_{i}$ and $f_{i}$ by $c_{i\sigma}$ and $f_{i\sigma}$ represent the spinons),
\begin{align}
&H_{t}= t\sum_{i,\alpha}s^{z}_{i}s^{z}_{i+\hat{\alpha}}(f_{i\alpha}-f^{\dag}_{i\alpha})(f_{i+\hat{\alpha},\alpha}+f^{\dag}_{i+\hat{\alpha},\alpha})\label{Ht}\\
&H_{U}=\frac{U}{2}\sum_{i}(n^{c}_{i}-1)^{2}-\frac{UN}{4}=\frac{U}{2}\sum_{i}N_{i}+\text{const.}
\end{align}
where $t=J_{x}/4$ and $U=J_{z}$. Since $H=H_{t}+H_{U}$ commutes with $P$, the partition function could be written as,
\begin{align}
Z=\text{Tr}[e^{-\beta H}P]=\text{Tr}[(e^{-\epsilon H}P)^{M}]\ \ \ \ \epsilon=\beta/M\label{eq-partition}
\end{align}
The partition function will now be expressed in terms of spinons ($f_{i\sigma}$) and $Z_{2}$ chargon ($s_{i}$) degrees of freedom. Slave-spin representation introduces $Z_{2}$ gauge redundancy: $s_{i}\rightarrow \epsilon_{i}s_{i}$ and $f_{i\sigma}\rightarrow \epsilon_{i}f_{i\sigma}$, with $\epsilon_{i}=\pm 1$ leave $c_{i\sigma}$ invariant. From this structure, it is expected that the low-energy effective action should involve emergent $Z_{2}$ gauge fields which will be minimally coupled to the gauge charges (i.e., spinons and chargons). We leave the details of this derivation for the supplemental section (see Appendix \ref{effective-theory}), and directly write the final result for the low-energy partition function, 
\begin{align}
Z\sim \int D[f,\bar{f}]\sum_{\lbrace \sigma_{ij}\rbrace=\pm 1}\sum_{\lbrace s_{i}\rbrace=\pm 1} e^{-S_{B}-S_{f}-S_{c}}\label{effective-action}
\end{align}
The different components of this effective Euclidean action (in $(2+1)d$) are given as follows, (1) Fermionic spinons ($f_{i\sigma}$) moving along crossed $1d$ chains coupled to $Z_{2}$ gauge fields ($\sigma_{ij}$) on the links,
\begin{align}
S_{f}=&\sum_{i,j=i+\tau}\sum_{\sigma=x,y}\bar{f}_{i\sigma}(\sigma_{ij}f_{j\sigma}-f_{i\sigma})\nonumber\\
&+t_{s}\sum_{i\sigma}\big[f_{i\sigma}\sigma_{i,i+\hat{e}_{\sigma}}f_{i+\hat{e}_{\sigma}}-\bar{f}_{i\sigma}\sigma_{i,i+\hat{e}_{\sigma}}f_{i+\hat{e}_{\sigma}}+h.c.\big]\label{spinon-action}
\end{align}
(2) $Z_{2}$ (bosonic) chargons ($s_{i}$) coupled to $Z_{2}$ gauge fields,
\begin{align}
S_{c}=-J_{c}\sum_{\langle ij\rangle}s_{i}\sigma_{ij}s_{j}\label{chargon-action}
\end{align}
(3) A Berry-phase term that arises solely from the fact that our microscopic model (Eq.\eqref{eq6}) is considered at half-filling (i.e., $\langle n^{c}_{i}\rangle=1$),
\begin{align}
S_{B}=\frac{i\pi}{2}\sum_{i,j=i+\tau}(1-\sigma_{ij})\label{berry-term}
\end{align}
In the slave-spin language, $S_{B}$ reflects the particular ``odd''-Ising structure of the local constraint (Eq.\eqref{const-slave}) \cite{MSF}. Another important term, although not explicitly present in Eq.\eqref{effective-action}, would always be generated when the high-energy part of the spinon (or chargon) spectrum (or action) is integrated out,
\begin{align}
S_{g}=-K\sum_{\square}\prod_{(ij)\in \square} \sigma_{ij}+\text{ larger plaquette terms}\label{gauge-term}
\end{align}
where $K$ is some unknown function of $t_{s}$ and $J_{c}$. The larger plaquette couplings are generally suppressed with their size (or surface area) and they mainly renormalize the leading-order gauge interaction $K$.\\
\indent Therefore, the Eqs.\eqref{spinon-action}-\eqref{gauge-term} constitute the low-energy effective theory of JW fermionized HQCM. Since the gauge group is $Z_{2}$, there must be a deconfined phase (with fractionalization) at $T=0$ for some range of $K$. We now discuss the general phase diagram of Eq.\eqref{effective-action} by treating $J_{c}$, $ t_{s}$ (and $K$) as arbitrary coupling parameters and then figure out the portion of the parameter space that matches closely with the expectations derived from the duality (between TFIM and Eq.\eqref{eq6}) and our previous analysis based on quantum fluctuating Cooper pairs.\\

\textit{Mott insulator with deconfinement}: When $J_{c}$ is small, the gapped chargons constitute the high-energy part of the exciation spectrum. After integrating them out of the low-energy theory, the resulting action just becomes $S_{g}$ (due to $Z_{2}$ gauge invariance). So, the total action is now the following,
\begin{align}
S=S_{s}+S_{g}+S_{B}
\end{align}
Away from the critical region, the spinon band is also gapped due to the $p$-wave pairing. The gapped nature of the dual TFIM spectrum further confirms it. If gapless fermionic excitations exist at all, they will only arise at the QCP. So, these gapped spinons could be integrated out of the low-energy theory. This will renormalize of the pure gauge coupling ($K$). Hence, we get the action of odd Ising gauge theory in $(2+1)d$ \cite{MSF}, 
\begin{align}
S=S_{g}+S_{B}
\end{align}
Using the quantum-classical mapping, one gets the corresponding quantum Hamiltonian,
\begin{align}
H_{g}=-\frac{1}{g}\sum_{\square}\prod_{(ij)\in \square}\sigma^{z}_{ij}-g\sum_{\langle ij\rangle}\sigma^{x}_{ij} \label{H-g}
\end{align}
with the constraint $\prod_{\langle ij\rangle\in +}\sigma^{x}_{ij}=-1$ at every site of the square lattice, that arises from the Berry-phase term ($S_{B}$) in the action \cite{MSF}. Here, the dimensionless coupling of the quantum model, $1/g^{2}$ is proportional to $K$. The $2d$ odd Ising gauge theory is dual to the fully-frustrated TFIM on a square lattice \cite{Senthil-Fisher},
\begin{align}
H_{I}=-g\sum_{\langle ij\rangle}v^{z}_{i}\Theta_{ij}v^{z}_{j}-\frac{1}{g}\sum_{i}v^{x}_{j}
\end{align}
with $\prod_{\langle ij\rangle\in \square} \Theta_{ij}=-1$ on every elementary plaquette of the dual square lattice. For $K>K_{c}$ (or $g<g_{c}$), the (dual) Ising model is in the paramagnetic phase. Since the Ising interaction is frustrated, the disordered phase is more stable. As a result, the deconfined phase (for $K>K_{c}$) of the dual odd Ising gauge theory should also be more stable (compared to its even counterpart). On the other hand, when $K<K_{c}$, the Ising spins ($v^{z}_{i}$) obtain long-range order. This ordering happens with the spontaneous breaking of translational and (discrete) rotational symmetry of the square lattice, which results in four-fold degenerate ground states \cite{Jalabert, MS}. In the gauge theory language, vison condensation ($\langle v^{z}_{i}\rangle \neq 0$) implies confinement of $Z_{2}$ gauge charges (spinons and chargons) and due to the Berry phase term ($S_{B}$), the resulting confined phase becomes a translation symmetry-broken valence bond solid (VBS) phase (one of the four degenerate columnar VBS states) formed by nearest neighbor $p$-wave paired spinons. From the strong-coupling $(t/U)$ expansion, we already know that the effective (low-energy) spin Hamiltonian is toric code (Eq.\eqref{Htc}), which is similar to the deconfined phase of the $Z_{2}$ gauge theory. From this connection, we expect that in the insulating phase ($\alpha>\alpha_{c}$ in HQCM), our microscopic Hamiltonian (Eq.\eqref{eq6}) will be rather in the $K>K_{c}$ (deconfined) regime of the low-energy gauge theory.\\
\indent What are the properties of the elementary excitations of this insulating phase? We have gapped $1d$ Bogoliubov spinons. There are also gapped $Z_{2}$ flux excitations, called visons, i.e. states in which $\prod_{(ij)\in \square}\sigma^{z}_{ij}=-1$ on a single plaquette and elsewhere remains one. These have a $2d$ energy-momentum dispersion relation, as one can easily check (perturbatively) from \eqref{H-g}. The visons and spinons (also the chargons, see below) are coupled by long-range statistical interaction: while encircling a vison, a spinon wavefunction obtains a non-trivial phase factor $e^{i\pi}$, due to the spinon-gauge field coupling in $S_{s}$. We now raise the chargon hopping, $J_{c}$ but continue staying inside the insulating phase. The chargons are now included in the full action, 
\begin{align}
S=S_{s}+S_{c}+S_{g}+S_{B}
\end{align}
In the Hamiltonian formulation, $S'=S_{c}+S_{g}+S_{B}$ is described by a $2d$ quantum Ising gauge theory coupled to Ising matter fields (denoting the chargons),
\begin{align}
H_{g+m}=-\frac{1}{\lambda}\sum_{i}s^{x}_{i}&-\lambda\sum_{\langle ij\rangle}s^{z}_{i}\Theta_{ij}\sigma^{z}_{ij}s^{z}_{j}\nonumber\\
&-\frac{1}{g}\sum_{\square}\prod_{(ij)\in \square}\sigma^{z}_{ij}-g\sum_{\langle ij\rangle}\sigma^{x}_{ij} \label{H-gm}
\end{align}
with modified gauge constraint, $s^{x}_{i}\prod_{\langle ij\rangle\in +}\sigma^{x}_{ij}=\prod'_{\langle ij\rangle\in \square}\Theta_{ij}=-1$ at every site of the square lattice (the prime denotes a product over the dual square lattice links). Here, the chargon-gauge coupling $(\sim \lambda^{2})$ is related to $J_{c}$ of Eq.\eqref{chargon-action}. From Eq.\eqref{H-gm}, we see from a simple perturbative calculation, that the chargons obtain a $2d$ band dispersion. This is most easily seen by taking large-$1/g$ limit, i.e. $g\rightarrow 0$ (or $K\rightarrow \infty$). The visons are then completely absent from the low-energy spectrum, and we can fix $\sigma_{ij}=+1$ on all the links. Thus \eqref{H-gm} reduces to a pure Ising model. Perturbatively, these chargons have the energy dispersion, $\epsilon^{c}_{k}\sim \epsilon^{c}_{0}-2\lambda^{2}(\cos{k_{x}}+\cos{k_{y}})+ \mathcal{O}(\lambda^{4})$. Like the spinons, chargons also interact with the visons via a long-range statistical interaction due to Eq.\eqref{chargon-action}.\\
\indent So, there are three different types of gapped quasiparticle excitations, as one expects in a $Z_{2}$ topologically ordered phase: (1) $1d$ fermionic spinons, (2) $2d$ bosonic (Ising) chargons, and (3) $2d$ flux excitations or visons. In the lore of toric code \cite{Kitaev}, these are called $\epsilon$, $e$, and $m$ particles respectively. Due to the long-range statistical interaction, these particles have non-trivial mutual (semionic) statistics, and follows the usual fusion rules of the toric code (for example, $\epsilon =e\times m$).\\

\textit{Superfluid or Higgs phase of the gauge theory}: Let us now make $J_{c}$ very large. We also keep $K>K_{c}$, so that the visons remain gapped. By increasing $J_{c}$ beyond a critical value, the chargons will start to proliferate, i.e., an off-diagonal order of chargons ($\langle s^{z}_{i}\rangle\neq 0$ in \eqref{H-gm}) will be obtained. As a result, the chargon action becomes, $S_{c}=-J_{c}\langle s_{i}\rangle^{2}\sum_{\langle ij\rangle}\sigma_{ij}$. Due to Anderson-Higgs type mechanism, chargon condensation breaks the local $Z_{2}$ gauge symmetry, which results in strong pairing (or confinement) of any two vison quasiparticles. The ``string'' attached to a single vison now costs $\sim 2J_{c}\langle s_{i}\rangle^{2}$ per unit length of the ``string'' and this results in a confining potential for any two $m$ particles. Once the visons are tightly paired and the length of the connecting string vanishes, these (gapped) paired visons become free to hop over the $2d$ lattice, with an energy dispersion, $\epsilon^{2m}_{k}\sim 2J_{c}\langle s_{i}\rangle^{2}-2g^{2}(\cos{k_{x}}+\cos{k_{y}})$.
\\

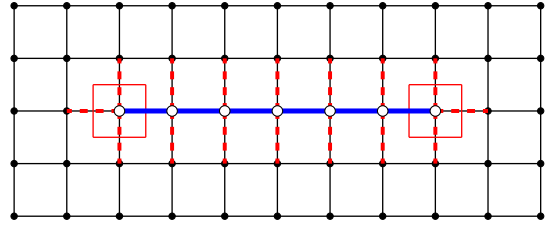
\begin{figure}
\centering
\begin{tikzpicture}[scale=0.7]
\foreach \i in {0,1,...,10}
{
\draw[line width=0.5] (\i,0) --(\i,4);
 }
\foreach \i in {0,1,...,4}
{
\draw[line width=0.5] (0,\i) --(10,\i);
 }
\foreach \i in {0,1,...,10}
\foreach \j in {0,1,...,4}
{
\draw[fill=black] (\i,\j) circle [radius=0.06];}
\draw[line width=0.5, red] (1.5,1.5) --(1.5,2.5); \draw[line width=0.5, red] (1.5,1.5) --(2.5,1.5);
\draw[line width=0.5, red] (2.5,1.5) --(2.5,2.5); \draw[line width=0.5, red] (1.5,2.5) --(2.5,2.5);
\draw[line width=0.5, red] (7.5,1.5) --(7.5,2.5); \draw[line width=0.5, red] (7.5,1.5) --(8.5,1.5);
\draw[line width=0.5, red] (8.5,1.5) --(8.5,2.5); \draw[line width=0.5, red] (7.5,2.5) --(8.5,2.5); 
\draw[dashed, red, line width=1.5] (2,2) --(1,2); \draw[dashed, red, line width=1.5] (2,2) --(2,3);
\draw[dashed, red, line width=1.5] (2,2) --(2,1); \draw[dashed, red, line width=1.5] (3,2) --(3,3);
\draw[dashed, red, line width=1.5] (3,2) --(3,1); \draw[dashed, red, line width=1.5] (4,2) --(4,3);
\draw[dashed, red, line width=1.5] (4,2) --(4,1); \draw[dashed, red, line width=1.5] (5,2) --(5,3);
\draw[dashed, red, line width=1.5] (5,2) --(5,1); \draw[dashed, red, line width=1.5] (6,2) --(6,1);
\draw[dashed, red, line width=1.5] (6,2) --(6,3); \draw[dashed, red, line width=1.5] (7,2) --(7,3);  
\draw[dashed, red, line width=1.5] (8,2) --(8,3); \draw[dashed, red, line width=1.5] (7,2) --(7,1); 
\draw[dashed, red, line width=1.5] (8,2) --(8,1); \draw[dashed, red, line width=1.5] (8,2) --(9,2); 
\draw[blue, line width=1.9] (2,2) --(8,2); 
\foreach \i in {0,1,...,6}
{\draw[fill=white] (2+\i,2) circle [radius=0.1];}
\end{tikzpicture}
\caption{A $Z_{2}$ vortex anti-vortex pair (shown by the red squares) in the chargon condensate separated by a string (blue line) or domain wall of oppositely aligned (matter) Ising spins (white circles) inside the perfectly ordered background.  In a pure Ising model, such domain walls cost energy proportional to their lengths (unhappy bonds are shown by dashed, red lines). Here, the coupling between Ising spins (chargons) and $Z_{2}$ gauge fields liberate the vortices or the domain walls.}\label{vortex}
\end{figure}

\indent The spinon part of the action mostly remains unaffected in this phase. The gapped spinons propagate along the $1d$ chains (due to the exact $1d$ parity symmetries). Since, the single visons are now completely absent at low energies (due to confinement), spinons no longer feel the effect of statistical interaction as described before; encircling a vison pair no longer generate any phase change (of $e^{i\pi}$) in the spinon wavefunction.\\
\indent Now, we discuss the possibility of having finite-energy topological excitations in the Ising chargon condensate. Let us assume for a moment that the gauge fields are totally absent. This can be implemented by taking $K\rightarrow \infty$ which completely prohibits the visons. Then, we make the gauge choice $\sigma_{ij}=1$ on every space-time links. The remaining $(2+1)d$ Ising model (or $2d$ TFIM, in the Hamiltonian description) has only droplet-like excitations that do not have any topological features. Now, consider a string (of length $l>>1$) of oppositely aligned Ising spins ($s^{z}_{i}$) inside the ferromagnetically ordered backgorund of $s_{i}$. Let us define a \textit{current} operator, $J_{i,i+\bar{\alpha}}=(1-s^{z}_{i}s^{z}_{i+\alpha})/2$ on the nearest neighbor links ($i,i+\bar{\alpha}$) of the dual lattice that bisect the original lattice links $i,i+\alpha$; the eigenvalues of $J_{ij}\in \lbrace 0,1\rbrace$ ($J_{ij}$ is analogous to the gradient of the condensate ``phase'', $\sim \nabla\phi$). For the above string of oppositely aligned spins, we can calculate the circulation of this current, defined as a lattice curl of $J_{ij}$: except at the ends of the string (where $\Delta\times J=\pm 1$), everywhere else the circulation or vorticity vanishes. We can think of this as two oppositely charged particles connected by a string. Now, for a pure Ising model, such strings (or domain walls) cost an energy proportional to $l$, so the charges (single vortices) are strictly confined in that case. In our case, since the gauge fields are also acting along the links, we can set $\sigma^{z}_{ij}=-1$ on the links that emanates from the lattice sites placed along the string (see fig.\ref{vortex}). This choice of gauge spins does not create any new visons, and as a result, any two $Z_{2}$ vortices (of the chargon condensate) are no longer confined. Loosely speaking, given that the condensate charge is $e$ (in the units of electronic charge), these single vortices must carry the ``flux quantum'' $hc/e$.\\
\indent In the previous subsection, we constructed a theory of fluctuating charge-$2e$ $p$-wave Cooper pairs and found that phase coherence of the Ising-like Cooper pair condensate demands pairing of $hc/2e$ vortices (of the Cooper pair condensate) that could then freely hop (no longer confined). Here, we see how the $Z_{2}$ gauge fields lead to finite energy Ising vortices (with ``flux quantum'' $hc/e$). In addition, we also discover that $Z_{2}$ visons, which are free (gapped) excitations of the insulating phase, become tightly paired in the superconducting phase. The present gauge theory approach is more microscopic (since we derived it directly from the Hamiltonian) and does not rely on any assumption of the size (coherence length) of the $p$-wave Cooper-pairs. We now describe the above chargon condensate in a dual vortex language by using a $Z_{2}$ version of canonical particle-vortex duality.\\

\textit{Dual representation of the spinon-chargon $Z_{2}$ gauge theory}: We now make use of a discrete ($Z_{2}$) version of the particle-vortex duality \cite{JKKN, Dasgupta} to convert the Ising chargons (particles) to the Ising vortex degrees of freedom. The underlying principle of these duality transformations lies in the (local) conservation of the particle current, which is a direct consequence of some global symmetry present in the matter sector of the theory. For example, in our case, the Ising (matter) interaction on every link ($ij$) could be written as follows \cite{Senthil-Fisher},
\begin{align}
\exp{\big(J_{c}s_{i}\sigma_{ij}s_{j}\big)}&=C\sum_{m_{ij}=0,1}\exp{\big(-2\tilde{J}_{c}m_{ij}\big)}\nonumber\\
&\times\exp{\big[i\frac{\pi}{2}m_{ij}(s_{i}-s_{j}+1-\sigma_{ij})\big]}
\end{align}
where, $C=\cosh(J_{c})$ and $\tanh{(\tilde{J}_{c})}=e^{-2J_{c}}$. Then the matter fields ($s_{i}$) can be easily traced out. After performing the trace, one gets
\begin{align}
\text{Tr}_{s_{i}}&\exp{\big(J_{c}s_{i}\sigma_{ij}s_{j}\big)}=\text{Tr}_{m_{ij}}\bigg[\prod_{i}(-1)^{\frac{\Delta\cdot m}{2}}\delta_{ (-1)^{\Delta\cdot m},1} \bigg]\nonumber\\
&\times \exp{\bigg(-2\tilde{J}_{c}\sum_{(ij)}m_{ij}+i\frac{\pi}{2}\sum_{(ij)} m_{ij}(1-\sigma_{ij})\bigg)}
\end{align}
where $\Delta\cdot m=\sum_{\alpha}(m_{i,i+\hat{\alpha}}-m_{i-\hat{\alpha},i})$ is lattice divergence. Also, $\sum_{i}(\Delta\cdot m)_{i}=0$. The Kronecker delta constraint implies $\Delta\cdot m=$ even integer. This represents a \textit{particle current} conservation law. The current $m_{ij}$ could be defined (in terms of matter fields) as $m_{ij}=[1-e^{i\pi(s_{i}-s_{j})/2}]/2$. Therefore, if the Ising spins are parallel on a given link, the particle current is zero; otherwise, it will be one. For every $m_{ij}$, one can define a $Z_{2}$ current, $(-1)^{m_{ij}}$ that takes values $\pm 1$. This $Z_{2}$ current can also be written as a $Z_{2}$ flux of some dual (vortex) variable, $\theta_{ij}$
\begin{align}
e^{i\pi(s_{i}-s_{j})/2}&=(-1)^{m_{ij}}\equiv \prod_{(ij)\in \square} \theta_{ij} \nonumber\\
&\ \ \Rightarrow\ \ \ m_{ij}=(1-\prod_{\square} \theta_{ij})/2
\end{align}
where the product is on the dual lattice plaquette, which is pierced by the link $(ij)$ of the original lattice. The above divergence-free condition on the particle current suggests that the $m_{ij}$ form closed loops of various sizes. The time component of the current, $m_{i,i+\hat{\tau}}$ denotes the number (density) of chargons (or net electric charge with respect to the average filling). So, a $Z_{2}$ flux (or vortex) generated by the spatial circulation (or curl) of vortex ``phase'' ($\theta_{ij}$) is nothing but a chargon with electric charge $e$. After replacing $m_{ij}$ by dual vortex variables, one gets the following dual action (where charges are replaced by the vortices),
\begin{align}
S_{dual}=-\tilde{J}_{c}\sum'_{\square}\prod_{(ij)\in \square} \theta_{ij}+\frac{i\pi}{4}\sum_{\langle ij\rangle} \bigg(1-\prod'_{\square}\theta_{ij}\bigg)\nonumber\\
\times (1-\sigma_{ij})+S_{s}+S_{B}
\end{align}
Here, the prime symbol over the above sum and product denotes that the corresponding operations are being carried out on the dual lattice plaquettes. For the second ($Z_{2}$ Chern-Simons (CS) like) term, dual lattice plaquettes are those that are pierced by the direct lattice link $(ij)$. Away from the critical point, fermionic spinons are gapped due to $p$-wave pairing. So, one could integrate them out, generating a pure gauge interaction term (Near the gapless critical point or line, one can still integrate out the high-energy part of the spinon band, contributing to the pure gauge term. But an action representing the low-energy gapless spinons must be considered in that case). So, we get an effective action of $Z_{2}$ vortices and $Z_{2}$ gauge fields with a mutual CS term coupling the vortex ``phase'' to the gauge fields (or vison ``phase''),
\begin{align}
S_{dual}[\theta,\sigma]&=-\tilde{J}_{c}\sum'_{\square}\prod_{(ij)\in \square} \theta_{ij}-K \sum_{\square}\prod_{(ij)\in \square}\sigma_{ij}\nonumber\\
&+\frac{i\pi}{4}\sum_{\langle ij\rangle} \bigg(1-\prod'_{\square}\theta_{ij}\bigg)(1-\sigma_{ij})+S_{B}
\end{align}
Physically, the CS term encodes the information of the statistical phase ($e^{i\pi}$) which a chargon (in the dual picture, a vortex in the vortex phase) accumulate by encircling a vison (or vice-versa). The Berry phase term, $S_{B}$ can also be written in CS form \cite{Senthil-Fisher}, 
\begin{align}
S_{B}=\frac{i\pi}{4}\sum_{j=i+\hat{\tau}}\bigg(1-\prod'_{\square}\Theta_{ij}\bigg)(1-\sigma_{ij})
\end{align}
where we introduce a new (non-dynamical) Ising variable, $\Theta_{ij}$ that satisfies $\prod_{\langle ij\rangle\in \square}\Theta_{ij}=-1 (+1)$ only on the spatial (temporal) plaquettes. Now the CS term (or more appropriately $e^{-S_{CS}}$) remains invariant under the interchange $\theta\leftrightarrow \sigma$ (similarly $\sigma\leftrightarrow \Theta$ for $S_{B}$). Then, following the inverse steps of the above discrete particle-vortex duality, we can exactly perform the trace over $\sigma$, which results in the following,
\begin{align}
S_{dual}[v,\theta]=-\tilde{J}_{c}\sum'_{\square}\prod_{(ij)\in \square} \theta_{ij}-\tilde{K}\sum'_{\langle ij\rangle}v_{i}\theta_{ij}\Theta_{ij}v_{j}\label{S-dual}
\end{align}
Here, all the summations and products are over the dual lattice links and sites. As already mentioned, on spatail (temporal) plaquettes $\prod_{\langle ij\rangle\in \square}\Theta_{ij}=-1 (+1)$. The dual coupling constant is given by $\tanh(\tilde{K})=e^{-2K}$. The dual action (Eq.\eqref{S-dual}) describes visons ($v_{i}$) coupled to $hc/e$ vortices ($\theta_{ij}$). Equivalently, we could also work with the $2d$ quantum Hamiltonian description of  $S_{dual}$. It can also be derived directly from the Eq.\eqref{H-gm}, using a set of unitary transformations (See Appendix \ref{gauge-matter-duality}). The final result is the following,
\begin{align}
H^{dual}=-\frac{1}{\lambda}&\sum_{\square}\prod_{(ij)\in \square} \theta^{z}_{ij}-\lambda\sum_{(ij)}\theta^{x}_{ij}\nonumber\\
&-\frac{1}{g}\sum_{i}v^{x}_{i}-g\sum_{(ij)}v^{z}_{i}\Theta_{ij}\theta^{z}_{ij}v^{z}_{j} \label{H-dual}
\end{align}
with the following local constraint, $v^{x}_{i}\prod_{(ij)\in +}\theta^{x}_{ij}=\prod_{\langle ij\rangle\in \square}\Theta_{ij}=-1$. \\

When $J_{c}>(J_{c})^{crit}$ ($\lambda>\lambda_{c}$) and $K>K_{c}$ ($g<g_{c}$), the dual gauge theory (in terms of $\theta_{ij}$) is in the confining phase. So, the $Z_{2}$ monopoles of this dual gauge theory (or vortices in the vortex-phase condensate) must have condensed. These monopoles are nothing but the chargons of the original theory. The Ising model of the visons ($v_{i}$) is in the paramagnetic phase (for $K>K_{c}$ or $g<g_{c}$); so, in the ground state $\langle v^{z}_{i}\rangle=0$. Now, consider two such visons separated by a distance $l$; to create this pair we need to act the following gauge-invariant operator, $O=v^{z}_{i}\prod_{\langle mn\rangle\in [r_i,r_j]}\theta^{z}_{mn}v^{z}_{j}$ on the ground state. Since the dual gauge field is in the confining phase, the ``string'' connected between two visons will cost energy proportional to its length. Therefore, the visons will be tightly paired (to cancel the string) as we discussed before. \\
\indent In the confining phase of this dual gauge theory, a gauge invariant ``box''-like excitations can be found: to clearly see it, consider the limit $\lambda\rightarrow \infty$. The ground state is $\prod_{(ij)}\ket{\theta^{x}_{ij}=+1}$. On acting with the plaquette term (proportional to $1/\lambda$), one creates a box or plaquette of oppositely aligned $\theta^{x}$ spins (of energy $8\lambda$). The box excitation could be created at any of the spatial plaquettes of the square lattice. From the first-order degenerate perturbation theory, energy dispersion of a single box excitation turns out to be the following,
\begin{align}
\epsilon_{k}^{b}-\epsilon_{0}^{b}\approx-2t_{b}(\cos{k_{x}}+\cos{k_{y}})+..., 
\end{align}
where $\epsilon_{0}^{b}=8\lambda$, $t_{b}=1/(24\lambda^{3})$. What do these box particles physically mean? In the dual gauge theory, $\theta^{z}_{ij}$ (vortex phase) is analogous to a vector potential, i.e. $\sim e^{iA_{ij}}$ with $A_{ij}\in [0,\pi]$ and similarly, the canonically conjugate $\theta^{x}_{ij}\sim e^{i\pi E_{ij}}$, with $E_{ij}\in [0,1]$ represents a discrete electric field. So, a flipped plaquette represents a closed electric loop or circulation. If the circulation of vector potential represents chargons (charge $e$), then from duality, a circulation of the conjugate electric field should correspond to the vortices (of flux $hc/e$). Confining phase of this dual gauge theory also helps in clustering the vortices (or elementary electric loops) as the energy of these loops depends on their length (or perimeter). Finally, when the dual gauge theory enters into the deconfined phase (i.e. $\lambda<\lambda_{c}$) and if $g<g_{c}$ (so visons are still gapped), electric loops of all sizes proliferate. This confirms that the insulating phase of the orginial theory is a condensate of $hc/e$ vortices. This provides a complementary picture of chargons and vortices, now in a dual description of the spinon-chargon gauge theory.
%%%%%%%%%%%%%%%%%%%%%%%%%%%%%%%%%%%%%%%%%%%%%%%%%%%%%%%%%%%%%%%%%%%%%%%%%%%%%%%%%%%%%%%%%%
%%%%%%%%%%%%%%%%%%%%%%%%%%%%%%%%%%%%%%%%%%%%%%%%%%%%%%%%%%%%%%%%%%%%%%%%%%%%%%%%%%%%%%%%%%%%
\subsection{Consequences in TMO with $t_{2g}$ orbital degeneracy}\label{t2g}
The $d=3$ Ising criticality in the orbital-only HQCM also has interesting consequences for magnetism within a Kugel-Khomskii (KK) framework \cite{KK} for Mott insulators with $t_{2g}$-orbital degeneracy.  Specifically, consider a brickwall (honeycomb) lattice with two, degenerate $d_{xz}, d_{yz}$ orbitals per site. An electron in the $d_{xz}$ orbital can only hop along the $x$-direction with strength $t_{xz}$, while that in the $d_{yz}$ orbital can only hop along the $ZZ$-bonds with strength $t_{yz}$. In the strong coupling limit, where the intra-orbital $U$, the inter-orbital $U'$ and the Hund coupling are large compared to $t_{xz}, t_{yz}$, (we ignore the much smaller inter-orbital $d_{xz}$-$d_{yz}$ hopping in this geometry) the effective KK Hamiltonian for each transition-metal ion in $d^{1}, d^{4}$ states reads
\begin{align}
H_{KK}=\sum_{i}[J_{xx}S^{x}_{i}S^{x}_{i+x}({\bf \sigma}_{i}.{\bf \sigma}_{i+x}) +        J_{zz}S^{z}_{i}S^{z}_{i+z}({\bf \sigma}_{i}.{\bf \sigma}_{i+z})]
\end{align}
where the {\it spin} operators are $|{\bf \sigma}_{i}|=1/2$ ($d^{1}$) and $|{\bf \sigma}_{i}|=1$ ($d^{4}$). When $\alpha <\alpha_{c}$, XX-orbital order (OO) is obtained (i.e., $\langle S^{x}_{i}S^{x}_{i+x}\rangle > \langle S^{z}_{i}S^{z}_{i+z}\rangle$). If we further consider the case where orbital and magnetic ordering scales are separated and OO precedes magnetic ordering, the {\it effective} spin-spin super-exchange is set by the OO: explicitly, 
\begin{align}
H_{\sigma}=J^{\sigma}_{x}\sum_{i}{\bf \sigma}_{i}.{\bf \sigma}_{i+x} +
           J^{\sigma}_{z}\sum_{i}{\bf \sigma}_{i}.{\bf \sigma}_{i+z}
\end{align}
with $J^{\sigma}_{x}=J_{x}\langle S^{x}_{i}S^{x}_{i+x}\rangle > J^{\sigma}_{z}=J_{z}\langle S^{z}_{i}S^{z}_{i+z}\rangle$. For $\alpha > \alpha_{c}$, however, we have $J^{\sigma}_{x} < J^{\sigma}_{z}$. In both cases, the discrete lattice rotational symmetry is spontaneously broken, but $SU(2)$   spin-rotational symmetry is left unbroken by the OO which sets in before magnetic order. Moreover, due to the symmetry-adapted orbital-lattice coupling, onset of OO causes a structural transition that either elongates ($\alpha < \alpha_{c}$) or shrinks ($\alpha > \alpha_{c}$) each ZZ-bond of the brickwall lattice. Phenomenologically, this is driven by coupling the change in bond length, $\delta_{i,i+a}=(x_{i}-x_{i+a})$ to the appropriate $S^{a}_{i}S^{a}_{i+a}$. The orbital-lattice coupling Hamiltonian is given by,
\begin{align}
H_{o-l}= g\sum_{i}[\delta_{i,i+x}S^{x}_{i}S^{x}_{i+x} + \delta_{i,i+z}S^{z}_{i}S^{z}_{i+z}]
\end{align}
Adding the elastic term, $H_{l}=\frac{1}{2}M\Omega^{2}\sum_{i,a}\delta^{2}_{i,i+a}$ with $M$ denoting the ionic mass and $\Omega$ the ionic vibrational frequency, this Hamiltonian is reminiscent of the well-known spin-Jahn-Teller (JT) model \cite{Tchernyshov}. Minimizing $H_{o-l}+H_{l}$ with respect to the bond lengths, we find that $\langle\delta_{i,i+z}\rangle = -\frac{g}{M\Omega^{2}}\langle S^{z}_{i}S^{z}_{i+z}\rangle$. This is re-expressible as
\begin{align}
\langle\delta_{i,i+z}\rangle = -\frac{g}{M\Omega^{2}}&[\langle (S^{z}_{i}S^{z}_{i+z}+S^{x}_{i}S^{x}_{i+x})\rangle \nonumber\\
&- \langle (S^{x}_{i}S^{x}_{i+x}-S^{z}_{i}S^{z}_{i+z})\rangle]
\end{align}
The first term transforms as a singlet of $C_{6v}$, and is just related to
the expectation value of the energy density.  It is the second term above, 
$\frac{g}{M\Omega^{2}}\langle (S^{x}_{i}S^{x}_{i+x}-S^{z}_{i}S^{z}_{i+z})\rangle$, transforming as one of the $E_{2u}$ irreps of $C_{6v}$, which causes the structural change in the OO states. As expected, $\delta_{i,i+z} >0$ for $\alpha < \alpha_{c}$ and $\delta_{i,i+z} < 0$ for $\alpha > \alpha_{c}$.
It is the shortening of the ZZ-bonds in the latter case that facilitates the nematic-valence bond solid (nVBS) order in the ${\bf \sigma}=1/2$ case.\\
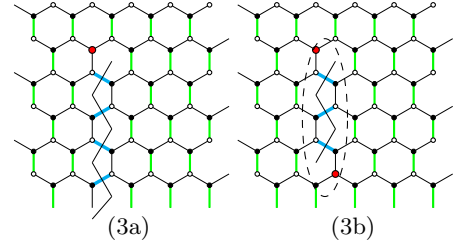
\begin{figure}
\begin{tikzpicture}[scale=0.3]
  \foreach \i in {0,...,4} 
  \foreach \j in {0,...,2} {
  \foreach \a in {30,150,-90} \draw[line width=0.2] (2*cos{30}*\i,3*\j) -- ++(\a:1);
  }
  \foreach \i in {0,2,3,4}
  \foreach \j in {0,1,2}{
  \draw[thick, green] (2*cos{30}*\i,3*\j) -- ++(-90:1);}
  \foreach \i in {-1,...,3} 
  \foreach \j in {0,...,2} {\foreach \a in {30,150,-90} \draw[line width=0.2] (2*cos{30}*\i+cos{30},3*\j+1+sin{30}) -- ++(\a:1);}
  \foreach \i in {-1,0,2,3} 
  \foreach \j in {0,1} {\foreach \a in {-90} \draw[thick, green] (2*cos{30}*\i+cos{30},3*\j+1+sin{30}) -- ++(\a:1);}
  \foreach \i in {-1,0,1,2,3} 
  \foreach \j in {2} {\foreach \a in {-90} \draw[thick, green] (2*cos{30}*\i+cos{30},3*\j+1+sin{30}) -- ++(\a:1);}
  \draw[thick, cyan, line width=1.3] (2*cos{30},5) --(3*cos{30},5-sin{30});
  \draw[thick, cyan, line width=1.3] (3*cos{30},4-sin{30}) --(2*cos{30},4-2*sin{30});
  \draw[thick, cyan, line width=1.3] (2*cos{30},3-2*sin{30}) --(3*cos{30},3-3*sin{30});
  \draw[thick, cyan, line width=1.3] (3*cos{30},2-3*sin{30}) --(2*cos{30},2-4*sin{30});
  \draw[line width=0.3] (3*cos{30},-0.5) --(2*cos{30},sin{30}+0.5);
  \draw[line width=0.3] (2*cos{30},sin{30}+0.5) --(3*cos{30},2.5);
  \draw[line width=0.3] (3*cos{30},2.5) --(2*cos{30},4.0);
  \draw[line width=0.3] (2*cos{30},4.0) --(3*cos{30}, 5.5);
  \draw[line width=0.3] (3*cos{30},-0.5) --(2*cos{30},-1.5);
  \foreach \i in {0,...,4} 
  \foreach \j in {0,3,6} { \draw[fill=black] (2*cos{30}*\i,\j) circle [radius=0.1];
  \draw[fill=white] (2*cos{30}*\i-cos{30},\j+sin{30}) circle [radius=0.1];
  \draw[fill=black] (2*cos{30}*\i-cos{30},\j+1+sin{30}) circle [radius=0.1];}
  \foreach \i in {0,...,4} 
  \foreach \j in {2,5,8} { \draw[fill=white] (2*cos{30}*\i,\j) circle [radius=0.1];}
  \draw[fill=red] (2*cos{30},6) circle [radius=0.15];
  \node [right, black] at (2.0,-1.9) {(3a)};
\end{tikzpicture}
\begin{tikzpicture}[scale=0.3]
  \foreach \i in {0,...,4} 
  \foreach \j in {0,...,2} {
  \foreach \a in {30,150,-90} \draw[line width=0.2] (2*cos{30}*\i,3*\j) -- ++(\a:1);
  }
  \foreach \i in {0,2,3,4}
  \foreach \j in {1,2}{
  \draw[thick, green] (2*cos{30}*\i,3*\j) -- ++(-90:1);}
  \foreach \i in {1,0,2,3,4}
  \foreach \j in {0}{
  \draw[thick, green] (2*cos{30}*\i,3*\j) -- ++(-90:1);}
  \foreach \i in {-1,...,3} 
  \foreach \j in {0,...,2} {\foreach \a in {30,150,-90} \draw[line width=0.2] (2*cos{30}*\i+cos{30},3*\j+1+sin{30}) -- ++(\a:1);}
  \foreach \i in {-1,0,2,3} 
  \foreach \j in {0,1} {\foreach \a in {-90} \draw[thick, green] (2*cos{30}*\i+cos{30},3*\j+1+sin{30}) -- ++(\a:1);}
  \foreach \i in {-1,0,1,2,3} 
  \foreach \j in {2} {\foreach \a in {-90} \draw[thick, green] (2*cos{30}*\i+cos{30},3*\j+1+sin{30}) -- ++(\a:1);}
  \draw[thick, cyan, line width=1.3] (2*cos{30},5) --(3*cos{30},5-sin{30});
  \draw[thick, cyan, line width=1.3] (3*cos{30},4-sin{30}) --(2*cos{30},4-2*sin{30});
  \draw[thick, cyan, line width=1.3] (2*cos{30},3-2*sin{30}) --(3*cos{30},3-3*sin{30});
 % \draw[thick, cyan, line width=1.3] (3*cos{30},2-3*sin{30}) --(2*cos{30},2-4*sin{30});
 % \draw[thick, dashed, line width=1.1] (3*cos{30},-0.5) --(2*cos{30},sin{30}+0.5);
  \draw[line width=0.3] (2*cos{30},sin{30}+0.5) --(3*cos{30},2.5);
  \draw[line width=0.3] (3*cos{30},2.5) --(2*cos{30},4.0);
  \draw[line width=0.3] (2*cos{30},4.0) --(3*cos{30}, 5.5);
  \foreach \i in {0,...,4} 
  \foreach \j in {0,3,6} { \draw[fill=black] (2*cos{30}*\i,\j) circle [radius=0.1];
  \draw[fill=white] (2*cos{30}*\i-cos{30},\j+sin{30}) circle [radius=0.1];
  \draw[fill=black] (2*cos{30}*\i-cos{30},\j+1+sin{30}) circle [radius=0.1];}
  \foreach \i in {0,...,4} 
  \foreach \j in {2,5,8} { \draw[fill=white] (2*cos{30}*\i,\j) circle [radius=0.1];}
  \draw[fill=red] (2*cos{30},6) circle [radius=0.15]; \draw[fill=red] (3*cos{30},sin{30}) circle [radius=0.15];
  \draw[dashed] (2*cos{30}+0.4,3.0) ellipse (1 and 3.5);
  \node [right, black] at (2.0,-1.9) {(3b)};
\end{tikzpicture}
\caption{The effect of doping one (left figure) and two (right figure) holes in the nVBS state. In low lying excited states, one hole creates a ``defect string" which has a staggered VBS character. Doping a second hole cures this string as shown, allowing the composite of two holes (as shown by the shaded ellipse in the right figure) to hop coherently in the nVBS background.}\label{rvb3}
\end{figure}
This nVBS ordered state is the $d=2$ analogue of the $d=1$ spin-Peierls phase widely studied in $S=1/2$ chains. There are a few $S=1/2$ honeycomb magnets, {\it e.g}, MgVO$_{3}$ (\cite{MgVO3}), which show an explicit nVBS ground state: however, in this case, it arises from cation dimerization. In our case, it arises from the ZZ-OO, which may now be understood as an OO-induced {\it spin dimerization}. Such states have been discussed earlier \cite{Khomskii}. In analogy with the OO-induced bond shortening, onset of nVBS order will result in further ZZ-bond shortening via the spin-phonon interaction, wherein the displacements $\delta_{i,i+z}$ couple to the VB ${\bf \sigma}_{i}.{\bf \sigma}_{i+z}$, on each $ZZ$-bond. Given the mathematical similarity to the JT effect in $d$-band oxides, one may further add ``cubic'' (anharmonic) terms to $H_{o-l}+H_{l}$ : this will drive the paramagnetic-nVBS transition first-order \cite{Maekawa-book}.  Inclusion   of further-neighbor Heisenberg spin-spin exchanges may also drive QPTs between the nVBS and plaquette-RVB states.  Such novel cases have first been studied in the context of the quantum dimer model on the honeycomb lattices \cite{Moessner-Sondhi}, but this requires $J^{\sigma}_{z}=J^{\sigma}_{x}$, which seems to be hard to achieve in our case.\\
\indent In the case of ${\bf \sigma}=1$ in the $d^{2}$ configration (realizable when the Hund coupling, $J_{H}$ exceeds a critical value in the multi-orbital Hubbard model; a classic example is the famed V$_{2}$O$_{3}$), one ends up with a spin-$1$ Heisenberg model. When $J^{\sigma}_{z}>>J^{\sigma}_{x}$, we
have a spin-$1$ nVBS state like the one for spin-$1/2$ case above, while in the opposite limit, we have almost decoupled spin-$1$ Heisenberg chains coupled by {\it staggered} Heisenberg ``interchain'' interactions. We expect that each of these phases will survive over a wide range of $y=J^{\sigma}_{z}/J^{\sigma}_{x}$. Thus, we expect a QPT from a Haldane-like topological phase of weakly coupled spin-$1$ chains to a nVBS state at a critical $y=y_{c}$ : however, details of such a QPT are beyond the scope of present work.\\

Onset of nVBS order has further enticing consequences. Consider doping one hole to this state.  It will be localized in the nVBS state because it can only propagate by disordering the ordered VB background, and this costs energy associated with breaking the VBs. However, the unpaired spin left behind as a ``partner'' of the doped hole acts like a spinon ``defect'' in the VBS background. In the ground state, in a $t-J$ model framework, we expect that this hole and it's unpaired spin-$1/2$ ``partner'' can propagate in the nVBS background with a much reduced kinetic energy, $O(J^{\sigma}_{x}<<t_{xz},t_{yz})$.  However, in low-lying excited states, the hole can also ``propagate'' by creating a ``defect string'' as illustrated by the zig-zag black line in Fig (\ref{rvb3}a).  Notice that this string has the character of a strip possessing staggered VBS ``order'' \cite{Moessner-Sondhi}, and acts as an extended defect separating two nVBS ``domains''. Thus, a single added hole fractionalizes into a spinless charge-$e$ {\it holon} and a neutral spin-$1/2$ {\it spinon}.\\
\indent Remarkably, this mechanism of fractionalization aids in hole pairing when we consider the fate of two holes doped into the nVBS state. If the second hole replaces the unpaired spinon above, the two-hole pair can coherently hop along the bonds surrounding the pair, without disturbing the nVBS background. In a ``plain vanilla'' RVB or Gutzwiller mean-field \cite{Anderson} framework, introducing a weak next-near neighbor hopping (of holes) will result in unconventional superconductivity with a $d_{x^{2}-y^{2}}+id_{xy}$ symmetry \cite{Chubukov,Vojta,Baskaran1,Baskaran2}. Moreover, as shown in Fig.(\ref{rvb3}b), if the second hole sits farther, on the sublattice opposite to that occupied by the first hole, it partially {\it cancels} the defect string, leading to the two holes being separated, in Fig.(\ref{rvb3}b), by an extended ``defect'' (of {\it two} unit cells in Fig.(\ref{rvb3}b)) {\it having ``staggered VBS'' pattern}.  It is easy to see that this {\it composite pair}, consisting of a bound pair of holes dressed by a cloud of intervening VB singlets from these two unit cells, can propagate without distrurbing the background nVBS pattern, along the three bonds surrounding each hole. Concomitantly, the two spinons at each end of the two defect strings ``recombine'' into a VB singlet, thereby healing the string fluctuations. Though the undoped ground state is a nVBS instead of a true RVB state of spin singlets, we find that the confinement of one-hole versus propagating character of the two-hole state is reminiscent of toplogical mechanisms of hole superconductivity in $t-J$ models \cite{Abanov}.  Since the nVBS background is not scrambled by the hopping of the composite pair (Fig.(\ref{rvb3}b)), unconventional superconductivity occurs in presence of a spin gap, so this is a rare instance of a $d=2$ version of the famed Luther-Emery scenario that is known in $d=1$. Depending upon the coherence length of the two-hole pair wavefunction, pairing with $s_{\pm}$ or $d_{x^{2}-y^{2}}\pm id_{xy}$ symmetry maybe possible, but this requires more work.\\
\indent Examples of real transition-metal oxides (TMO) with brickwall lattice structure are rare. Remarkaby, the $O$-deficient cuprate Ba$_{2}$CuO$_{3+x}$ with squashed CuO$_{6}$ octahedra but no orbital degeneracy is a high-T$_{c}$ SC with T$_{c}=73$~K~\cite{zhang}.  This gives rise to hopes that early-TMO with squashed octahedra and $O$-vacancy ordering may exhibit enhanced T$_{c}$ upon hole doping the OO and nVB solid we have proposed here.
%%%%%%%%%%%%%%%%%%%%%%%%%%%%%%%%%%%%%%%%%%%%%%%%%%%%%%%%%%%%%%%%%%%%%%%%%%%%%%%%%%%%%%%%%%%%%%%%%%
%%%%%%%%%%%%%%%%%%%%%%%%%%%%%%%%%%%%%%%%%%%%%%%%%%%%%%%%%%%%%%%%%%%%%%%%%%%%%%%%%%%%%%%%%%%%%%%%%%
\section{Fermionic Quantum Compass Models in $d=2$} \label{fQCM}

\indent  Fermionic QCM without $p$-wave pairing terms are of interest in their own right.  As mentioned in the introduction, there may be several ways to generate
such models.  One way is to consider a cold atomic
system of atoms with $s$ and $p$ orbitals \cite{Vincent} and two electrons per atom, in the
confinement limit, wherein $s$- and $p_{z}$ states can be ignored.  With one
electron in two degenerate $p_{x},p_{y}$ ``orbitals'', neglecting inter-site
$\pi\pi$ hoppings between $p_{x}-p_{y}$ orbitals leads to the $d=2$ fermionic
QCM (relabelling $p_{x}=d$ and $p_{y}=f$ for brevity),

\begin{align}
H_{\text{fQCM}} = t\sum_{i}(d_{i}^{\dag}d_{i+x} + f_{i}^{\dag}f_{i+y} + h.c) +
U\sum_{i}n_{i,d}n_{i,f}\label{fqcm}
\end{align}

\indent The second way to generate $H_{\text{fQCM}}$ is to revert to the brickwall
lattice, but replace the $S^{x}_{i}S^{x}_{i+x}$ couplings along the horizontal
bonds by an XY coupling, $(S^{x}_{i}S^{x}_{i+x} + S^{y}_{i}S^{y}_{i+x})$.
Upon following the same sequence of JW transformations and Mattis dualities as
 before for the honeycomb QCM, we are again led to the $H_{\text{fQCM}}$. While it is hard to conceive of such a XY-coupling in real $d$-band
oxides on a honeycomb lattice, such term(s) can be readily engineered in the
Josephson junction arrays, by exploiting the freedom involved in coupling two
JJ qubits \cite{Wendin}.  In this avatar, such a JJ array is another
option to emulate the fermionic QCM.  This should also be of interest in the
context of simulating one- (via implementing single-``spin'' rotations) and two-(via implementation of two-spin rotations) qubit gates needed for a universal quantum computer.
\noindent  Here, we investigate the ``hidden'' order(s) and possible novel quantum and thermal
criticality in $H_{\text{fQCM}}$ using TPSC and strong-coupling (large $U/t$) expansion. 
%TPSC has been successfully used in
%studies of QPTs between antiferromagnetic (AF) and paramagnetic phases, as well as
%in studies of $d$-wave superconductivity in the $d=2$ Hubbard model ~\cite{Tremblay}.

\section{Two-Particle Self-Consistent Approach to the Fermionic QCM}\label{tpsc-section}

\indent The two-particle self-consistent (TPSC) approach is a non-trivial extension of the simple random phase approximation (RPA) that exploits the Pauli's exclusion principle and certain exact many-body sum rules \cite{Tremblay}. The latter relate to demanding that the momentum-averaged static susceptibilities that include a renormalized but static vertex be equal to their local counterparts.  It is thus similar to, but a more involved form of the self-consistent renormalization theory by Moriya \cite{Moriya}. Owing to use of the (different) renormalized charge ($U_{ch}$) and spin ($U_{sp}$) vertices via, for example $U_{sp}=U\langle n_{\uparrow}n_{\downarrow}\rangle/\langle n_{\uparrow}\rangle\langle n_{\downarrow}\rangle$, a negative feedback is generated in the RPA spin susceptibility which cuts off its unphysical finite-$T$ singularities (at a suitable $\mathbf{q}$) in $2d$ models with continuous $SU(2)$ spin rotational symmetry, in accordance with the famous Mermin-Wagner's theorem \cite{Mermin-Wagner}.  Moreover, TPSC accurately determines the boundary separating the ``renormalized classical'' and the ``quantum disordered'' regions of finite-$T$ phase diagram of (mostly) one-orbital, locally interacting (Hubbard-like) microscopic models. The low-T correlation lengths (of various collective fluctuations) it yeilds are in good accord with the field-theoretic non-linear sigma model predictions, making it an excellent choice in cases where we are interested in the fate of ordered/disordered states in $d=2$.  It's main shortcoming is that it cannot describe Mott transitions accompanied by non-trivial correlation-induced spectral weight transfers. Though many studies have applied TPSC (or its extended version) to investigate anti-ferromagnetic (AF) or superconducting (SC) instabilities in $d=2$ \cite{Allen-tremblay, Davoudi-tremblay, Bergeron-tremblay}, we are not aware of attempts to study order-disorder transitions in interacting models with $1d$ band structure. In contrast to AF or SC transitions in the original Hubbard model, orbital or ``nematic'' ordered states (of Eq.\eqref{fqcm}) break only the discrete (Ising) symmetries of the Hamiltonian, and thus exhibit finite-$T$ Ising-like order in $d=2$ (without violating the constraints by Mermin-Wagner).\\

\begin{figure}
\includegraphics[height=50mm,width=70mm]{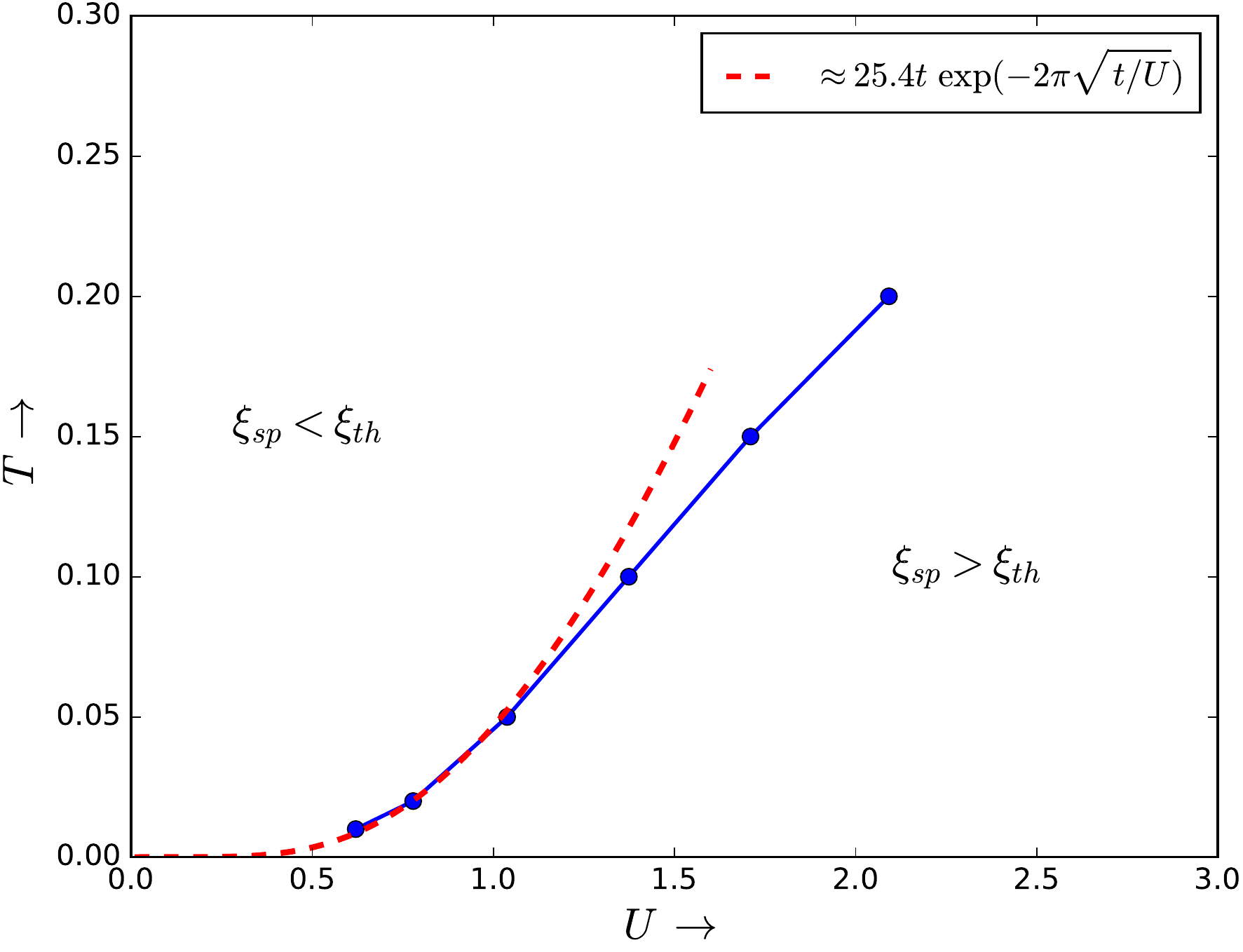}
\caption{$T$-$U$ phase diagram for $H_{fQCM}$ at half-filling. On the blue curve, $\xi\approx \xi_{th}=v_{F}/\pi T$, which separates two phases, where (1) $\xi\sim e^{C/T}$ or $\xi>\xi_{th}$ (corresponding to the orbital-ordered state) and (2) $\xi<\xi_{th}$ (orbitally disordered state). The $2d$ Hartree-Fock fit to the phase separation curve at low $(U,T)$ is shown by red, dashed curve.}\label{fig3}
\end{figure}

\begin{figure}
\includegraphics[height=50mm,width=70mm]{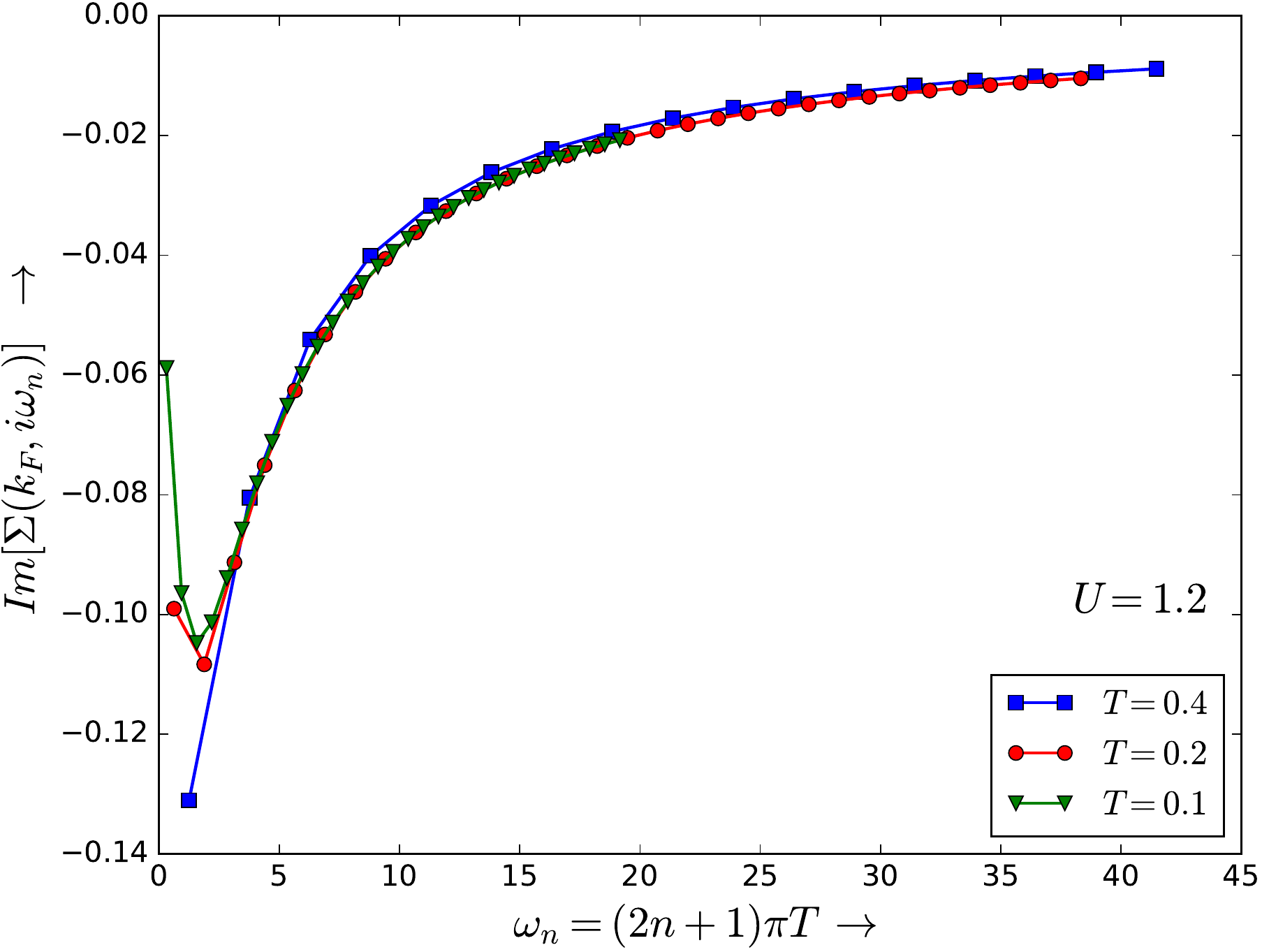}
\caption{Imaginary part of Matsubara self-energy, $\text{Im}\Sigma(k_{F},i\omega_{n})$ at Fermi wave vector at $U=1.2$ and $T=0.1,0.2,0.4$. As a function of decreasing $T$, clear insulator to bad-metal crossover is clearly seen.}\label{fig4}
\end{figure}

\indent We now describe our results, relegating details of computation of the TPSC susceptibilities and self-energy to Appendices \ref{A} and \ref{B}.  In Fig.~\ref{fig3}, we show the crossover line demarcating regions where the $T$-dependent (TPSC) spin correlation length, $\xi_{sp}$ exceeds (or is less than) the thermal correlation length, $\xi_{th}\sim v_{F}/\pi T$ ($v_{F}$=Fermi velocity). The corresponding divergence of the ``spin'' correlation length ($\xi_{sp}$) and the static spin-structure factor ($\chi_{sp}(\mathbf{q},q_{n}=0)$) are shown in the appendix sections (see Fig.\ref{xisp} and Fig.\ref{chis}). In TPSC, this is a signal for a finite-$T$ crossover between magnetically ordered and disordered phases. Since the ``pseudo-spin'' susceptibility, $\chi_{sp}({\bf q},i\nu_{n}) \simeq \int d\tau e^{i\nu_{n}\tau}\langle T_{\tau}T^{z}_{\bf q}(\tau)T^{z}_{-\bf q}(0)\rangle$, with $T^{z}=(n_{d}-n_{f})/2$ shows singularity at ${\bf q}=(\pi,\pi)$ (see Fig.\ref{chis}), the regime with $\xi_{sp} >\xi_{th}$ represents an Ising-like anti-ferromagnetic ``orbital''-ordered (AFOO) state. This AFOO will couple to an appropriate symmetry-adapted lattice mode by a dynamic JT process (for cold-atom lattices, this could be done by introducing an additional super-lattice potential that is different for $d$ and $f$-fermions). Thus, one expects a spontaneous ``antiferrodistortive'' structural change to accompany electronically driven AFOO. While it is hard to experimentally observe such a manifestation of AFOO, this might be possible in future (see \cite{humming}). If it be feasible to tailor phonon effects in future, one should observe an anti-ferrodistortive tilting of the lattice accompanying AFOO.\\
\indent Since onset of AFOO breaks the discrete symmetry of $H_{\text{fQCM}}$, having a finite-$T$ 
  transition at $T_{oo}$ (blue line in Fig.~\ref{fig3}) does not contradict the 
  Mermin-Wagner theorem.  At small $U$, we find that a Hartree-Fock (HF) fit 
  for $T_{oo}(U/t)$ works very well, as expected.  Up to intermediate $U/t$, 
  we find a correct trend in $T_{oo}(U)$.  However, we are unable to access the
  large $U/t$ limit using TPSC, where one expects that $T_{oo}(U)\simeq t^{2}/U$. A separate 
  argument from a perturbative-in-$(t/U)$ calculation for large $U$ indeed shows this (see below).\\
  \indent The TPSC one-electron self-energy (see Fig.\ref{fig4}) shows very interesting behavior.  At high-$T$ without any order, the imaginary part of the self-energy exhibits an insulator-like dependence on Matsubara frequency, $i\omega_{n}=(2n+1)\pi T$. As $T$ is lowered, a crossover to a ``bad-metallic'' state (at $T=0.2$) is initially found and, at lower $T$ (but slightly above $T_{oo}$, where $\xi/\xi_{th}\sim \mathcal{O}(1)$), this slowly evolves into Im$\Sigma(i\omega_{n}\rightarrow 0)\simeq -B(T)$ (at $T=0.1$), where $B(T)\sim T$ (see Appendix \ref{Self-energy-pseudogap}). This represents a $T$-dependent {\it dimensional crossover} as follows: at high $T$, $\xi_{sp} <\xi_{th}$ is short, and the carrier dynamics is predominantly one-dimensional (the $d$- and $f$- band carriers hop incoherently along two, $d=1$ chains. In such a situation, a finite $U$ will tend to localize $d,f$ carriers, simply because one is now dealing with two, effectively decoupled orthogonal chains at half-filling ($n_{d}=1/2=n_{f}$). As $T$ reduces, we find a first crossover from the $1d$-like insulator to an incoherent metal with resistivity $\rho(T)\sim T$. With a further reduction in $T$, $\xi_{sp}$ increases, ultimately far exceeding $\xi_{th}$: this signals the onset of anti-ferro orbital order (in TPSC, this occurs, when $\xi_{sp}>>\xi_{th}$). Around this crossover scale, the (now $d=2$) correlations couple to the $d,f$ carriers, driving a crossover to a pseudo-gap state due exponentially divergent anti-ferromagnetic ``spin" fluctuations (see Appendix \ref{Self-energy-pseudogap} for the analytical expression of the self-energy in this regime). Upon closer scrutiny, we observe that the incoherent metallicity (and the $1d$-to-$2d$ crossover) occurs at temperatures $T^{*}$ systematically larger than $T_{oo}$.  This implies that the dimensional crossover ($U/t$-dependent) from the high-$T$ ``insulator'' to the low-$T$ pseudo-gap regime occurs via an intermediate-$T$ regime of incoherent metallicity. It is a direct consequence of coupling fermionic carriers to $T$-dependent spin correlations in TPSC, and is {\it not} associated with Mott physics, which should become relevant beyond $U=4$ (this would give a pole at $\omega=0$, and is out of scope of TPSC). This incoherence-coherence crossover will also be manifested in the one-fermion spectral function, as a crossover from a pseudogapped state at high $T$ to one with a clear gap at the Fermi surface ($\omega =0$) at low $T$: this can experimentally be tested via time-of-flight spectroscopy and Bragg spectroscopy (\cite{Baym}, \cite{Trivedi}).\\
    \indent Armed with these findings, we propose that orbital-order and dimensional crossover physics could be observed in artificially 
    engineered cold-atom lattices with active $p_{x,y}$-like fermionic states 
    (which are well-separated from lower $s$-like states).  Though there will 
    also be small $\pi$-type hopping,  $t_{\pi}$, between $p_{x}(p_{y})$-orbitals along $y(x)$-axes, these are small compared to the $\sigma$-type hopping
    integrals.  This will break the strict one-dimensionality of the 
    one-electron hopping: nevertheless, we expect that the qualitative picture 
    obtained above should ``analytically continue'' into the $t_{\pi}<<t$
    regime.
%%%%%%%%%%%%%%%%%%%%%%%%%%%%%%%%%%%%%%%%%%%%%%%%%%%%%%%%%%%%%%%%%%%%%%%%%%%%%%%%%%%%%%%%%%%%%
%%%%%%%%%%%%%%%%%%%%%%%%%%%%%%%%%%%%%%%%%%%%%%%%%%%%%%%%%%%%%%%%%%%%%%%%%%%%%%%%%%%%%%%%%%%%%%
\section{Strong-Coupling Limit} \label{str-section}
We now turn to the strong coupling regime.  For brevity, we choose a fermionic model with a ratio of pairing to hopping processes taken to be $\lambda$, allowing one to interpolate between $H_{fQCM}$ ($\lambda=0$) and the JW fermionized HQCM, Eq.\eqref{eq6} ($\lambda=1$). In the limit $J_{x}<<J_{z}$ (or $t<<U$), charge fluctuations are completely absent at low-energies. The spin-exchange interaction is given as follows (see Appendix \ref{C} for details),   
\begin{align}
H_{eff}=&\frac{(1-\lambda^{2})t^{2}}{8U}\sum_{\langle ij \rangle}\sigma^{z}_{i}\sigma^{z}_{j}\nonumber\\
&+\frac{3t^{4}}{32U^{3}}(1-\lambda^{4})\sum_{i}\big( \sigma^{+}_{i}\sigma^{-}_{i+\hat{x}}\sigma^{+}_{i+\hat{x}+\hat{y}}\sigma^{-}_{i+\hat{y}} +h.c. \big)\nonumber\\
&-\frac{3t^{4}}{32U^{3}}(3\lambda^{2}+\lambda^{4})\sum_{i}\sigma^{x}_{i}\sigma^{y}_{i+\hat{x}}\sigma^{x}_{i+\hat{x}+\hat{y}}\sigma^{y}_{i+\hat{y}}\label{Heff}
\end{align}
For $\lambda=1$, we obtain the Wen's plaquette model \cite{Wen2}, which is equivalent to Kitaev's toric code (TC) \cite{toric}. As discussed earlier (see the section on the HQCM above), the ground state of its JW fermionized version, Eq.\eqref{eq6}, in the limit $J_{z}>>J_{x}$, was argued to be the Gutzwiller-projected BCS wave function (Eq.\eqref{Gutzwiller}). This bares a far from obvious connection between
topologically ordered state TC and a $p$-wave RVB wavefunction. Interestingly, such a connection is also 
known for the honeycomb Kitaev model \cite{Chen-Nussinov}.\\
\indent Adding external field terms gives rise to additional QPTs. One can add field terms like $-h_{\alpha}\sum_{i}\sigma^{\alpha}_{i}$ ($\alpha=x,y,z$) to $H_{Wen}$, obtaining
\begin{align}
H_{eff}(\lambda=1) = -J_{w}\sum_{i}\sigma^{x}_{i}\sigma^{y}_{i+\hat{x}}\sigma^{x}_{i+\hat{x}+\hat{y}}\sigma^{y}_{i+\hat{y}}\nonumber\\
- \sum_{a=x,y,z}h_{a}\sum_{i}\sigma^{a}_{i}
\end{align}
where $J_{w}=\frac{3t^{4}}{8U^{3}}$.  Such terms can arise from intra-island 
hybridization of Majorana zero modes. When $h_{x}=h_{y}=0$ but 
$h^{z}\neq 0 $, this is equivalent to the Xu-Moore model, itself dual to the 
quantum compass model on a square lattice \cite{Chen1}.
\begin{align}
H_{QC} = -J_{w}\sum_{r}\tau^{x}_{r}\tau^{x}_{r+x} - h_{y}\sum_{r}\tau^{z}_{r}\tau^{z}_{r+z}
\end{align}
This model exhibits a first-order transition separating two, self-dual, Ising-nematic ordered states at the self-dual point, $J_{w}=h_{z}$. In TC lore, this is a transition between topologically
ordered and trivially ordered phases \cite{Vidal}.  In addition to this, the square lattice
QCM also exhibits another exotic first-order transition between two, rigorously dual orders: a ``hidden'' plaquette order, characterized by $\Theta_{ij}=\langle\tau^{x}_{i}\tau^{x}_{i+z}\tau^{x}_{i+x}\tau^{x}_{i+x+z}\rangle$ and a next-near-neighbor correlation, $\Omega_{z}=\langle\tau^{z}_{i}\tau^{z}_{i+2z}\rangle$ \cite{MSlaad}, \cite{Brzezicki}.\\
\indent One may also engineer a staggered zeeman field term, $h_{z}\sum_{i}(-1)^{i}S^{z}_{i}$, in the original Eqn.\eqref{eq1}.  For instance, in the orbital-only context,
  an {\it effectively} staggered magnetic field can arise from an alternating
  $g$-tensor in an applied uniform (real) magnetic field.  Due to spin-orbital coupling, this can induce a staggered {\it orbital}
  polarization in the orbital sector.  In the TC (or Wen) context, such a term can 
  be engineered in an appropriate Josephson junction array realization of the
  $d=2$ QCM by a generalization of a scheme proposed for the quantum compass
  chain by You {\it et al.} \cite{W.You}.  As we detail in Appendix \ref{D},
  the large-$J_{z}$ (JW fermionized) honeycomb QCM with a staggered zeeman field maps onto the
  Wen model in a field along $y$-axis \cite{J.You} :
\begin{align}
H_{Wen +h_{z}} = -J_{w}\sum_{i}\sigma^{x}_{i}\sigma^{y}_{i+x}\sigma^{x}_{i+x+y}\sigma^{y}_{i+y} - h_{z}\sum_{i}\sigma^{y}_{i}
\end{align}
Remarkably, thanks to a duality mapping, $\sigma^{x}_{i}\sigma^{y}_{i+x}\sigma^{x}_{i+x+y}\sigma^{y}_{i+y} = \tau^{x}_{i+\frac{1}{2}}, \sigma^{y}_{i}=\tau^{z}_{i-\frac{1}{2}}\tau^{z}_{i+\frac{1}{2}}$, it is {\it exactly} dual to the $d=1$ quantum Ising chains along diagonals of the
square lattice \cite{J.You}.  The topological-non-topological QPT is characterized by duality between 
open string (with order parameter $\phi_{1}\sim \langle\tau^{z}_{a,1/2}\tau^{z}_{a,j+1/2}\rangle$) and closed string (with order parameter $\phi_{2}\sim\langle\prod_{j=1}^{i}\tau^{x}_{a,j+1/2}\rangle$) condensation.  Here, it is the condensation of {\it both}, the 
$Z_{2}$ vortex and $Z_{2}$ charge, that destabilizes the topological order, and the QPT from
the topological to the polarized phase falls into the $d=2$ Ising universality class.\\

On the other hand, when $\lambda=0$, $H_{eff}(\lambda)$ reduces to an Ising model with a 4-spin ring exchange term
\begin{align}
H^{(\lambda=0)} =&\frac{t^{2}}{8U}\sum_{<i,j>}\sigma^{z}_{i}\sigma^{z}_{j} \nonumber\\
&- \frac{3t^{2}}{32U^{3}}\sum_{i}(\sigma^{+}_{i}\sigma^{-}_{i+x}\sigma^{+}_{i+x+y}\sigma^{-}_{i+y}+h.c)
\end{align}
Since $U>>t$, the first term will always dominate over the ring exchange, yielding the AFOO
  ground state.  This is just the strong-coupling version of the AFOO uncovered by TPSC in the weak-to-intermediate coupling limit. It is interesting to think of realizing (engineered) perturbations that suppress the above Ising term.
In a more realistic scenario (cold-atoms with active $p$-orbitals), fermions can hop in both directions, i.e. $t_{\mu\nu}=t_{||}\delta_{\mu\nu}+t_{\perp}(1-\delta_{\mu\nu})$, with $t_{||}>>t_{\perp}$. In certain cold-atom setups, $t_{\perp}/t_{||}$ ratio is also tunable. Thus, we get additional terms beside the Ising and ring-exchange terms, i.e. $H_{eff}=J_{z}$ (n.n. antiferromagnetic Ising)+ $J_{xy}$ (n.n. XY exchange)+ $J_{\times}$ (n.n.n. diagonal XY coupling) + $J_{p}$ (ring exchange), with the coupling strengths, $J_{z}\sim (t_{||}^{2}+t_{\perp}^{2})/U$, $J_{xy}\sim 2t_{||}t_{\perp}/U$, $J_{\times}\sim - t^{2}_{||}t^{2}_{\perp}/U$, and $J_{p}\sim - [O(t^{4}_{||}/U^{3})  +O(t^{2}_{||}t^{2}_{\perp}/U^{3})]$. Here n.n. and n.n.n. denote nearest neighbour and next-nearest neighbour couplings respectively. \\
\indent Following the proposal envisioned in Ref.\cite{Bloch} and recently realized in cold-atom experiments (on a single plaquette level) \cite{JWPan}, we suppress the nearest neighbour superexchange terms by placing Zeeman field gradient $\Delta$ fields in a suitable manner (see Fig.\ref{figfield}) on the square lattice. Such a choice of field gradients also suppress the $XY$ exchange along one diagonal but the other diagonal interaction remains untouched. Finally, the resultant effective Hamiltonian becomes (in the limit $\Delta>> t^{2}/U$),
\begin{figure}
\begin{tikzpicture}[scale=1.0]
 \draw[thick] (0,0) --(4,0); \draw[thick] (0,1) --(4,1); \draw[thick] (0,2) --(4,2); \draw[thick] (0,3) --(4,3); \draw[thick] (0,4) --(4,4); \draw[thick] (0,0) --(0,4); \draw[thick] (1,0) --(1,4); \draw[thick] (2,0) --(2,4); \draw[thick] (3,0) --(3,4); \draw[thick] (4,0) --(4,4);
 
 \draw [->-=.5, line width=1.0, black] (0,0) to (1,0); \draw [->-=.5, line width=1.0, black] (2,0) to (1,0); \draw [->-=.5, line width=1.0, black] (2,0) to (3,0); \draw [->-=.5, line width=1.0, black] (4,0) to (3,0);
\draw [->-=.5, line width=1.0, black] (0,0) to (0,1); \draw [->-=.5, line width=1.0, black] (1,0) to (1,1); \draw [->-=.5, line width=1.0, black] (2,0) to (2,1); \draw [->-=.5, line width=1.0, black] (3,0) to (3,1); \draw [->-=.5, line width=1.0, black] (4,0) to (4,1); 

 \draw [->-=.5, line width=1.0, black] (0,1) to (1,1); \draw [->-=.5, line width=1.0, black] (2,1) to (1,1); \draw [->-=.5, line width=1.0, black] (2,1) to (3,1); \draw [->-=.5, line width=1.0, black] (4,1) to (3,1);
 
\draw [->-=.5, line width=1.0, black] (0,2) to (0,1); \draw [->-=.5, line width=1.0, black] (1,2) to (1,1); \draw [->-=.5, line width=1.0, black] (2,2) to (2,1); \draw [->-=.5, line width=1.0, black] (3,2) to (3,1); \draw [->-=.5, line width=1.0, black] (4,2) to (4,1);

\draw [->-=.5, line width=1.0, black] (0,2) to (1,2); \draw [->-=.5, line width=1.0, black] (2,2) to (1,2); \draw [->-=.5, line width=1.0, black] (2,2) to (3,2); \draw [->-=.5, line width=1.0, black] (4,2) to (3,2);

\draw [->-=.5, line width=1.0, black] (0,2) to (0,3); \draw [->-=.5, line width=1.0, black] (1,2) to (1,3); \draw [->-=.5, line width=1.0, black] (2,2) to (2,3); \draw [->-=.5, line width=1.0, black] (3,2) to (3,3); \draw [->-=.5, line width=1.0, black] (4,2) to (4,3);

\draw [->-=.5, line width=1.0, black] (0,3) to (1,3); \draw [->-=.5, line width=1.0, black] (2,3) to (1,3); \draw [->-=.5, line width=1.0, black] (2,3) to (3,3); \draw [->-=.5, line width=1.0, black] (4,3) to (3,3);

\draw [->-=.5, line width=1.0, black] (0,4) to (1,4); \draw [->-=.5, line width=1.0, black] (2,4) to (1,4); \draw [->-=.5, line width=1.0, black] (2,4) to (3,4); \draw [->-=.5, line width=1.0, black] (4,4) to (3,4);

\draw [->-=.5, line width=1.0, black] (0,4) to (0,3); \draw [->-=.5, line width=1.0, black] (1,4) to (1,3); \draw [->-=.5, line width=1.0, black] (2,4) to (2,3); \draw [->-=.5, line width=1.0, black] (3,4) to (3,3); \draw [->-=.5, line width=1.0, black] (4,4) to (4,3);

\draw[dashed] (1,0) --(0,1); \draw[dashed] (1,0) --(2,1); \draw[dashed] (2,1) --(3,0); 
\draw[dashed] (3,0) --(4,1);
\draw[dashed] (0,1) --(1,2); \draw[dashed] (1,2) --(2,1); \draw[dashed] (2,1) --(3,2); 
\draw[dashed] (3,2) --(4,1);
\draw[dashed] (1,2) --(0,3); \draw[dashed] (1,2) --(2,3); \draw[dashed] (2,3) --(3,2); 
\draw[dashed] (3,2) --(4,3);
\draw[dashed] (1,4) --(0,3); \draw[dashed] (1,4) --(2,3); \draw[dashed] (2,3) --(3,4); 
\draw[dashed] (3,4) --(4,3);

 \draw[fill=white] (0,0) circle [radius=0.1]; \draw[fill=blue] (1,0) circle [radius=0.1]; \draw[fill=white] (2,0) circle [radius=0.1]; \draw[fill=blue] (3,0) circle [radius=0.1]; \draw[fill=white] (4,0) circle [radius=0.1]; 
 
 \draw[fill=blue] (0,1) circle [radius=0.1]; \draw[fill=red] (1,1) circle [radius=0.1]; \draw[fill=blue] (2,1) circle [radius=0.1]; \draw[fill=red] (3,1) circle [radius=0.1]; \draw[fill=blue] (4,1) circle [radius=0.1]; 

 \draw[fill=white] (0,2) circle [radius=0.1]; \draw[fill=blue] (1,2) circle [radius=0.1]; \draw[fill=white] (2,2) circle [radius=0.1]; \draw[fill=blue] (3,2) circle [radius=0.1]; \draw[fill=white] (4,2) circle [radius=0.1]; 
 \draw[fill=blue] (0,3) circle [radius=0.1]; \draw[fill=red] (1,3) circle [radius=0.1]; \draw[fill=blue] (2,3) circle [radius=0.1]; \draw[fill=red] (3,3) circle [radius=0.1]; \draw[fill=blue] (4,3) circle [radius=0.1]; 
 
 \draw[fill=white] (0,4) circle [radius=0.1]; \draw[fill=blue] (1,4) circle [radius=0.1]; \draw[fill=white] (2,4) circle [radius=0.1]; \draw[fill=blue] (3,4) circle [radius=0.1]; \draw[fill=white] (4,4) circle [radius=0.1]; 
 
 \draw[fill=white] (4.6,2.0) circle [radius=0.1]; \node [right, black] at (4.7,2.0) {$=0$};
 \draw[fill=blue] (4.6,1.6) circle [radius=0.1]; \node [right, black] at (4.7,1.6) {$=\Delta$};
 \draw[fill=red] (4.6,1.2) circle [radius=0.1]; \node [right, black] at (4.7,1.2) {$=2\Delta$};
\end{tikzpicture}
\caption{Arrangement of Zeeman field gradients on a square lattice to suppress n.n. spin exchange terms, the arrows indicate the direction of these gradients and the colors (on the sites) are used to denote the effective field strengths at the corresponding sites. The XY exchange that survives the field gradients are acting along the dashed lines.}\label{figfield}
\end{figure}
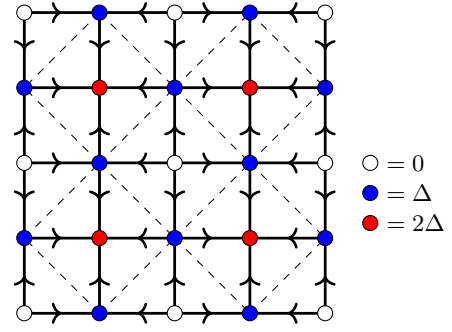
\begin{align}
H_{eff}\sim -J_{\times}\sum_{i\in blue}&(\sigma^{+}_{i}\sigma^{-}_{i\pm x+y}+h.c.)\nonumber\\
&-J_{p}\sum_{p}(\sigma^{+}_{1}\sigma^{-}_{2}\sigma^{+}_{3}\sigma^{-}_{4}+h.c.)
\end{align}
Representing the spins as hard-core bosons $S^{\pm}\sim b^{\dag} (\sim e^{i\phi})$, $b (\sim e^{-i\phi})$, $S^{z}=b^{\dag}b-1/2=n-1/2$, the Hamiltonian becomes
\begin{align}
H_{eff}\sim -J_{\times}\sum_{i\in blue}\cos{(\phi_{i}-\phi_{i\pm x+y})}+\frac{U}{2}\sum_{i}(n_{i}-\langle n\rangle)^{2}\nonumber\\
-J_{p}\sum_{i}\cos{(\phi_{i}-\phi_{i+x}-\phi_{i+y}+\phi_{i+x+y})}
\end{align}
with $\langle n\rangle=1/2$ and $U\rightarrow \infty $ (the hard-core limit).\\
\indent Now, two possibilities result: the single boson hopping ($\sim J_{\times}$) term ``stiffens'' the phase degrees of freedom. When $J_{\times}>J_{p}$, it will result in a $(2+1)d$ superfluid with stripe (with wave-vector $(\pi,0)$ or ($0,\pi$)) order. Although here $U\rightarrow \infty$, the $T=0$ superfluidity survives because the filling is $\langle n\rangle=1/2$. One can think of this situation in a following way: when $U=\infty$, we have a highly degenerate ground state manifold (when $\langle n\rangle=1/2$), the presence of one-boson hopping lifts this degeneracy and leads to superfluidity. In the case of integer filling, $U\rightarrow \infty$ ground state will of course be in a Mott state.\\
\indent The hopping of two-boson particle-hole pairs (or Bose-excitons $\sim J_{p}$) allows mobility of these excitons (fluctuation in the number of excitons on links) and Bose condenses them (when $J_{p}>J_{\times}$). Now, the elementary bosons remain uncondensed because the ring exchange allows significant amount of phase fluctuations at low energies. The phase fluctuations generate ``vortex-dipoles" (see below) that are gapless. When these dipoles proliferate, they completely destroy the (boson) phase coherence. The onsite repulsion $U$ tries to freeze $n_{i}$, and for large $U$, generates violent phase fluctuations which induce creation and hopping of vortices. The single vortex excitaions now have finite energy due to second derivative structure of the ring term, in stark contrast to what happens in the pure $XY$ model, where vortex energy cost is logarithimic (in system size).\\
\indent If we consider the ``soft-core" case, the local constraint is in a sense irrelevant. There exists a parameter regime where single vortex creation and single vortex hopping terms are irrelavent as shown from renormalization group arguments \cite{Paramekanti}. In this case [$J_{p}>J_{\times}$, $(U/J_{p})<(U/J_{p})_{c}$], the effective low energy theory could be expressed either as a theory of freely hopping Bose excitons or an effective theory of freely hopping vortex-dipoles (in dual variables) \cite{Paramekanti},
%\begin{widetext}
\begin{align}
 S=\int_{0}^{\beta}d\tau &\bigg[\sum_{\mathbf{r}}\frac{i}{\pi}(\partial_{\tau}\varphi_{\mathbf{r}}) (\delta_{xy}\mathcal{\theta}_{\mathbf{r}'})\nonumber\\
 &+\sum_{\mathbf{k}}\bigg\lbrace \frac{\tilde{K}(\mathbf{k})}{2}|(\delta_{xy}\varphi)_{\mathbf{k}}|^{2}+\frac{\tilde{U}(\mathbf{k})}{2\pi^{2}}|(\delta_{xy}\theta)_{\mathbf{k}}|^{2}\bigg\rbrace\bigg]\label{eq3}
\end{align} 
%\end{widetext}
Here $\mathbf{r}$ is the direct lattice and $\mathbf{r}'$ is the dual lattice (plaquette centers of original square lattice). The $\mathbf{k}$ is summed over $2d$ Brillouin zone. The $\varphi$ and $\theta$ represent coarse-grained boson and dual ``vortex'' variables respectively. Like $e^{\pm i\varphi}$, the single boson creation and annihilation operatores, $e^{\pm i\theta}$ denotes single vortex creation and annihilation operators. Here $\delta_{xy}\varphi\equiv (\phi_{i}-\phi_{i+x}-\phi_{i+y}+\phi_{i+x+y})$ denotes the Bose-exciton hopping and similarly,  $\delta_{xy}\theta$ represents vortex-dipole hopping. By vortex-dipole, we mean vortex-antivortex pair (just like particle-hole pair). If we integrate one of the d.o.f between $\varphi$ or $\theta$, we get a gaussian theory for the rest. The $\tilde{U}$ and $\tilde{K}$ are effective momentum dependent parameters characterizing the phase (like in Fermi liquid theory), in principle these are non-trivial functions of the bare, microscopic parameters: $J_{\times}, J_{p}$ and $U$. \\
\indent What are the correlations of this special phase? Using the above gaussian fixed point theory,
 \\
(1) single boson correlation:
\begin{align}
\langle e^{i\varphi_{r}(\tau)}e^{-i\varphi_{0}(0)}\rangle\sim \delta_{r,0} e^{-\alpha |\ln(\omega_{0}\tau)|^{2}}
\end{align}
Here $\alpha=\frac{1}{4\pi^{2}}\sqrt{\tilde{U}_{0}/\tilde{K}_{0}}$, $\tilde{U}_{0}=U(\mathbf{k}=0)$, $\tilde{K}_{0}=K(\mathbf{k}=0)$, $\omega_{0}=\sqrt{\tilde{U}_{0}\tilde{K}_{0}}$. So, the correlation is ultra-local and decays faster than power-law (in time), resulting absence of superfluidity. The reason behind this locality lies in the subsystem symmetries present in the ring exchange term: the number of bosons along each row and column are conserved separately, i.e. the Hamiltonian Eq.\eqref{eq3} is invariant under $\varphi\rightarrow \varphi+f(x)+g(y)$, where $f(x)$, $g(y)$ are functions of either $x$ or $y$. The superfluidity or effective phase stiffness vanishes due to proliferation of vortex-dipoles.\\
(2) single vortex correlation: Since the fixed point theory is quadratic both in terms of bosons ($\varphi$) and vortices ($\theta$), the single vortex correlation has same local feature as single boson correlation, 
 \begin{align}
 \langle e^{i\theta_{r}(\tau)}e^{-i\theta_{0}(0)}\rangle\sim \delta_{r,0} e^{-\alpha_{v}|\ln(\omega_{0}\tau)|^{2}}
 \end{align}
Here, $\alpha_{v}=\frac{1}{4}\sqrt{\tilde{K}_{0}/\tilde{U}_{0}}$. So, vortices are also uncondensed. This means that the system is not an insulator either. In the dual picture, conservation of vortices along rows and columns results in the above.
 \\
\indent What is the fate of the Bose exciton correlations? We consider an operator $\eta^{\dag}_{r}(y)=e^{i\varphi_{r}}e^{-i\varphi_{r+y\hat{y}}}$ which creates a particle-hole pair (bose exciton) separated by distance $y$ (finite) along y axis. The equal-time correlation
\begin{align}
\langle \eta^{\dag}_{r}(y)\eta_{r+x\hat{x}}(y)\rangle\sim |x|^{-\beta(y)}\ \ |x|>>|y|
\end{align}
with a positive exponent dependent on $y$ and the ratio of $\tilde{U}/\tilde{K}$. This implies a quasi-long range order of excitons (thus called exciton liquid). Similarly, vortex-dipole correlation also shows a quasi-long range order. Therefore, we can call this exciton liquid also as vortex-dipole liquid (in the dual description).
 \\
\indent However, in the case of hard-core bosons, single vortex hopping is a relevant perturbation since we must impose the constraint of $U \rightarrow \infty$ \cite{Sandvik},\cite{Paramekanti}, \cite{Motrunich}. In this situation, proliferation and condensation of vortices destabilizes the exciton Bose liquid and results in a ($\pi,\pi$) charge density wave (CDW) insulator when $J_{p}>J_{\times}$.
\\
\indent Thus, the emergent $T=0$ phase diagram (in the hard-core case) shows two symmetry broken phases: (a) when $J_{\times}>J_{p}$, one gets a striped superfluid, and (b) when $J_{\times}<J_{p}$, one gets a CDW insulator. In the spin language, this is a transition between stripe ($\pi,0$) (or ($0,\pi$)) order and ($\pi,\pi$) staggered Ising order. The diagonal and ring exchange interactions strongly frustrate each other when $J_{p}/J_{\times}\sim O(1)$, even in the hard-core limit. $J_{\times}$ tries to destabilize the ($\pi,\pi$) CDW order, while $J_{p}$ frustrates the superfluidity. The transition between these two phases can happend in three possible ways (1) a first order transition, both orders abruptly vanish at $J_{p}/J_{\times}\sim O(1)$, (2) there could be a regime of co-existence of two phases, or some new symmetry broken phases (in this case there will be two different phase transitions). (3) Finally, a second order critical point separating two symmetry broken phases. This kind of criticality is beyond the Landau paradigm (called deconfined criticality) where system becomes gapless and hosts emergent, fractional excitations coupled to non-compact gauge fields. The second order critical point (if it exists at all) is descibed by a low energy field theory similar to \eqref{eq3}. Indeed, this field theory is shown to be dual to a deconfined gauge theory \cite{Pretko} with gapped charges (spinons) which corresponds to vortices of $\varphi$ fields. The critical theory (also called Lifshitz criticality) hosts gapless modes with a quadratic dispersion, $\omega\sim k_{x}k_{y}$. Thus, it may still be possible to have an exotic exciton-Bose liquid at finite temperatures above such a critical point. Which of these scenarios finally reamins, however, is an open question.\\ 

\indent This suggests that a quantum phase transition hosting a critical Bose-excitonic liquid could be realized and probed in cold atom lattices by $(i)$ engineering a large $U$ by tuning the system close to a Feshbach resonance, $(ii)$ by engineering a sizable potential gradient along one of the diagonals of each 4-site plaquettes to suppress the two-spin superexchange, and $(iii)$ measuring the static spin structure factor in each phase by Bragg spectroscopy. Given the advances in cold-atom technology, these conditions should be implementable experimentally.
%\indent This suggests that a quantum phase transition from AFOO to an Bose-excitonic condensate can be driven in cold atom lattices with partially filled
%  $p_{x}, p_{y}$ states at each site, by $(i)$ engineering a large $U$ by
%  tuning the system close to a Feshbach resonance,
%  $(ii)$ by engineering a sizable potential gradient along one of the
%  diagonals of each 4-site plaquettes to suppress the two-spin superexchange,
%  and $(iii)$ measuring the static spin structure factor in each phase
%  by Bragg spectroscopy.\\
%\indent Given the advances in cold-atom technology, these conditions should be implementable experimentally.  Hence, we have proposed a specific, experimentally viable protocol to simulate the elusive {\it critical} Bose liquid state.
%%%%%%%%%%%%%%%%%%%%%%%%%%%%%%%%%%%%%%%%%%%%%%%%%%%%%%%%%%%%%%%%%%%%%%%%%%%%%%%%%%%%%%%%%%
%%%%%%%%%%%%%%%%%%%%%%%%%%%%%%%%%%%%%%%%%%%%%%%%%%%%%%%%%%%%%%%%%%%%%%%%%%%%%%%%%%%%%%%%%%%
\section{Conclusion}\label{conc}
To conclude, we have investigated two varieties of quantum compass models in
  this work.  Using a direct Jordan-Wigner transformation combined with the
  Mattis trick, the orbital-only honeycomb QCM is found to host a $d=3$ Ising
  criticality from an XX- to ZZ-ordered phases in accord with earlier work
  \cite{Nussinov-vdbrink}. We have studied the strong coupling version
  of this model with $J_{z}>> J_{x}$ in presence of ``external Zeeman fields''
  as well.  Such fields can be engineered in both, multi-orbital systems and
  Josephson junction arrays.  We uncover additional QPTs between topological
  and trivial field-polarized states or first-order transitions between
  self-dual nematic states.  In the multi-orbital context, consideration of
  transition-metal oxides within a Kugel-Khomskii framework will lead to
  novel QPTs between different {\it magnetically} ordered phases.\\
\indent We also studied the {\it fermionic} QCM on a square lattice, both in the
   weak-to-intermediate as well as strong coupling limits using complementary
   methods.  Such a model is relevant to cold-atomic lattices where the relevant states of interest are degenerate and strongly-directional $p_{x}$ and $p_{y}$
   orbitals.  Such features are also widely encountered in $t_{2g}$-orbital
   systems, most notably in Sr$_{2}$RuO$_{4}$.
   In the itinerant limit, we used TPSC to uncover an Ising-like AFOO
   phase at low $T$.  While this was already well known in extant literature,
   a new feature of the present work is that we uncover a non-trivial
   temperature-dependent dimensional crossover, from a high-$T$ insulator-like
   state to an AFOO state with a low energy pseudo-gap feature via an incoherent
   metal. This arises from the one-dimensionality of the unperturbed band
   structure, as well as from the coupling of
   fermions to the strongly $T$-dependent ``spin'' correlations in the TPSC.
   In the strong coupling ``Mott'' regime, we show how engineering a potential
   gradient in a specific way in cold-atom context can facilitate investigation
   of a novel QPT between Bose superfluid and CDW insulator via an
   intervening critical liquid of Bose-excitons.  This novel Bose
   liquid has long been of great interest, and our work may open up avenues for simulating such an exotic state of matter in the cold-atom setting.\\
\indent Several interesting issues call for more work.  First, the detailed
   investigation of magnetic phase transitions within the Kugel-Khomskii
   framework (where orbital correlations dictate the magnitude and signs
   of the effective Heisenberg couplings for a magnetic model) may be
   expected to lead to additional novelties, depending on the value (integer
   or half-integer) of the spin.  A related aspect pertains to the possibility
   of fractionalization of fundamental excitations as well as orbital-spin
   decoupling and re-confinement across orbital- and magnetically ordered
   phases in this context.  Consideration of these issues is left for future.
\section{Acknowledgements}
\indent We thank the Institute of Mathematical Sciences and Department of Atomic Energy, Government of India for the financial support. Arya Subramonian acknowledges financial support from the Carl Tryggers Stiftelse (project CTS 19:79).
%%%%%%%%%%%%%%%%%%%%%%%%%%%%%%%%%%%%%%%%%%%%%%%%%%%%%%%%%%%%%%%%%%%%%%%%%%%%%%%%%
%%%%%%%%%%%%%%%%%%%%%%%%%%%%%%%%%%%%%%%%%%%%%%%%%%%%%%%%%%%%%%%%%%%%%%%%%%%%%%%%%%%%%

\appendix
\section{Derivation of the chargon-spinon $Z_{2}$ gauge theory}\label{effective-theory}
In order to construct the Euclidean path integral for Eq.\eqref{eq-partition}, we make use of the Grassmann fields ($f_{i\sigma},\bar{f}_{i\sigma}$) for the fermionic spinons and the eigenstates of $s^{z}_{i}$ (eigenvalues are labelled by $s_{i}$) for the chargons. The partition function is obtained by tracing over these eigenstates, 
\begin{align}
Z=\int d&\bar{f}_{\sigma}(\tau_{1})df_{\sigma}(\tau_{1})\sum_{s(\tau_{0})=\pm 1}e^{-\bar{f}_{\sigma}f_{\sigma}}\nonumber\\
&\times \bra{-\bar{f}_{\sigma}(\tau_{1}), s_{0}}(e^{-\epsilon H}P)^{M}\ket{f_{\sigma}(\tau_{1}),s_{0}}
\end{align}
We next insert the following identity $M$ times, i.e., $[\bra{-\bar{f}_{1}, s_{0}}(.)e^{-\epsilon H}P (.)e^{-\epsilon H}P\cdot\cdot\cdot (.)e^{-\epsilon H}P\ket{f_{1},s_{0}}]$ inside the first brackets denoted by $(.)$,
\begin{align}
\int d\bar{f}_{m}df_{m+1} \sum_{s_{m}=\pm 1}e^{-\bar{f}_{m}f_{m+1}} \ket{f_{m+1},s_{m}}\bra{\bar{f}_{m}, s_{m}}=1
\end{align}
we get,
\begin{align}
Z=\prod_{j=1}^{M}\int d\bar{f}_{\tau_{j}\alpha}& df_{\tau_{j}\alpha}\sum_{s_{j}=\pm 1}e^{-\bar{f}_{\tau_{j}}f_{\tau_{j+1}}}\nonumber\\
&\times\bra{\bar{f}_{\tau_{j}}, s_{\tau_{j}}}e^{-\epsilon H}P\ket{f_{\tau_{j}}, s_{\tau_{j-1}}}\label{mat-elem}
\end{align}
The terms which involve $N_{i}$, i.e., $e^{-\epsilon N_{i}}$ are off-diagonal in $s^{z}_{i}$ basis. Therefore,
\begin{align}
\bra{s_{i}}e^{-\epsilon N_{i}}\ket{s_{j}}=\sum_{N_{i}=0,1}\bra{s_{i}}\ket{N_{i}}\bra{N_{i}}e^{-\epsilon N_{i}}\ket{N_{i}}\bra{N_{i}}\ket{s_{j}}
\end{align}
The eigenstates of $N_{i}$ are following: $\ket{N=0}\sim (\ket{s=+1}+\ket{s=-1})$ and $\ket{N=1}\sim (\ket{s=+1}-\ket{s=-1})$. For our convenience, we use the normalization convention $\langle s_{i}|N_{j}\rangle\sim e^{i\frac{\pi}{2}N_{j}s_{i}}$. Hence,
\begin{align}
\bra{s_{\tau_{i}}}e^{-\epsilon N_{i}}\ket{s_{\tau_{j}}}=\sum_{N_{i}}e^{i(\pi/2)N_{i}(s_{i}-s_{j})}e^{-\epsilon N_{i}}
\end{align} 
We can think of the integer $N_{i}(\tau)$ as a link variable connecting nearest neighbor time links $(i,i+\tau)$, labeled as $N_{i}\equiv N_{i\tau}$. Collecting all such terms that involve $N_{i\tau}$, we get (for every time-link),
\begin{align}
\sum_{N_{i\tau}=0,1}&\exp{\big[-\epsilon\frac{U}{2}N_{i\tau}+i\frac{\pi}{2}N_{i\tau}(s_{i}-s_{i+\tau}+1-\sigma_{i\tau})\big]}\nonumber\\
&= A e^{J_{\tau}s_{i}\sigma_{i\tau}s_{i+\tau}}\label{eqNsum}
\end{align}
where $J_{\tau}=-\frac{1}{2}\ln{(\tanh(U\epsilon/4))}$. Here, $A$ is some constant prefactor unimportant to our discussion since it does not modify the structure of the low energy theory apart from some overall shift in the free energy. Next, we use the following identity,
\begin{align}
e^{i\frac{\pi}{2}(1-\sigma_{r})\hat{f}^{\dag}_{r}\hat{f}_{r}}\ket{f_{r}}&=e^{i\frac{\pi}{2}(1-\sigma_{r})\hat{f}^{\dag}_{r}\hat{f}_{r}}(\ket{0}-f_{r}\ket{1})\nonumber\\
&=\ket{0}-f_{r}e^{i\frac{\pi}{2}(1-\sigma_{r})}\ket{1}\nonumber\\
&=\ket{0}-\sigma_{r}f_{r}\ket{1}\equiv \ket{\sigma_{r}f_{r}}\label{eqfcoh}
\end{align}
Then, using \eqref{eqNsum} and \eqref{eqfcoh}, the matrix element in Eq.\eqref{mat-elem} becomes
%\begin{widetext} 
\begin{align}
&\bra{\bar{f}_{\tau_{j}}, s_{\tau_{j}}}e^{-\epsilon H}P\ket{f_{\tau_{j}}, s_{\tau_{j-1}}}\nonumber\\
&\sim \sum_{\sigma_{j}} e^{\frac{i \pi}{2}(1-\sigma_{j})} e^{J_{\tau}s_{\tau_{j}}\sigma_{j}s_{\tau_{j-1}}}
\bra{\bar{f}_{\tau_{j}},s_{\tau_{j}}}e^{-\epsilon H_{t}}\ket{\sigma_{j}f_{\tau_{j}}, s_{\tau_{j-1}}}\nonumber\\
&=\sum_{\sigma_{j}}e^{i(\pi/2)(1-\sigma_{j})}\exp{\big(J_{\tau}s_{\tau_{j}}\sigma_{j}s_{\tau_{j-1}}\big)} \nonumber\\
&\ \ \ \ \ \ \ \times\exp{(\bar{f}_{j}\sigma_{j}f_{j})}\exp{[-\epsilon H_{t}(s_{j}, \bar{f}_{j}, \sigma_{j}f_{j})]}
\end{align}
%\end{widetext}
Next, we make the change of variables $\sigma_{j}f_{j}\rightarrow f_{j}$. The partition function now becomes,
\begin{align}
Z&\sim \int \bigg(\prod_{j=1}^{M} d\bar{f}_{\tau_{j}}df_{\tau_{j}}\bigg)\sum_{\lbrace s_{j}\rbrace=\pm 1}\sum_{\lbrace \sigma_{j}\rbrace=\pm 1}\bigg[e^{i\frac{\pi}{2}\sum_{j=1}^{M}(1-\sigma_{j})}\nonumber\\
&\times e^{-\sum_{j=1}^{M}\bar{f}_{j}(\sigma_{j+1}f_{j+1}-f_{j})}e^{\sum_{j=1}^{M}J_{\tau}s_{\tau_{j}}\sigma_{j}s_{\tau_{j-1}}}\nonumber\\
&\ \ \ \ \ \ \ \ \ \ \times e^{-\epsilon \sum_{j=1}^{M}H_{t}(s_{j}, \bar{f}_{j}, f_{j})}\bigg]\label{a9}
\end{align}
For the sake of simplicity, spatial indices are not explicitly stated in any of the aforementioned expressions. These labels will be restored at the end. Now, the Ising variables $\sigma_{j}$ act along the temporal links, whereas $s_{i}$ are placed on the sites of the space-time lattice. The first (exponential) term  in Eq.\eqref{a9} leads to the Berry phase term, Eq.\eqref{berry-term}. The second and third expressions represent respectively the temporal coupling between spinon (chargon) and gauge fields, and the bare chargon-spinon interaction (Eq.\eqref{Ht}).\\
\indent Next, we insert the following identity along every spatial link $(ij)$ and at all time slices,
\begin{align}
&C^{-1}\int d\chi_{ij}d\chi_{ij}^{*}\nonumber\\
&\exp{\bigg(-\epsilon \sum_{(ij)}\sum_{\tau}J_{s}[|\chi_{ij}|^{2}-(\chi_{ij} U_{ij}+h.c.)]\bigg)}=1\label{a10}
\end{align} 
Where, $U_{i,i+\hat{\alpha}}=(f_{i\alpha}-\bar{f}_{i\alpha})(f_{i+\hat{\alpha},\alpha}+\bar{f}_{i+\hat{\alpha},\alpha})$ is a function of Grassmann variables. Hence, $(U_{ij})^{2}=0$. The primary reason behind plugging such a ``fat'' identity into our action (and the specific choice of the fermionic billinear) comes from Wen's parton ``mean-field'' theory, Eq.\eqref{Hspinon} for the strong-coupling (large $J_{z}/J_{x}$) toric code Hamiltonian, Eq.\eqref{Htc}. The order of the spinon coupling, $J_{s}$ can be estimated from the strong-coupling expansion, which to the leading order is $\mathcal{O}(J^{4}_{x}/J^{3}_{z})$.  Next, we perform the following shift of variables, 
\begin{align}
\chi_{ij}\rightarrow \chi_{ij}+\frac{t}{2J_{s}}s_{i}s_{j}
\end{align}
where $s_{i}, s_{j}$ are ising variables (which take values $\pm 1$). This change of variables eliminates the bare spinon-chargon interaction in $H_{t}$ (Eq.\eqref{Ht}) and also generates a kinetic term for the chargon fields (Ising variables $s_{i}$) (see Ref.\cite{Senthil-Fisher}),
\begin{align}
H_{c}=\epsilon t\sum_{\langle ij\rangle}(\chi_{ij}+\chi_{ij}^{*})s_{i}s_{j}
\end{align}
We then flip the Ising spins $s_{i}\rightarrow -s_{i}$ on a particular sub-lattice (for every time slice). This will make the spatial Ising interaction ferromagnetic (without changing the signature of the temporal coupling strengths). Now, the amplitude fluctuations of $\chi_{ij}$ are massive (notice the $e^{-\alpha|\chi|^{2}}$ term in the action), the only low-energy degrees of freedom here are $Z_{2}$ gauge fluctuations (sign change of $\chi$). So, we make the approximation, $\chi_{ij}\approx \chi_{0}\sigma_{ij}$. Therefore, the chargon part of the action is described by a $(2+1)d$ Ising model minimally coupled to $Z_{2}$ gauge fields,
\begin{align}
S_{c}=-J_{\tau}\sum_{i,j=i+\hat{\tau}}s_{i}\sigma_{ij}s_{j}-\epsilon t_{c}\sum_{i,j=i+\hat{r}}s_{i}\sigma_{ij}s_{j}
\end{align}
This is in contrast to Senthil and Fisher's general model \cite{Senthil-Fisher}, where the chargon action also has a global $U(1)$ symmetry. The spatial hopping of the chargons and spinons are given by the following relations, $t_{s}=J_{s}(\chi_{0}+\chi^{*}_{0})$, $t_{c}=t(\chi_{0}+\chi^{*}_{0})$, where $J_{s}$ is roughly proportional to the spin-exchange coupling of the strong-coupling effective spin Hamiltonian (Eq.\eqref{Htc}), i.e., $J_{s}\approx \alpha t^{4}/U^{3}$. Now, the discrete time step $\epsilon$ introduces a high-energy cut-off $\hbar/\epsilon$ in the present theory. Since we are interested only in the low-energy spectrum, we keep it finite (but large) by choosing the spatial and temporal chargon hopping to be the same, i.e. $J_{\tau}=\epsilon t_{c}$ (following \cite{Senthil-Fisher}). This will fix the value of $\epsilon$ in terms of $t,U$ (the bare couplings of the theory); we get $\epsilon^{-1}\approx \sqrt{Ut \chi^{r}_{0}}$ for $\epsilon \rightarrow 0$. Thus, our chargon action becomes exactly the Eq.\eqref{chargon-action}. The temporal part of spinon-only action, $S_{f}$ (Eq.\eqref{spinon-action}) comes from the second line of Eq.\eqref{a9} and its spatial component arises totally from inserting the identity, Eq.\eqref{a10} into the partition function. This completes the derivation of Eq.\eqref{eq-partition}.

\section{Duality between Ising gauge and matter fields}\label{gauge-matter-duality}
To begin with, consider the toric code Hamiltonian with external Zeeman field terms,
\begin{align}
H_{TC}=&-\frac{1}{\lambda}\sum_{+}\bigg(\prod_{\langle ij\rangle\in +}\Theta_{ij}\bigg)\prod_{\langle ij\rangle\in +}\sigma^{x}_{ij} -\lambda\sum_{\langle ij\rangle}\Theta_{ij}\sigma^{z}_{ij}\nonumber\\
&-\frac{1}{g}\sum_{\square}\prod_{\langle ij\rangle\in \square}\sigma^{z}_{ij}-g\sum_{\langle ij\rangle}\sigma^{x}_{ij}\label{toric-mag2}
\end{align}
This is equivalent to a $Z_{2}$ gauge theory coupled to $Z_{2}$ matter fields (see below). We have introduced here an additional classical Ising variable, $\Theta_{ij}$ along the links; its purpose will be explained shortly. Next, we transform Eq.\eqref{toric-mag2} to its dual Hamiltonian where the coupling parameters $\lambda$ and $g$ interchange their roles. To derive it, we perform the following unitary transformations; (1): $\pi$ rotation about the $\sigma_{y}$ axis, i.e., $\sigma^{x}_{ij}\rightarrow \sigma^{z}_{ij}$ , $\sigma^{z}_{ij}\rightarrow -\sigma^{x}_{ij}$; (2): $2\pi$ rotation about the $\sigma_{z}$-axis, i.e., $\sigma^{x}_{ij}\rightarrow -\sigma^{x}_{ij}$, $\sigma^{y}_{ij}\rightarrow -\sigma^{y}_{ij}$. In Eq.\eqref{toric-mag2}, $A_{s}=\prod_{\langle ij\rangle\in +}\sigma^{x}_{ij}$ and $B_{p}=\prod_{\langle ij\rangle\in \square}\sigma^{z}_{ij}$ are centered at the original ($r$) and dual ($\tilde{r}=r+\frac{\hat{x}}{2}+\frac{\hat{y}}{2}$) lattice sites respectively. Now, an elementary plaquette ($\square$) of a square lattice can be thought of as a ``star'' ($+$)-like structure of the dual square lattice, centered at the mid-point of the original lattice plaquette. Therefore, $A_{s}(r)\rightarrow \tilde{B}_{p}(r)$ and $B_{p}(\tilde{r})\rightarrow \tilde{A}_{s}(\tilde{r})$. Eq.\eqref{toric-mag2} now becomes
\begin{align}
H^{dual}_{TC}&=-g\sum_{\langle ij\rangle}\sigma^{z}_{ij}-\frac{1}{g}\sum_{+}\prod_{\langle ij\rangle\in +}\sigma^{x}_{ij}\nonumber\\
&-\frac{1}{\lambda}\sum_{\square}\bigg(\prod_{\langle ij\rangle\in +}\Theta_{ij}\bigg)\prod_{\langle ij\rangle\in \square}\sigma^{z}_{ij}-\lambda\sum_{\langle ij\rangle}\sigma^{x}_{ij}\Theta_{ij}
\end{align}
Here, the $\sigma^{\alpha}_{ij}$ ($\Theta_{ij}$) are defined on the dual (original) lattice links. We now rescale $\sigma^{\alpha}_{ij}$ by $\Theta_{ij}$, by $\sigma^{\alpha}_{ij}\rightarrow \Theta_{ij}\theta^{\alpha}_{ij}$ (where $\theta^{\alpha}_{ij}$ are another set of Pauli spins). This scaling transformation, keeps the spin commutation relations intact. The Hamiltonian transforms accordingly,
\begin{align}
H^{dual}_{TC}=-\frac{1}{g}\sum_{+}\bigg(\prod_{\langle ij\rangle\in +}&\Theta_{ij}\bigg)\prod_{\langle ij\rangle\in +}\theta^{x}_{ij}-\frac{1}{\lambda}\sum_{\square}\prod_{\langle ij\rangle\in \square}\theta^{z}_{ij}\nonumber\\
&-\lambda\sum_{\langle ij\rangle}\theta^{x}_{ij}-g\sum_{\langle ij\rangle}\Theta_{ij}\theta^{z}_{ij}\label{toric-mag3}
\end{align}
Notice that the $\Theta_{ij}$ fields were defined on the links (or the center of the links) of original lattice. We can also think of these as sitting at the center of the dual lattice links, since both are at the same location. Hence, for the dual Toric code, $\Theta$ fields are now redefined on the dual lattice links. Therefore, $H^{dual}_{TC}(g,\lambda)=H_{TC}(\lambda,g)$.\\

\noindent $\ast${\it Transforming the toric code model with external Zeeman fields (Eq.\eqref{toric-mag2}) to a $Z_{2}$ gauge Hamiltonian coupled with $Z_{2}$ matter fields (Eq.\eqref{H-gm})}:
\\

\indent Although the local $Z_{2}$ symmetry ($[A_{s},H_{TC}]=0$) of the pure toric code is explicitly broken in the presence of Zeeman fields, a modified local invariance now appears. Notice that $\sigma^{z}_{ij}$ creates a $Z_{2}$ charge-pair ($A_{s}=-1$) at the lattice sites $i$ and $j$, beside flipping $\sigma^{x}_{ij}$ (i.e., creating an electric field line connecting the two $Z_{2}$ charges). We define the charge (number) operator $s^{x}_{i}$ which has the eigenvalue $-1$, when there is a $Z_{2}$ charge sitting at site $i$. So, in the presence of external (longitudinal) magnetic fields, we have the following local $Z_{2}$ constraints on the physical Hilbert space \cite{Tupitsyn},
\begin{align}
s^{x}_{i}A_{i}\ket{\text{Phys}}=\bigg(\prod_{\langle ij\rangle\in +}\Theta_{ij}\bigg)\ket{\text{Phys}} \label{constraint}
\end{align}
The role of $\Theta_{ij}$ fields are now clear; it measures the static background charge configuration of the model. To rewrite Eq.\eqref{toric-mag2} in terms of $s_{i}^{\alpha}$ and $\sigma^{\alpha}_{ij}$, we consider the following transformations (under the constraint Eq.\eqref{constraint})\cite{Tupitsyn}
\begin{align}
A_{i}\rightarrow s^{x}_{i}\bigg(\prod_{\langle ij\rangle\in +}\Theta_{ij}\bigg)\ \ \ \ \sigma^{z}_{ij}\rightarrow s^{z}_{i}\sigma^{z}_{ij}s^{z}_{j}
\end{align}
The last transformation reflects the fact that $\sigma^{z}_{ij}$ creates a charge-pair at lattice sites $i$ and $j$ ($s^{x}_{i,j}=-1$), and it keeps $B_{p}$ unchanged. So, in the enlarged Hilbert space of  gauge fields ($\sigma^{\alpha}_{ij}$) and matter ($s^{\alpha}_{i}$), the Hamiltonian (Eq.\eqref{toric-mag2}) becomes
\begin{align}
H_{g+m}&=-\frac{1}{\lambda}\sum_{i}s^{x}_{i}-\lambda\sum_{\langle ij\rangle}s^{z}_{i}\Theta_{ij}\sigma^{z}_{ij}s^{z}_{j}\nonumber\\
&-g\sum_{\langle ij\rangle}\sigma^{x}_{ij}-\frac{1}{g}\sum_{\square}\prod_{\langle ij\rangle\in \square}\sigma^{z}_{ij} \label{matter-gauge1}
\end{align}
with the local constraint Eq.\eqref{constraint}. Following a similar set of steps, we can convert the dual toric code,  Eq.\eqref{toric-mag3} to the following Hamiltonian,
\begin{align}
H^{dual}_{g+m}=-\frac{1}{g}\sum_{i}v^{x}_{i}&-g\sum_{\langle ij\rangle}v^{z}_{i}\Theta_{ij}\theta^{z}_{ij}v^{z}_{j}\nonumber\\
&-\lambda\sum_{\langle ij\rangle}\theta^{x}_{ij}-\frac{1}{\lambda}\sum_{\square}\prod_{\langle ij\rangle\in \square}\theta^{z}_{ij} \label{matter-gauge2}
\end{align}
now with the following local gauge constraint, 
\begin{align}
v^{x}_{i}\prod_{(ij)\in +}\theta^{x}_{ij}=\prod_{(ij)\in +}\Theta_{ij}\equiv \prod'_{(ij)\in \square}\Theta_{ij}
\end{align}
The prime on the product denotes that this product is performed on the dual lattice links. In Eq.\eqref{matter-gauge2}, the $v^{\alpha}_{i}$ denotes vison (or $Z_{2}$ flux) degrees of freedom of the original gauge theory Eq.\eqref{matter-gauge1}.
\\
Now, consider the limiting case, $g\rightarrow 0$ in Eq.\eqref{matter-gauge1} (pure Ising model limit), The gauge fields are completely frozen, i.e. $\prod_{\langle ij\rangle\in \square}\sigma^{z}_{ij}=+1$ for all plaquettes. One can choose the gauge, $\sigma^{z}_{ij}=1$ for all the links. 
\begin{align}
H_{g+m}(g\rightarrow 0)= H_{I}=-\frac{1}{\lambda}\sum_{i}s^{x}_{i}-\lambda\sum_{\langle ij\rangle}s^{z}_{i}\Theta_{ij}s^{z}_{j}
\end{align}
On the other hand, 
\begin{align}
H^{dual}_{g+m}(g\rightarrow 0)=H_{gauge}=-\lambda\sum_{\langle ij\rangle}\theta^{x}_{ij}-\frac{1}{\lambda}\sum_{\square}\prod_{\langle ij\rangle\in \square}\theta^{z}_{ij}
\end{align}
with the constraint, $\prod_{\langle ij\rangle\in +}\theta^{x}_{ij}=\prod'_{\langle ij\rangle\in \square}\Theta_{ij}$. This is exactly the duality between the TFIM with frustrated plaquettes and the Ising gauge theory with background charges. Frustrated plaquettes of the Ising model map to the (static) background charges of the $Z_{2}$ gauge theory.
%%%%%%%%%%%%%%%%%%%%%%%%%%%%%%%%%%%%%%%%%%%%%%%%%%%%%%%%%%%%%%%%%%%%%%%%%%%%%%%%%%%%%%%%%%%%%%%
%%%%%%%%%%%%%%%%%%%%%%%%%%%%%%%%%%%%%%%%%%%%%%%%%%%%%%%%%%%%%%%%%%%%%%%%%%%%%%%%%%%%%%%%%%%%%%%
\section{Brief description of TPSC equations}\label{A}
The equations for charge, spin susceptibilities, and single-particle self-energy used in the main text are derived in this section. With the exception of the fact that in our model pseudo-spin rotational symmetry is not preserved (because of the directional hopping of orbitals), the derivation is almost identical to that presented in Ref.\cite{Tremblay}.\\

\textit{Expressions for the charge and longitudinal spin susceptibilities}:\\

For three-point susceptibility in the particle-hole channel, the Bethe-Salpeter equation is given by \cite{Tremblay},
\begin{align}
\chi_{\sigma\sigma'}(1,3;2)=&-G_{\sigma}(1,2)G_{\sigma}(2,3)\delta_{\sigma\sigma'}\nonumber\\
 &+G_{\sigma}(1,\bar{2})\Gamma_{\sigma\bar{\sigma}}^{ir}(\bar{2},\bar{3};\bar{4},\bar{5})\chi_{\bar{\sigma}\sigma'}(\bar{4},\bar{5};2)G_{\sigma}(\bar{3},3)
\end{align}
One of the main assumptions of TPSC theory is the replacement of the momentum and frequency-dependent vertex functions by constant parameters, that are determined self-consistently from the sum-rules.
\begin{align}
\Gamma_{\sigma\bar{\sigma}}^{ir}(\bar{2},\bar{3};\bar{4},\bar{5})=\Gamma_{\sigma\bar{\sigma}}\delta(\bar{2}-\bar{5})\delta(\bar{3}-\bar{4})\delta(\bar{4}^{+}-\bar{5})\label{A2}
\end{align}
The two-point susceptibility is defined as $\chi_{\sigma\sigma'}(1,3\rightarrow 1^{+};2)\equiv\chi_{\sigma\sigma'}(1,2)$, that satisfies the following equation,
\begin{align}
\chi_{\sigma\sigma'}(1,2)&=-G_{\sigma}(1,2)G_{\sigma}(2,1^{+})\delta_{\sigma\sigma'}\nonumber\\
&+G_{\sigma}(1,\bar{2})\Gamma_{\sigma\bar{\sigma}}\chi_{\bar{\sigma}\sigma'}(\bar{2},2)G_{\sigma}(\bar{2},1^{+})
\end{align}
Hence,
\begin{align}
&\chi_{dd}(1,2)=-G_{d}(1,2)G_{d}(2,1^{+})+G_{d}(1,\bar{2})\Gamma_{dd}\chi_{dd}(\bar{2},2)\nonumber\\
&\times G_{d}(\bar{2},1^{+})+G_{d}(1,\bar{2})\Gamma_{df}\chi_{fd}(\bar{2},2)G_{d}(\bar{2},1^{+})\label{A4}
\end{align}
and
\begin{align}
\chi_{fd}(1,2)=G_{f}(1,\bar{2})\Gamma_{ff}\chi_{fd}(\bar{2},2)G_{f}(\bar{2},1^{+})\nonumber\\
+G_{f}(1,\bar{2})\Gamma_{fd}\chi_{dd}(\bar{2},2)G_{f}(\bar{2},1^{+})\label{A5}
\end{align}
In the Hamiltonian Eq.\eqref{fqcm}, the anisotropy arises due to the different spatial dependence of single particle Green's functions. If we consider any local quantity (like $\Gamma_{\sigma\bar{\sigma}}$ in TPSC), that should have the following symmetry: $\Gamma_{dd}=\Gamma_{ff}$, $\Gamma_{df}=\Gamma_{fd}$. This happens as the local Green's functions are identical $G^{loc}_{d}(i\omega)=G^{loc}_{f}(i\omega)$, although $k$-dependent Green's functions are (numerically) different. The pseudo-spin rotational symmetry is preserved at the local level as there is no magnetic field term in the Hamiltonian (or any kind of  spontaneously broken symmetry).\\
\indent From Eqs. \eqref{A4}, \eqref{A5}, we arrive (in the Fourier space) at the following expression
\begin{align}
&\chi_{\sigma\sigma}(q)=\nonumber\\
&\ \frac{[1+\Gamma_{\sigma\sigma}\chi^{0}_{\bar{\sigma}}(q)]\chi^{0}_{\sigma}(q)}{1+\Gamma_{\sigma\sigma}(\chi^{0}_{\sigma}(q)+\chi^{0}_{\bar{\sigma}}(q))+(\Gamma^{2}_{\sigma\sigma}-\Gamma^{2}_{\sigma\bar{\sigma}})\chi^{0}_{\sigma}(q)\chi^{0}_{\bar{\sigma}}(q)}
\end{align}
Here, we have defined $\chi^{0}_{\sigma}(1,2)=-G_{\sigma}(1,2)G_{\sigma}(2,1^{+})$, the so-called ``particle-hole bubble" or Lindhard function. Similarly,
\begin{align}
&\chi_{\sigma\bar{\sigma}}(q)=\nonumber\\
&\frac{-\Gamma_{\sigma\bar{\sigma}}\chi^{0}_{\sigma}(q)\chi^{0}_{\bar{\sigma}}(q)}{1+\Gamma_{\sigma\sigma}(\chi^{0}_{\sigma}(q)+\chi^{0}_{\bar{\sigma}}(q))+(\Gamma^{2}_{\sigma\sigma}-\Gamma^{2}_{\sigma\bar{\sigma}})\chi^{0}_{\sigma}(q)\chi^{0}_{\bar{\sigma}}(q)}
\end{align}
The charge and (longitudinal) spin susceptibilities are defined as $\chi_{\alpha}(q)=\chi_{dd}(q)+\chi_{ff}(q)\pm 2\chi_{df}(q)$, where $\alpha= [\text{ch},\ \text{sp}]$, 
\begin{align}
&\chi_{\alpha}(q)=\nonumber\\
&\frac{\big[\chi^{0}_{d}(q)+\chi^{0}_{f}(q)\mp 2 \Gamma_{df}\chi^{0}_{d}(q)\chi^{0}_{f}(q)\big]+2\Gamma_{dd}\chi^{0}_{f}(q)\chi^{0}_{d}(q)}{\big[1+\Gamma_{dd}(\chi^{0}_{d}(q)+\chi^{0}_{f}(q))+(\Gamma^{2}_{dd}-\Gamma^{2}_{df})\chi^{0}_{d}(q)\chi^{0}_{f}(q)\big]}
\end{align}
Now, using the canonical definitions of the charge and spin vertices, $\Gamma_{ch}=\Gamma_{df}+\Gamma_{dd}$, $\Gamma_{sp}=\Gamma_{df}-\Gamma_{dd}$, we get the following expressions for charge and (longitudinal) spin susceptibilities
\begin{widetext}
\begin{align}
\chi_{\alpha}(q)=\frac{\big[\chi^{0}_{d}(q)+\chi^{0}_{f}(q)\mp (\Gamma_{ch}+\Gamma_{sp}) \chi^{0}_{d}(q)\chi^{0}_{f}(q)\big]+(\Gamma_{ch}-\Gamma_{sp})\chi^{0}_{f}(q)\chi^{0}_{d}(q)}{\bigg[1+\frac{(\Gamma_{ch}-\Gamma_{sp})}{2}[\chi^{0}_{d}(q)+\chi^{0}_{f}(q)]-\Gamma_{ch}\Gamma_{sp}\chi^{0}_{d}(q)\chi^{0}_{f}(q)\bigg]}\label{A9}
\end{align}
\end{widetext}
($-$ve sign for charge and $+$ve sign for spin ($z$ component) susceptibility).
In the absence of pseudospin rotational symmetry, $\Gamma_{ch}$ and $\Gamma_{sp}$ get mixed in $\chi_{ch}$, $\chi_{sp}$. We can check the correctness of the above expression in the RPA limit, where $\Gamma_{ch}=\Gamma_{sp}=U$, the previous equation reduces to,
\begin{align*}
\chi^{\text{RPA}}_{\alpha}(q)=\frac{\chi_{d}^{0}(q)+\chi_{f}^{0}(q)\mp 2U\chi_{d}^{0}(q)\chi_{f}^{0}(q)}{1-U^{2}\chi_{d}^{0}(q)\chi_{f}^{0}(q)}
\end{align*}
which matches exactly with the RPA formula, which could be easily derived from the Feynman diagrammatic calculations.\\
\indent In TPSC theory, the unknown charge and spin vertex $\Gamma_{ch}\equiv U_{ch}$ and $\Gamma_{sp}\equiv U_{sp}$ are determined from the following sum rules,
\begin{align}
\frac{T}{L^{2}}\sum_{\mathbf{q}}\sum_{iq_{n}}[\chi_{ch}(\mathbf{q},iq_{n})+\chi_{sp}(\mathbf{q},iq_{n})]=2n-n^{2}\label{A10}
\end{align}
here $n=\langle n_{d}\rangle+\langle n_{f}\rangle$. To determine $U_{sp}, U_{ch}$, we solve the following two integral equations,
\begin{align}
&\frac{T}{L^{2}}\sum_{\mathbf{q}}\sum_{iq_{n}}\chi_{sp}(\mathbf{q},iq_{n},U_{ch},U_{sp})=n-2\langle n_{d}n_{f}\rangle\label{A11}\\
&\frac{T}{L^{2}}\sum_{\mathbf{q}}\sum_{iq_{n}}\chi_{ch}(\mathbf{q},iq_{n},U_{ch},U_{sp})=n+2\langle n_{d}n_{f}\rangle-n^{2}\label{A12}
\end{align}
for a given value of the double occupancy $\langle n_{d}n_{f}\rangle$. We need one more equation for $\langle n_{d}n_{f}\rangle$ to complete the above set. In TPSC, that equation turns out to be a local equation which does not change in the presence of pseudo-spin dependent hopping,
\begin{align}
U_{sp}=U\frac{\langle n_{d}n_{f}\rangle}{\langle n_{d}\rangle\langle n_{f}\rangle}\label{A13}
\end{align}
Solving \eqref{A11}, \eqref{A12}, and \eqref{A13}, we determine the self-consistent solutions for $U_{sp}, U_{ch}$, and $\langle n_{d}n_{f}\rangle$.\\

\textit{Expression for the transverse spin susceptibility}:\\

Next, we derive the expression for transverse spin susceptibility. To derive it, we at first consider it's RPA-level expression. The RPA longitudinal susceptibilities are found by summing the infinite set of ``bubble" diagrams and for the transverse ones ($\chi_{\pm\mp}\sim\langle S^{\pm}_{i}(\tau)S^{\mp}_{j}(0)\rangle$), one has to sum a specific set of ``ladder''-like diagrams in the particle-hole channel upto infinite order.
\begin{align}
&\chi^{+-}(q)=\nonumber\\
&\frac{1}{L^{2}}\sum_{\mathbf{k},\mathbf{k}'}\int_{0}^{\beta}d\tau e^{iq_{n}\tau}\langle T_{\tau} d^{\dag}_{\mathbf{k}}(\tau)f_{\mathbf{k}+\mathbf{q}}(\tau)f^{\dag}_{\mathbf{k}'}(0)d_{\mathbf{k}'-\mathbf{q}}(0)\rangle\label{A14}
\end{align}
After summing the infinite set of particle-hole ladder diagrams, one finds
\begin{align}
\chi^{+-}_{RPA}(q)=\frac{\Phi^{0}_{df}(q)}{1-U\Phi^{0}_{df}(q)}\label{A15}
\end{align}
Here 
\begin{align}
\Phi^{0}_{df}(\mathbf{q},iq_{n})&=-\frac{1}{\beta L^{2}}\sum_{\mathbf{k}, ik_{n}}G^{0}_{f}(\mathbf{k},ik_{n})G^{0}_{d}(\mathbf{k}+\mathbf{q},ik_{n}+iq_{n})\nonumber\\
&=-\frac{1}{\beta}\sum_{ik_{n}}G^{0,\text{loc}}_{f}(ik_{n})G^{0,\text{loc}}_{d}(ik_{n}+iq_{n})
\end{align}
The non-interacting transverse spin susceptibility turns out to be a local quantity. The reason behind this locality lies in the {\it phase space decoupling} of the momentum summation, which arises from the directional nature of the fermionic band structure. From Eq.\eqref{A14}, it becomes clear that a particle-hole pair of opposite pseudo-spins created at site $i$ and $\tau=0$ will never meet again at any later time $\tau>0$ and for any $j\neq i$ due to orthogonal band motion of the fermions, when they are non-interacting. 
\\
\indent In TPSC, the bare $U$ in \eqref{A15} is replaced by a renormalized $U_{t}$. The transverse susceptibility is expected to satisfy the following sum rule,
\begin{align}
\frac{1}{\beta}\sum_{iq_{n}}\frac{\Phi^{0}_{df}(iq_{n})}{1-U_{t}\Phi^{0}_{df}(iq_{n})}=\langle d^{\dag}_{i}f_{i}f^{\dag}_{i}d_{i}\rangle=\langle n_{d}\rangle-\langle n_{d}n_{f}\rangle
\end{align}
We find that $U_{t}$ is negligible compared to $U_{ch}$ and $U_{sp}$ (in the fermionic QCM). Due to momentum independence of $\chi^{x(y)}(iq_{n})$, we don't observe any singularity in this particular particle-hole channel. The reason lies in the absence of pseudo-spin rotational symmetry. The $\chi^{x(y)}(iq_{n})$ has some quantitative effects on the single-particle self-energy (will give a non-singular, frequency dependent contribution), but not in the description of ordering instabilities.\\
\begin{figure}
\centering
\includegraphics[width=65mm]{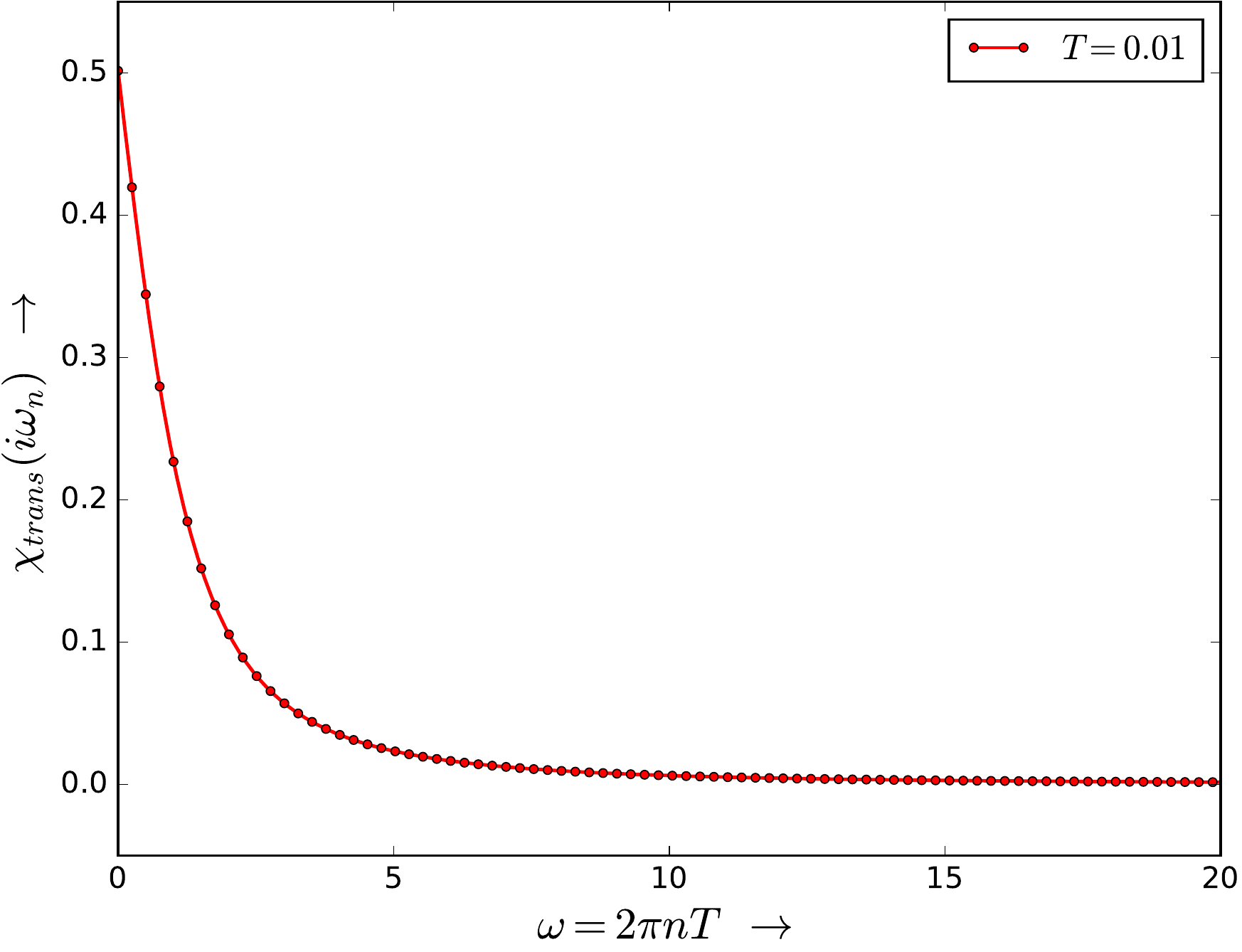}
\caption{Transverse spin susceptibility, $\chi_{trans}(i\omega_{n})$ vs. $\omega_{n}=2\pi nT$ for $T=0.01$. Compared to $\chi_{sp}$ (see Fig.\ref{chis}), $\chi_{trans}$ does not show any singularity even at lower temperatures.}\label{chit}
\end{figure}

\textit{Expression for the self-energy}:\\

The longitudinal part of the single-particle self-energy could be formally expressed by the following relation,

\begin{align}
\Sigma^{l}_{\sigma}(1,2)=&U n_{-\sigma}\delta(1-2)\nonumber\\
&+UG_{\sigma}(1-\bar{2})\Gamma^{ir}_{\sigma\bar{\sigma}}(\bar{2},2;\bar{4},\bar{5})\chi_{\bar{\sigma},-\sigma}(\bar{4},\bar{5},1)
\end{align}
Using the local (real space) approximation \textit{ansatz} for the vertex functions \eqref{A2}, we find after some straightforward algebra the following,
\begin{align}
&\Sigma^{l}_{\sigma}(k)=\nonumber\\
& U n_{\bar{\sigma}}+\frac{U}{4\beta L^{2}}\sum_{q}\big[U_{ch}\chi_{ch}(q)+U_{sp}\chi_{sp}(q)\big]G_{\sigma}(k+q)\nonumber\\
&-\frac{U}{4\beta L^{2}}(U_{ch}+U_{sp})\sum_{q}[\chi_{\sigma\sigma}(q)-\chi_{\bar{\sigma}\bar{\sigma}}(q)]G_{\sigma}(k+q)
\end{align}
In the above, the last term in $\Sigma^{l}$ arises due to absence of spin rotational invariance. We find that this term has no significant effect in the renormalized-classical regime where only $\chi_{sp}$ controls the leading (singular) behavior.\\
\indent The transverse component of the self-energy is local due to the momentum independence of transverse susceptibility. Its expression is given as follows
\begin{align}
\Sigma^{(t)}_{\sigma}(ik_{n})=\frac{U}{\beta }\sum_{iq_{n}}\frac{U_{t}\Phi^{0}_{df}(iq_{n})}{1-U_{t}\Phi^{0}_{df}(iq_{n})}G^{0,loc}_{-\sigma}(ik_{n}+iq_{n})
\end{align}
\section{Supplementary results from TPSC calculations}\label{B}
\indent The non-interacting susceptibility of the individual orbitals $(d,f)$ show logarithmic $T$-divergence due to the completely nested one dimensional Fermi surface, i.e., $\chi^{\sigma}_{0}(Q)\sim N(0)\ln{(E_{0}/T)}$; here, $N(0)=1/2\pi t$ is equal to the density of states at the Fermi surface (here Fermi points) and $E_{0}\sim 2t$, the half-bandwidth. As a result, one expects instabilities to occur at the corresponding wave-vectors. Here, we show the interacting longitudinal spin susceptibility, $\chi_{sp}(\mathbf{q},i\nu=0)$ (see Eq.\eqref{A9}) for certain $(T,U)$ values (one slightly inside the $\xi>>\xi_{th}$ regime and the other one for the opposite regime).
\begin{figure}
\centering
\includegraphics[width=65mm]{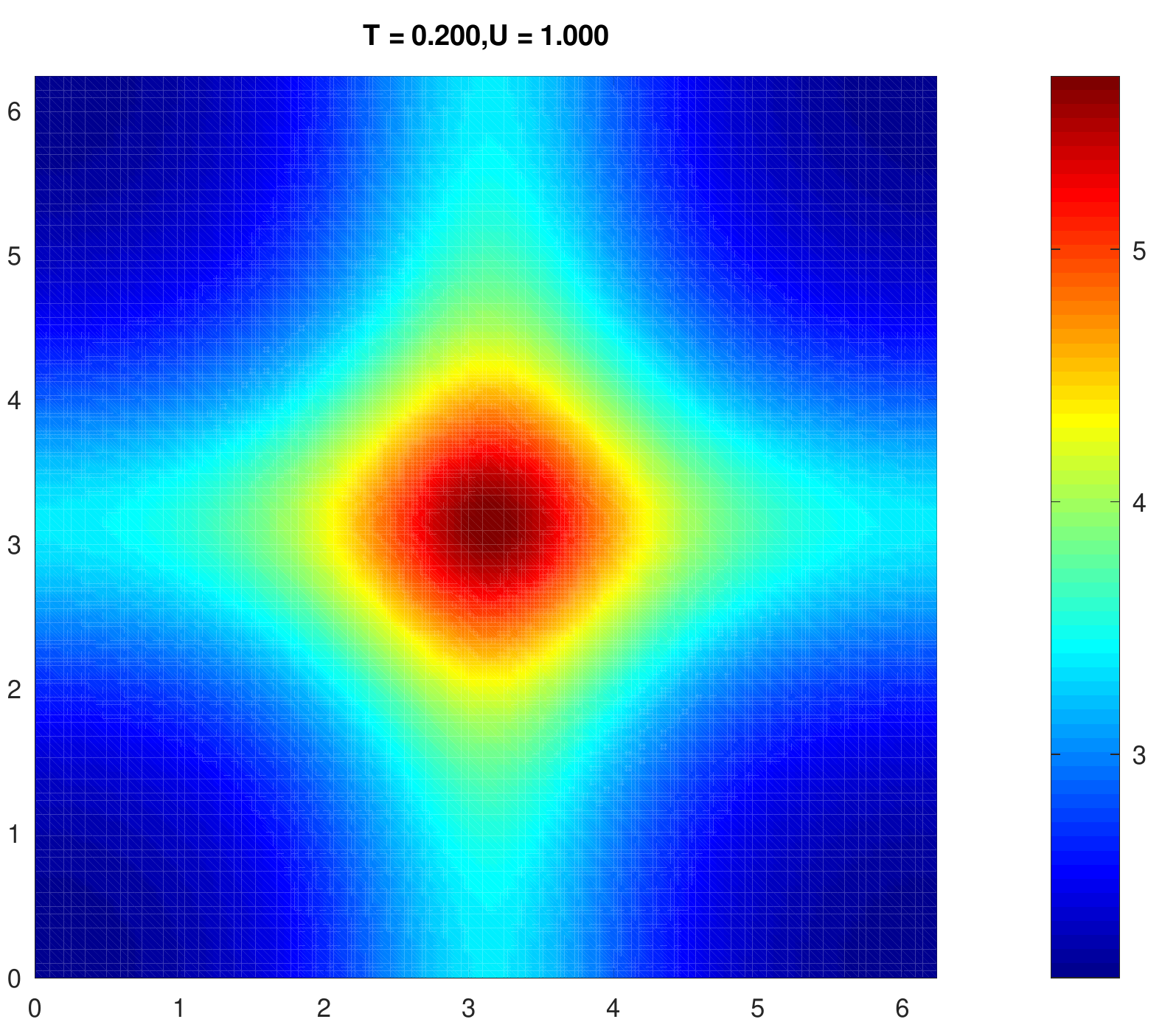}
\includegraphics[width=65mm]{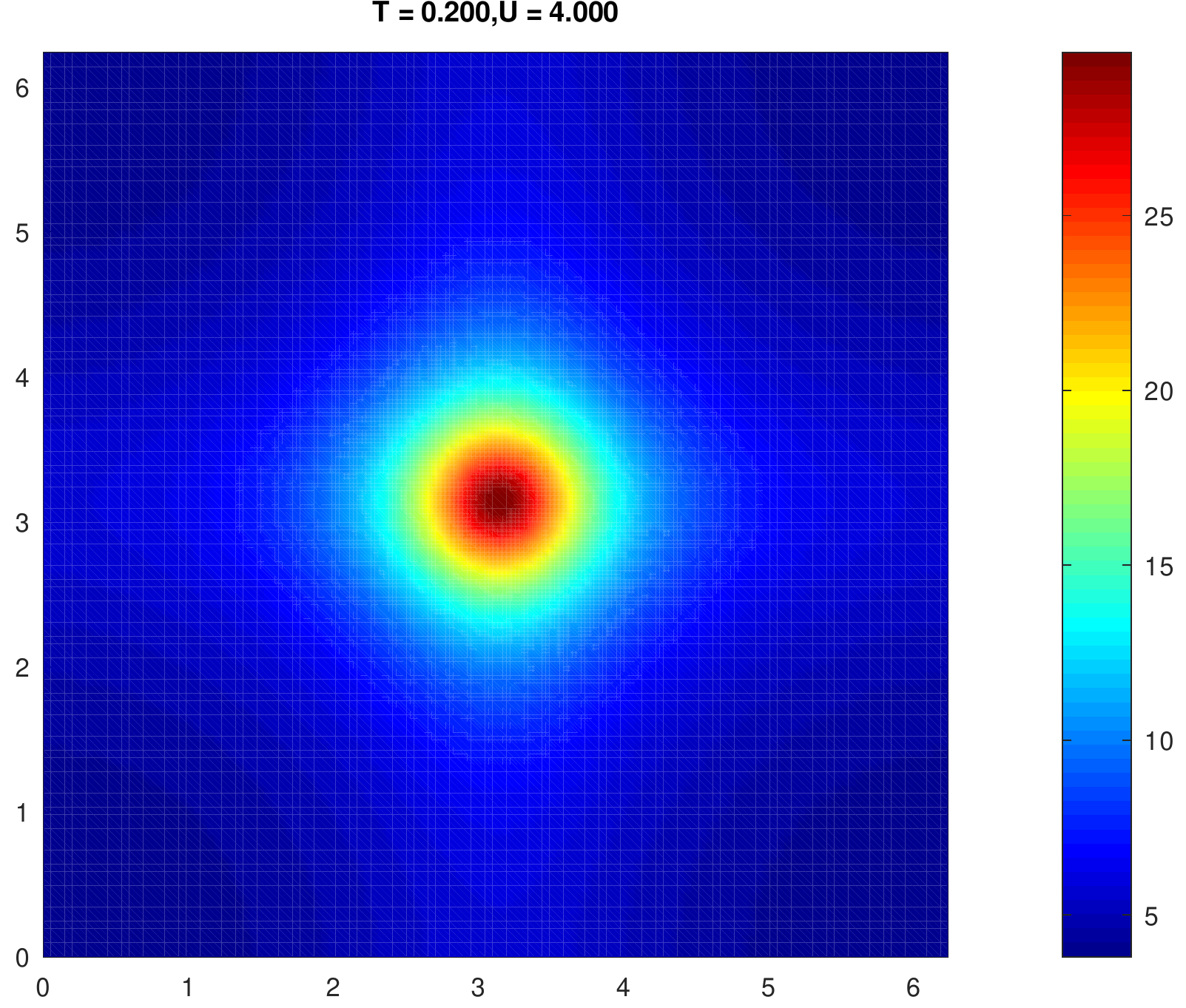}
\caption{Interacting $\chi_{sp}(q_{x},q_{y},i\nu=0)$ calculated by solving TPSC self-consistency equations (Eqs. \eqref{A11}-\eqref{A13}). (a) $T=0.2$ and $U=1.0$, in the $\xi<\xi_{th}$ regime (upper figure), (b) $T=0.2$ and $U=4.0$, in the $\xi>\xi_{th}$ regime (lower figure). In the high-$T$ regime, a broad diffused maximum in the scattering structure factor indicates a disordered state with short correlation length (upper). At low-$T$, this maximum sharpens up at $\mathbf{Q}=(\pi,\pi)$, testifying to onset of orbital order (lower).}\label{chis}.  
\end{figure}
As one can see from Fig.\ref{chis} (lower one) that $\chi_{sp}$ obtain very large numerical values near $(\pi,\pi)$ inside the renormalized classical regime ($\xi>>\xi_{th}$). This means the system develops a strong nearest-neighbour $2d$ anti-ferro orbital correlations in this regime.\\
\indent Next, We observe that $U_{ch}$ and $U_{sp}$ are almost same for very small $U$. Then, as $U$ increases, the charge vertex rises rapidly, whereas the spin vertex saturates to a finite value (see Fig.\ref{figUch}). The rise of $U_{ch}$ corresponds suppression of charge collective modes with the increase in $U$. The saturation of $U_{sp}$ is in accord with Kanamori-Bruckner renormalization phenomenon.\\
\begin{figure}[h!]
\centering
%\subfigure{\label{Fig.S1a}\includegraphics[width=60mm]{Uch_vs_U_lambda_0-cropped.pdf}}
%\subfigure{\label{Fig.S1b}\includegraphics[width=60mm]{Usp_vs_U_lambda_0-cropped.pdf}}
\includegraphics[width=65mm]{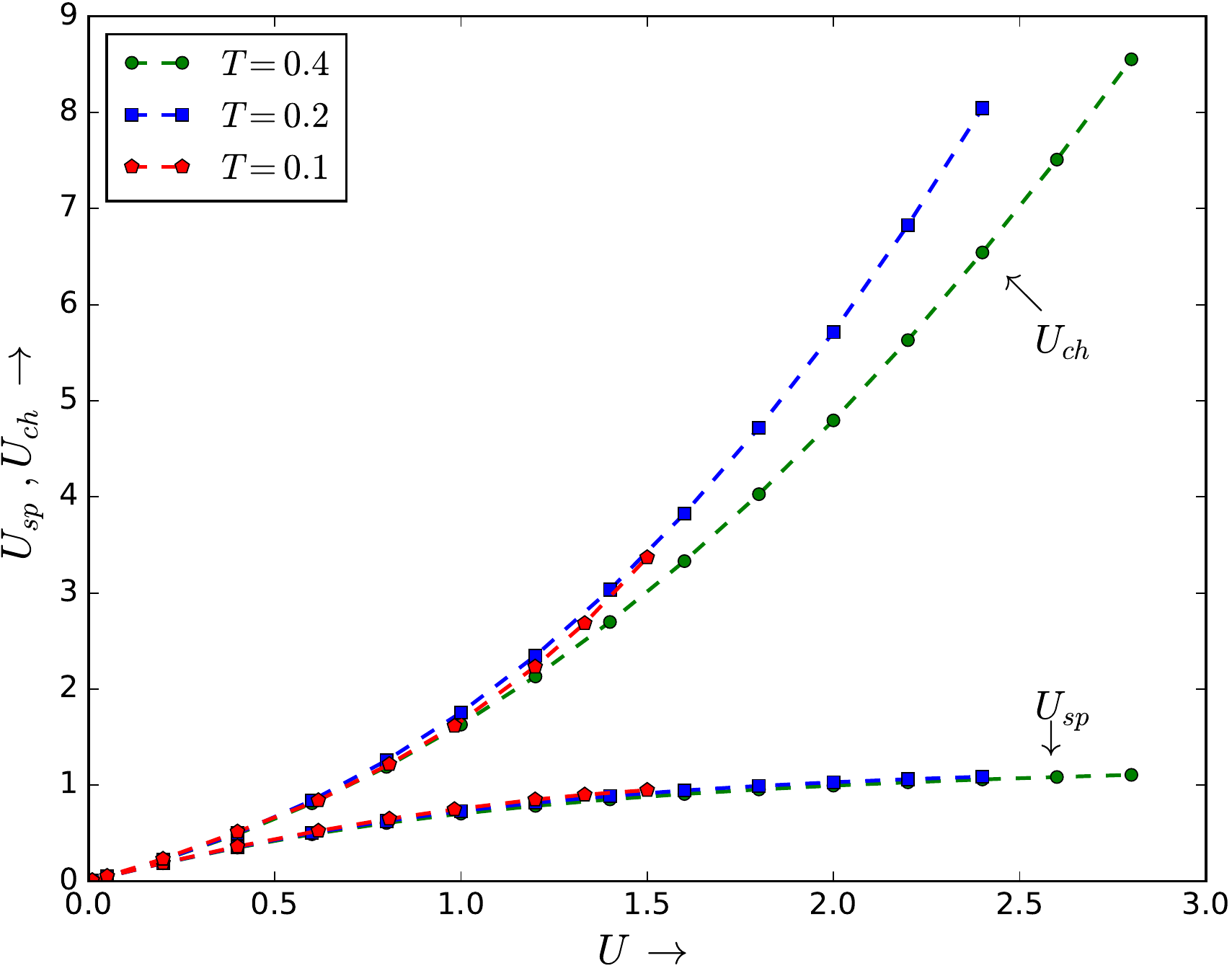}
\caption{$U_{ch}$, $U_{sp}$ vs. $U$ calculated from self-consistency equations (Eqs.\eqref{A10}-\eqref{A12}) for different $T$ values.}\label{figUch}
\end{figure}
We calculate the charge and spin correlation lengths, defined by $\xi_{ch}\sim \sqrt{\chi_{ch}(\mathbf{Q},i\nu=0)/\chi^{0}_{ch}(\mathbf{Q},i\nu=0)}$ and $\xi_{sp}\sim \sqrt{\chi_{sp}(\mathbf{Q},i\nu=0)/\chi^{0}_{sp}(\mathbf{Q},i\nu=0)}$, where $\mathbf{Q}=(\pi,\pi)$ is the vector where $\chi(q,\nu=0)$ is maximum, at half-filling. We see the suppression of charge fluctuations as $U$ is increased, and also as temperature is lowered (see Fig.\ref{xich}). Although, TPSC cannot capture metal-insulator phase transition (which involves frequency dependent vertex corrections), it correctly produces the low-$T$ collective behaviors.\\
\begin{figure}
\centering
\includegraphics[width=65mm]{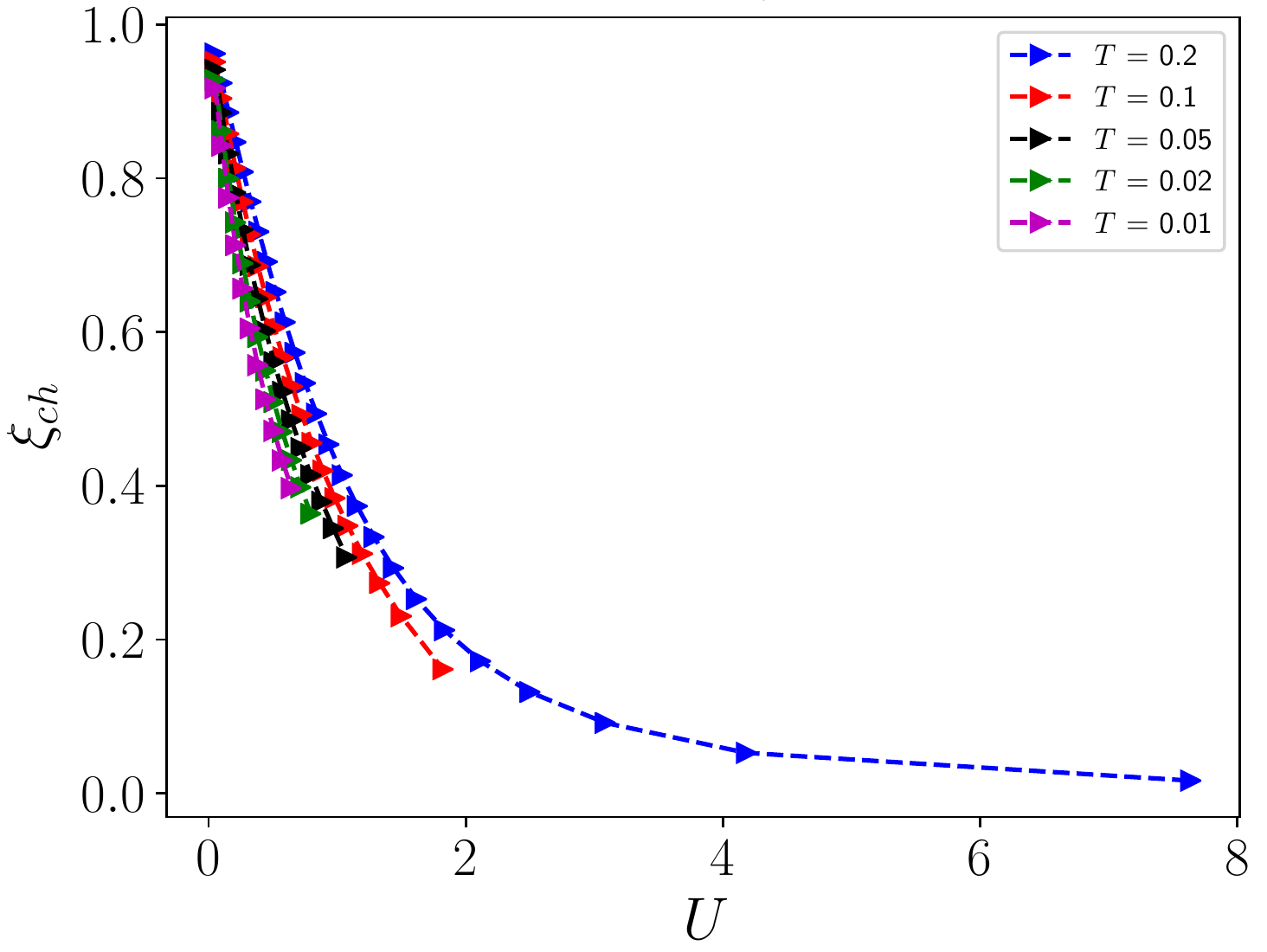}
\caption{charge correlation length, $\xi_{ch}$ vs. $U$ at half-filling for different $T$s, showing almost no sensitivity to $T$.}\label{xich}
\end{figure}
\indent On the other side, $\chi_{sp}$ shows opposite behavior (see Fig.\ref{xisp}), it grows exponentially as $T$ is lowered. The system enters into the so-called renormalized classical regime where the wavelength of orbital pseudo-spin fluctuations becomes larger than thermal de-Broglie wavelength of the excitations. One expects around this temperature, at half-filling, some crossover happens between a Fermi liquid at high-$T$ to a low-$T$ pseudo-gap phase driven by anti-ferro ``orbital'' correlations.\\
\begin{figure}[h!]
\centering
\includegraphics[width=65mm]{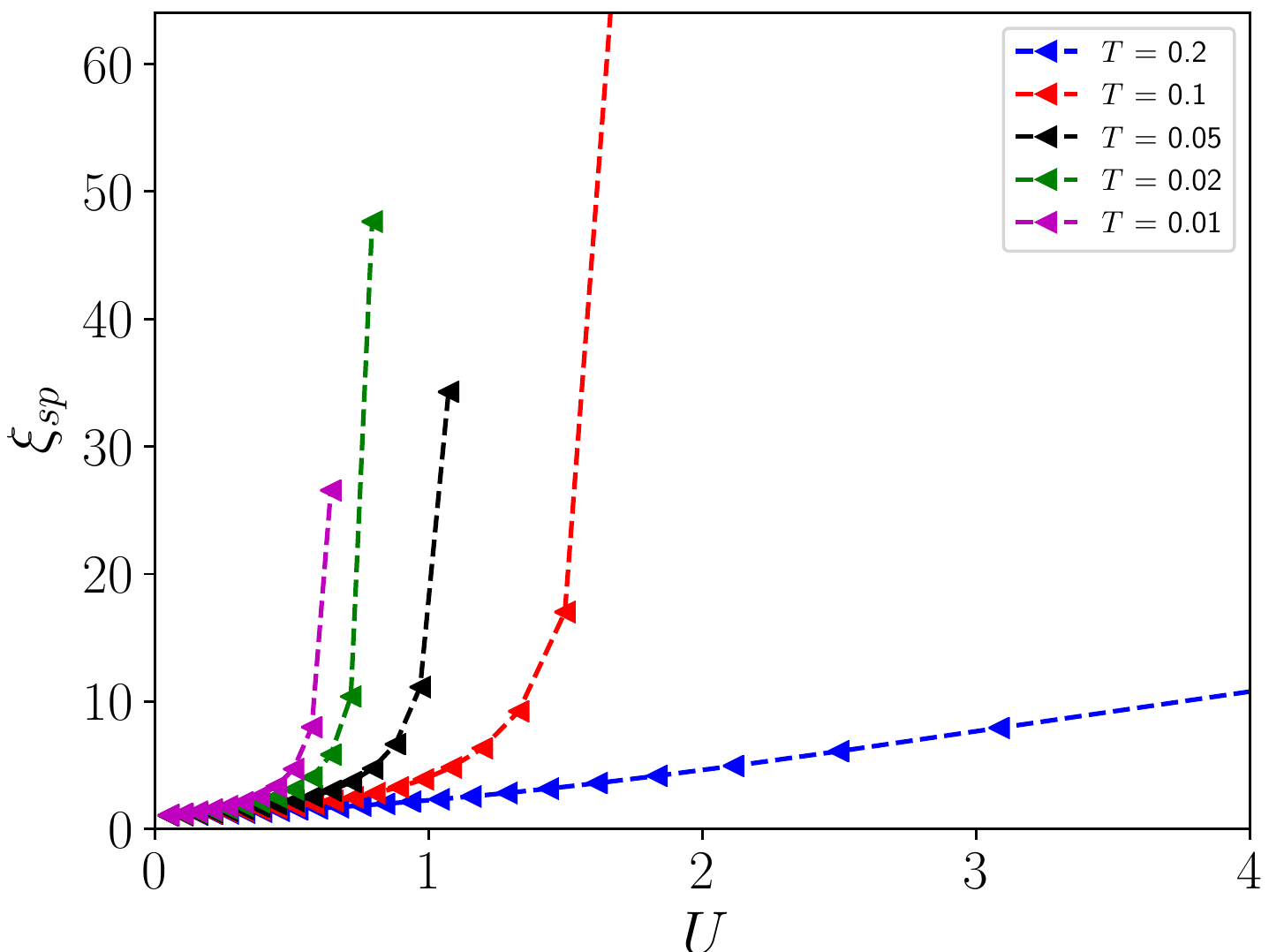}
\caption{longitudinal spin correlation length, $\xi_{sp}$ vs. $U$ at half-filling, for different $T$s. Onset of orbital order within TPSC is signalled by a finite $T$ divergence of $\xi_{sp}$. The $T-U$ phase diagram in the main text is constructed from this observation.}\label{xisp}
\end{figure}
We have used $128\times 128$ spatial lattice grid and $N_{\tau}=8192$ time grid to perform the above numerical calculations. As $\xi_{sp}$ diverges exponentially at low-$T$, it becomes increasingly difficult to correctly capture some low-$T$ data points with the above lattice size. The quantitative accuracy could be increased by choosing much larger lattice grid without much change in the qualitative behavior.\\
\indent Finally, we show the double occupancy, $\langle n_{d}n_{f}\rangle$ vs. $U$, for different $T$ (see Fig.\ref{docc}). It decreases continuously from $n^{2}/4$ as $U$ is raised. Suppression of $\langle n_{d}n_{f}\rangle$ at higher-$T$ values ($\sim T=0.4$) compared to low-$T$ is consistent with the self-energy results (shown in the main text) which finds incoherent, insulator-like features (suppression of charge fluctuations) at higher-$T$ values. Here, the ($T,U$) points in Fig.\ref{docc} lie on the $\xi<\xi_{th}$ part of the phase diagram.
\begin{figure}
\centering
\includegraphics[width=65mm]{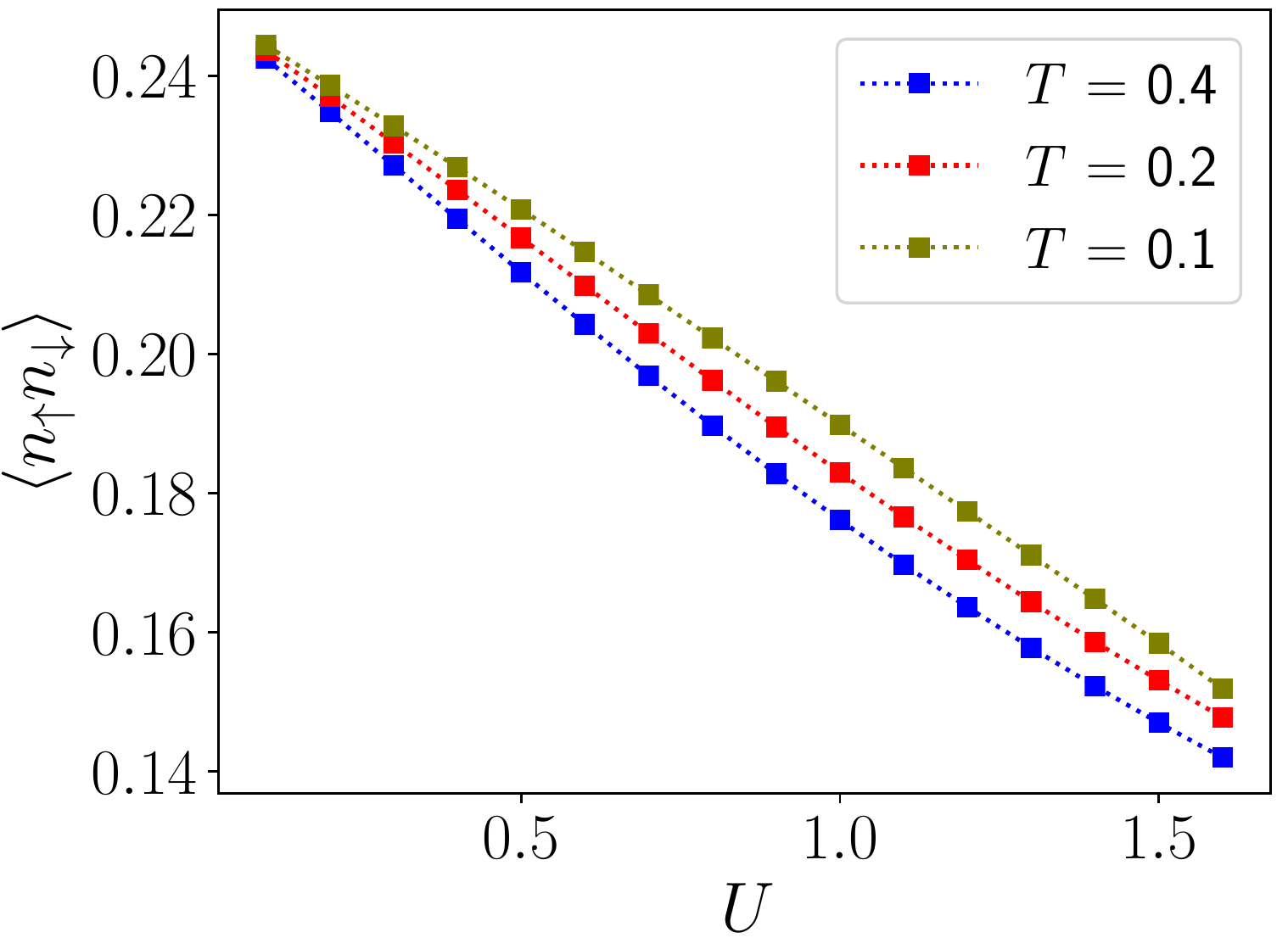}
\caption{Double occupancy $\langle n_{\uparrow}n_{\downarrow}\rangle$ vs. $U$ for different $T=0.1,0.2,0.4$, at half-filling (here $\uparrow \equiv d$, and $\downarrow \equiv f$)}\label{docc}
\end{figure}
\section{Self-energy in the $\xi>\xi_{th}$ regime}\label{Self-energy-pseudogap}
In this section, we will derive the analytical expression of the single-particle self-energy in the renormalized-classical (RC) regime, i.e. $\xi>>\xi_{th}$ following Ref. \cite{Tremblay}. In the present context, the $G^{(0)}$ has one dimensional $k$-dependence, although the susceptibilities are two dimensional functions. Close to this RC regime, spin correlations dominate, and the long wavelength and low-energy longitudinal spin susceptibility obtains the following form \cite{Tremblay2},
\begin{align}
\chi_{sp}(\mathbf{q},i\omega)\approx\frac{2\xi^{2}}{U_{sp}\xi_{0}^{2}}\frac{1}{1+\mathbf{(q-\mathbf{Q})}^{2}\xi^{2}-i\omega/\omega_{sp}}
\end{align}
Here, $\omega_{sp}=D/\xi^{2}$, the characteristic spin relaxation frequency and $D=(\tau_{0}/\xi^{2}_{0})^{-1}$, the microscopic diffusion co-efficient. The microscopic correlation length is $\xi_{0}=\frac{-1}{2\chi_{0}(\mathbf{Q})}\frac{\partial^{2}\chi_{0}(\mathbf{q})}{\partial q^{2}}\big|_{q=\mathbf{Q}}$ and microscopic correlation time is $\tau_{0}=\frac{1}{\chi_{0}(\mathbf{Q})}\frac{\partial\chi_{0}(\mathbf{Q,\omega})}{\partial \omega}\big|_{\omega=0}$\\

Close to the RC regime, $\xi\sim \exp\big[\pi\xi^{2}_{0}\sigma^{2}(T)U_{sp}/T\big]$, where $\sigma(T)$ is the (weakly) temperature dependent spin-stiffness. This implies $\omega_{sp}<<T$ and the susceptibility is dominated by the classical ($i\omega_{n}=0$) part, $\chi_{sp}(\mathbf{q},0)$. The self-energy is mostly affected by $\chi_{sp}$ compared to $\chi_{ch}$, which could be expressed as following,
\begin{align}
&\Sigma_{\sigma}(\mathbf{k},ik_{n})\approx\nonumber\\
& Un_{-\sigma}+\frac{U}{4}\frac{1}{\beta N}\sum_{\mathbf{q}}U_{sp}\chi_{sp}(\mathbf{q},0)G^{(0)}_{\sigma}(k_{\sigma}+q_{\sigma},ik_{n})\nonumber\\
&+ \frac{U}{4}\frac{1}{\beta N}\sum_{\mathbf{q}}\sum_{q_{n}\neq 0}U_{sp}\chi_{sp}(\mathbf{q},iq_{n})G^{(0)}_{\sigma}(k_{\sigma}+q_{\sigma},ik_{n}+iq_{n})\label{C2}
\end{align}
As argued above, we neglect the $iq_{n}\neq 0$ terms, as $\omega_{sp}<<T$ and the classical (thermal) spin correlations dominate. 
\begin{align*}
&\Sigma_{\sigma}(\mathbf{k},ik_{n})\approx \\
&\ \ \ \frac{UT}{2\xi^{2}_{0}}\int \frac{d^{2}q}{(2\pi)^{2}}\frac{1}{q^{2}+\xi^{-2}}\frac{1}{ik_{n}-\xi_{\mathbf{k}+\mathbf{Q}}-q_{\sigma}v^{\sigma}_{\mathbf{k}+\mathbf{Q}}}
\end{align*}
Close to the Fermi surface, $k\approx k_{F}$, $\mathbf{Q}=(\pi,\pi)$, so $\xi_{\mathbf{k}_{F}+\mathbf{Q}}=0$ and $v^{\sigma}_{\mathbf{k}_{F}+\mathbf{Q}}=v_{F}$,
%\begin{widetext}
\begin{align*}
&\Sigma_{\sigma}(\mathbf{k},ik_{n})\\
&=\frac{UT}{2\xi^{2}_{0}v_{F}}\int_{-\infty}^{\infty}\frac{dq_{\sigma}}{2\pi}\frac{1}{i(k_{n}/v_{F})-q_{\sigma}}\int_{-\infty}^{\infty}\frac{dq_{\bar{\sigma}}}{2\pi}\frac{1}{q_{\bar{\sigma}}^{2}+q_{\sigma}^{2}+\xi^{-2}}\\
&=\frac{UT}{2\xi^{2}_{0}v_{F}}\int_{-\infty}^{\infty}\frac{dq_{\sigma}}{2\pi}\frac{1}{\sqrt{q_{\sigma}^{2}+\xi^{-2}}[i(k_{n}/v_{F})-q_{\sigma}]}
\end{align*}
%\end{widetext}
Now, $\int_{-\infty}^{\infty}\frac{dx}{\sqrt{x^{2}+a^{2}}(ib-x)}=-\frac{i}{\sqrt{a^{2}-b^{2}}}\cos^{-1}(b/a)$, $\cos^{-1}{x}=\tan^{-1}(\sqrt{1-x^{2}}/x)$, and $\tan^{-1}(z)=-\frac{1}{2i}\ln{[(i+z)/(i-z)]}$, using these identities, we obtain 
%\begin{widetext}
\begin{align}
&\Sigma_{\sigma}(\mathbf{k}_{F},ik_{n})=\nonumber\\
&\ \ \ -i\frac{UT}{8\pi\xi^{2}_{0}\sqrt{k_{n}^{2}-v^{2}_{F}\xi^{-2}}}\ln{\bigg[\frac{k_{n}+\sqrt{k_{n}^{2}-v^{2}_{F}\xi^{-2}}}{k_{n}-\sqrt{k_{n}^{2}-v^{2}_{F}\xi^{-2}}}\bigg]}\label{C3}
\end{align}
%\end{widetext}
Now, $k_{n}$ is linear in $T$, but $\xi^{-1}\sim e^{-C/T}$ ($C>0$), so $k_{n}>>\xi^{-1}$ whenever $\xi>>\xi_{th}$. The self-energy in Matsubara frequency can be also expressed as
%\begin{widetext}
\begin{align}
\Sigma_{\sigma}(\mathbf{k}_{F},ik_{n})=-i&\frac{U}{8(2n+1)\pi^{2}\xi^{2}_{0}}\bigg[1-\frac{\xi^{2}_{th}}{\xi^{2}}\frac{1}{(2n+1)^{2}}\bigg]^{-\frac{1}{2}}\nonumber\\
&\ \ \ \times\ln{\bigg[\frac{1+\sqrt{1-\frac{\xi^{2}_{th}}{\xi^{2}}\frac{1}{(2n+1)^{2}}}}{1-\sqrt{1-\frac{\xi^{2}_{th}}{\xi^{2}}\frac{1}{(2n+1)^{2}}}}\bigg]}
\end{align}
%\end{widetext}
so, when $\xi_{th}<<\xi$, $\sqrt{1-\frac{\xi^{2}_{th}}{\xi^{2}}\frac{1}{(2n+1)^{2}}}\approx 1-\frac{\xi^{2}_{th}}{2\xi^{2}}\frac{1}{(2n+1)^{2}}$. Thus,
\begin{align}
\Sigma_{\sigma}(\mathbf{k}_{F},ik_{n})\approx-i\frac{U}{8(2n+1)\pi^{2}\xi^{2}_{0}}\ln{\big[4(2n+1)\xi/\xi_{th}\big]}
\end{align}
Hence, for $k_{n}=\pi T$, $\Sigma_{\sigma}(\mathbf{k}_{F},ik_{n=0})\sim -i\frac{U}{8\pi^{2}\xi^{2}_{0}}\ln{(4\xi/\xi_{th})}$ which diverges as $T\rightarrow 0$.
When $\xi_{th}/\xi\sim O(1)$, the matsubara self-energy at lowest Matsubara frequency $k_{n}=\pi T$, we may have a different behavior. We want to see what happens if $k_{n=0}\rightarrow v_{F}/\xi$ in \eqref{C3} (which is same as $\xi\rightarrow \xi_{th}$), we find $\Sigma_{\sigma}(k_{F},ik_{n=0})=-i\frac{UT}{2\pi\xi^{2}_{0}}+f(i\omega_{n})$, where $f(\mathbf{k}_{F},i\omega_{n})$ is the regular part of the self-energy which comes from the $iq_{n}\neq 0$ contributions of $\chi$, in Eq.\eqref{C2}. So, expanding $f(k_{F},i\omega_{n})\approx -C$ near $\omega_{n}=0$,
\begin{align}
-\text{Im}\Sigma_{\sigma}(k_{F},i\omega_{n}\approx 0)\bigg|_{\xi/\xi_{th}\sim O(1)}\approx C+\frac{UT}{2\pi\xi^{2}_{0}}
\end{align}
%%%%%%%%%%%%%%%%%%%%%%%%%%%%%%%%%%%%%%%%%%%%%%%%%%%%%%%%%%%%%%%%%%%%%%%%%%%%%
%%%%%%%%%%%%%%%%%%%%%%%%%%%%%%%%%%%%%%%%%%%%%%%%%%%%%%%%%%%%%%%%%%%%%%%%%%%%%%
\section{Derivation of the strong-coupling effective spin Hamiltonian from fermionic QCM}\label{C}
We consider the following generalized version of fermionic QCM which has the $p$-wave Cooper pairing of strength $\lambda$ (here, $|\lambda|\leq 1$),
\begin{align}
H=&t\sum_{i}(\lambda d_{i}d_{i+\hat{x}}-d^{\dag}_{i}d_{i+\hat{x}}+h.c.)\nonumber\\
&+t\sum_{i}(\lambda f_{i}f_{i+\hat{y}}-f^{\dag}_{i}f_{i+\hat{y}}+h.c.)\nonumber\\
&+U\sum_{i}(2d^{\dag}_{i}d_{i}-1)(2f^{\dag}_{i}f_{i}-1)
\end{align}
We write the complex fermions in the Majorana basis, $f_{i}=(A^{b}_{i}+iA^{w}_{i})/2$, $d_{i}=(B^{w}_{i}+iB^{b}_{i})/2$. Here, $(A^{b}_{i})^{2}=(A^{w}_{i})^{2}=(B^{b}_{i})^{2}=(B^{w}_{i})^{2}=1$. We get the following Majorana Hamiltonian, $H=H_{0}+V_{x}+V_{y}$, where
\begin{align}
&H_{0}=-UA^{w}_{i}A^{b}_{i}B^{b}_{i}B^{w}_{i}\\
&V_{x}=\frac{it}{2}(1+\lambda)(B^{b}_{i}B^{w}_{i+\hat{x}}+ A^{w}_{i}A^{b}_{i+\hat{y}}),\\
&V_{y}=-\frac{it}{2}(1-\lambda)(B^{w}_{i}B^{b}_{i+\hat{x}}+A^{b}_{i}A^{w}_{i+\hat{y}})
\end{align}
We take the limit, $U>>t$ and develop a strong-coupling perturbative expansion. The ground state subspace, in the limit, $U>>t$, corresponds to either $\ket{n_{id}=0,n_{if}=1}$ or $\ket{n_{id}=1,n_{if}=0}$ at each lattice site, $i$. It also satisfies, $A^{w}_{i}A^{b}_{i}B^{b}_{i}B^{w}_{i}=1$, i.e. even Majorana parity sector. One can treat the local subspace as some effective spin-1/2 degrees of freedom, which in the Majorana basis could be represented by the following relations, $\sigma^{z}_{i}=iB^{b}_{i}B^{w}_{i}=iA^{b}_{i}A^{w}_{i},\sigma^{x}_{i}=iB^{b}_{i}A^{w}_{i}=iB^{w}_{i}A^{b}_{i},
\sigma^{y}_{i}=iA^{w}_{i}B^{w}_{i}=-iA^{b}_{i}B^{b}_{i}$.\\
\indent The above representation is a faithful one under the constraint, $A^{w}_{i}A^{b}_{i}B^{b}_{i}B^{w}_{i}=1$, which has to be satisfied in the large $U$ limit. We define the projector onto the $2^{N}$ dimensional ground state subspace, $P=\frac{1}{2^{N}}\prod_{i=1}^{N}(1+A^{w}_{i}A^{b}_{i}B^{b}_{i}B^{w}_{i})$. Under the action of $V_{x}$ and $V_{y}$, the Majorana fermion parities flip on both the ends of the link $(i,i+\alpha)$.  In order to determine the effective low energy Hamiltonian, we have to figure out all possible ways of restoring the GS parity, at various orders of the perturbation theory. One can verify that the odd order terms don't contribute because the GS parity restriction is not maintained. In the second order, there are two contributions, (1) which just gives a constant shift, this arise when we apply the same link hopping twice to return to the GS parity sector. (2) Another contribution gives an anti-ferromagnetic Ising coupling. For example, consider the following process along the x-directions,
\begin{align*}
\frac{1}{4}i(-i)(1-\lambda^{2})B^{b}_{i}B^{w}_{i+\hat{x}}B^{w}_{i}B^{b}_{i+\hat{x}}=-(1-\lambda^{2})\sigma^{z}_{i}\sigma^{z}_{i+x}
\end{align*}
The excitation energy to flip the parities at any two site is $4U$. So, in the second order,
\begin{align}
H_{eff}^{(2)}=\frac{(1-\lambda^{2})t^{2}}{8U}\sum_{i}\big(\sigma^{z}_{i}\sigma^{z}_{i+\hat{x}}+\sigma^{z}_{i}\sigma^{z}_{i+\hat{y}}\big)
\end{align}
Similarly, the fourth-order effective Hamiltonian can be derived from
\begin{align}
H^{(4)}_{eff}=&PV(1-P)\frac{1}{E_{0}-H_{0}}V(1-P)\frac{1}{E_{0}-H_{0}}\nonumber\\
&\times V(1-P)\frac{1}{E_{0}-H_{0}}(1-P)VP
\end{align}
Here, there are several contributions; these come from terms like (1): $\tilde{V}_{x}^{4}$, (2): $\tilde{V}_{x}^{3}\tilde{V}_{y}$, (3): $\tilde{V}_{x}\tilde{V}^{3}_{y}$, (4): $\tilde{V}_{x}^{2}\tilde{V}_{y}^{2}$, (5): $\tilde{V}_{y}^{4}$, where the tilde denotes the projected operators. The appropriate energy denominator will be multiplied later.
\begin{align*}
&(1)=\frac{t^{4}(1+\lambda)^{4}}{16}\sigma^{x}_{i}\sigma^{y}_{i+\hat{x}}\sigma^{x}_{i+\hat{x}+\hat{y}}\sigma^{y}_{i+\hat{y}}\\
&(2)+(3)=-\frac{2}{16}t^{4}(1-\lambda^{4})[ \sigma^{y}_{i}\sigma^{x}_{i+\hat{x}}\sigma^{x}_{i+\hat{x}+\hat{y}}\sigma^{y}_{i+\hat{y}}\\
&\ \ +\sigma^{x}_{i}\sigma^{x}_{i+\hat{x}}\sigma^{y}_{i+\hat{x}+\hat{y}}\sigma^{y}_{i+\hat{y}}+\sigma^{y}_{i}\sigma^{y}_{i+\hat{x}}\sigma^{x}_{i+\hat{x}+\hat{y}}\sigma^{x}_{i+\hat{y}}\\
&\ \ +\sigma^{x}_{i}\sigma^{y}_{i+\hat{x}}\sigma^{y}_{i+\hat{x}+\hat{y}}\sigma^{x}_{i+\hat{y}}]\\
&(4)=-\frac{2}{16}t^{4}(1-\lambda^{4})[ \sigma^{y}_{i}\sigma^{y}_{i+\hat{x}}\sigma^{y}_{i+\hat{x}+\hat{y}}\sigma^{y}_{i+\hat{y}}\\
&\ +\sigma^{x}_{i}\sigma^{x}_{i+\hat{x}}\sigma^{x}_{i+\hat{x}+\hat{y}}\sigma^{x}_{i+\hat{y}}-\sigma^{y}_{i}\sigma^{x}_{i+\hat{x}}\sigma^{y}_{i+\hat{x}+\hat{y}}\sigma^{x}_{i+\hat{y}} ]\\
&(5)=\frac{t^{4}(1-\lambda)^{4}}{16}\sigma^{x}_{i}\sigma^{y}_{i+\hat{x}}\sigma^{x}_{i+\hat{x}+\hat{y}}\sigma^{y}_{i+\hat{y}}
\end{align*}
After adding all the above contributions (with the appropriate energy denominators) and multiplying a combinatorial factor, which counts the different possible orders of the link operators ($V_{x}, V_{y}$) in the above (1)-(5) terms, we finally arrive at the effective low-energy Hamiltonian (in the large $U/t$) given by \eqref{Heff} in the main text. Notice that the link operators all commute with each other, so their actual order of operation is unimportant. This is in contrast to the Kitaev model, where such simplifications do not happen.\\
\indent When $\lambda\rightarrow 1$ (equal hopping and pairing limit), we obtain the Wen's plaquette model (which is the same as toric code).\\
\indent To transform Wen's plaquette model to the canonical toric code form (see Eqn. \eqref{Htc}), we apply some local spin basis rotations. To perform it, we divide the square lattice into the union of alternate plaquettes, which are labeled as $a$ and $b$. On the ``$a$'' plaquettes , we rotate the spins on the sites $i+\hat{x}$ and $i+\hat{z}$ by $R_{z}(\pi/2)$ and leave the two others, i.e. $\sigma^{x}_{i+\hat{\alpha}}\rightarrow \sigma^{y}_{i+\hat{\alpha}}$ and $\sigma^{y}_{i+\hat{\alpha}}\rightarrow -\sigma^{x}_{i+\hat{\alpha}}$, where $\alpha=x,y$. Similarly, on the ``$b$'' plaquettes, we apply $R_{z}(\pi/2)$ on the sites $i$ and $i+\hat{d}$. Basically, the rotations have to be performed on two diagonally opposite points. Therefore, 
\begin{align}
H_{TC}=-J_{w}\sum_{a}\prod_{i\in a}\sigma^{x}_{i}-J_{w}\sum_{b}\prod_{i\in b}\sigma^{y}_{i}
\end{align}
Where $J_{w}=\frac{3t^{4}}{32U^{3}}(3\lambda^{2}+\lambda^{4})$.\\
\indent When, $\lambda\rightarrow 0$ (zero pairing), The $U(1)$ symmetry is restored,
\begin{align*}
H_{\lambda=0}=J_{I}\sum_{\langle ij \rangle}\sigma^{z}_{i}\sigma^{z}_{j}-J_{p}\sum_{i}\big( \sigma^{+}_{i}\sigma^{-}_{i+\hat{x}}\sigma^{+}_{i+\hat{x}+\hat{y}}\sigma^{-}_{i+\hat{y}} +h.c. \big)
\end{align*}
Here, $J_{I}=\frac{t^{2}}{8U}$, and $J_{p}=\frac{3t^{4}}{32U^{3}}$. The sign of the ring-exchange coupling term has been flipped without affecting the signature of the Ising coupling. We have rotated the spins on alternate sub-lattices by $\pi/2$ with respect to $z$ axis. So, $\sigma^{+}_{i}\rightarrow -i\sigma^{+}_{i}$ and $\sigma^{+}_{i+\hat{x}+\hat{y}}\rightarrow -i\sigma^{+}_{i+\hat{x}+\hat{y}}$.
\section{HQCM with staggered Zeeman field}\label{D}
Consider the HQCM Hamiltonian with a staggered Zeeman magnetic field coupling,
\begin{align}
H=J_{x}\sum_{i\in b}(S^{x}_{i,b}S^{x}_{i+\hat{x},w}+&S^{x}_{i,b}S^{x}_{i-\hat{x},w})+J_{z}\sum_{i\in b}S^{z}_{i,b}S^{z}_{i+\hat{z},w}\nonumber\\
&+h_{z}\sum_{i,\alpha=b,w}(-1)^{i_{\alpha}}S^{z}_{i,\alpha}\label{D1}
\end{align}
Here, we choose the negative sign for the Zeeman term on the $w$ sites, i.e. $i_{w}=1, i_{b}=0$. On the dual square lattice (joined by zz bonds of the brick wall lattice), the Hamiltonian could be re-written as follows,
\begin{align}
H=J_{x}\sum_{i}(S^{x}_{i,b}S^{x}_{i+\hat{e}_{x},w}+S^{x}_{i,w}S^{x}_{i+\hat{e}_{y},b})\nonumber\\
\ \ \ +J_{z}\sum_{i}S^{z}_{i,b}S^{z}_{i,w}+h_{z}\sum_{i}(S^{z}_{i,b}-S^{z}_{i,w})
\end{align}
Under the JW transformation and the subsequent Majorana fermion decompositions (as described at the beginning of the main text, see Eqs.\eqref{eq2}-\eqref{eq4}), $S^{z}_{i,b}=-(i/2)B^{b}_{i}A^{b}_{i}$ and $S^{z}_{i,w}=-(i/2)A^{w}_{i}B^{w}_{i}$. So, the Zeeman term transforms to the following,
\begin{align}
H_{z}=\frac{h_{z}}{2}\sum_{i}(iA^{w}_{i}B^{w}_{i}-iB^{b}_{i}A^{b}_{i})
\end{align}
It commutes with the $J_{z}\rightarrow \infty$ ground state subspace projectors $P\sim \prod_{i=1}^{N}(1+A^{w}_{i}A^{b}_{i}B^{b}_{i}B^{w}_{i})$, i.e. $[P,H_{z}]=0$. Therefore, it should not affect the structure of the strong-coupling ($J_{z}>>J_{x}$) Hamiltonian in the absence of $h_{z}$. The effective strong-coupling Hamiltonian after including $h_{z}$ now becomes,
\begin{align}
H^{s}_{TC}=-J_{w}\sum_{P}\sigma^{x}_{i}\sigma^{y}_{i+x}\sigma^{x}_{i+x+y}\sigma^{y}_{i+y}-h\sum_{i}(-1)^{i}\sigma^{y}_{i}
\end{align}
Here, the effective spins are defined as earlier (see Appendix \ref{C}), and the Zeeman term act as a staggered field along $y$. We now rotate the spins of one sub-lattice (by $\pi$), $\sigma^{y}\rightarrow -\sigma^{y}$ and $\sigma^{z}\rightarrow -\sigma^{z}$; this transforms the staggered coupling to a translationally symmetric term,
\begin{align}
H^{s}_{TC}=-J_{w}\sum_{P}\sigma^{x}_{i}\sigma^{y}_{i+x}\sigma^{x}_{i+x+y}\sigma^{y}_{i+y}-h\sum_{i}\sigma^{y}_{i}\label{D5}
\end{align}
This model belongs to the critical universality class of $1d$ TFIM. \\
%We get a topological Hamiltonian like Wen's plaquette model from a non-topological Hamiltonian \eqref{D1} by performing strong-coupling perturbation theory, because the intermediate step involves a non-local Jordan-Wigner transformation between spins to fermions. Such non-local dualities can transform a classically ordered Hamiltonian to a topological one \cite{Chen1}.\\
\indent Instead of using the JW fermionization in the intermediate step, we can perform the strong-coupling expansion directly on the spin Hamiltonian \eqref{D1}. The conclusion should not differ. The ground state (GS) subspace, when $J_{z}>>J_{x},h_{z}$ belongs to $\lbrace \ket{\uparrow\downarrow}_{i}, \ket{\downarrow\uparrow}_{i} \rbrace$. Since the model has local symmetries $S^{x}_{i,w}S^{x}_{i,b}$ (when $h_{z}=0$), we choose the symmetric and anti-symmetric combinations, $\ket{\phi^{\pm}_{i}}\sim \ket{\uparrow\downarrow}_{i}\pm \ket{\downarrow\uparrow}_{i}$ as our local basis states. The projector onto this subspace is $P=2^{-N_{b}}\prod_{i\in N_{b}}(1-\sigma^{z}_{i,b}\sigma^{z}_{i,w})$. Notice that $[H_{z},P]=0$ (so it will remain unchanged) but $[H_{x},P]\neq 0$. Here, we call the Hamiltonian involving $J_{x}$ couplings as $H_{x}$. When $H_{x}$ acts on the GS subspace, it creates transitions between $\lbrace \ket{\phi^{\pm}_{i}} \rbrace$ and $\lbrace \ket{\psi^{\pm}_{i}}\sim (\ket{\uparrow\uparrow}_{i}\pm \ket{\downarrow\downarrow}_{i})\rbrace$ on the two neighbouring zz links, $(i,i+\alpha)$, ($\alpha=x,y$). All the odd-order perturbative corrections are zero. The second-order correction is just a constant, because one has to unflip the flipped spins to return to the same GS space.\\
The fourth order process has the following structure (we show here one particular case): Take the following initial states on a plaquette, $\ket{\phi^{+}}_{i}$, $\ket{\phi^{-}}_{i+x}$, $\ket{\phi^{+}}_{i+x+y}$, and $\ket{\phi^{-}}_{i+y}$. Then,
%Here $\ket{\uparrow\downarrow}_{i}$ is equivalent to $\ket{\uparrow}_{b}, \ket{\downarrow}_{w}$, where $w,b$ are the original sub-lattice indices of the Honeycomb lattice.
\\

$(1)\ \sigma^{x}_{i,b}\sigma^{x}_{i+x,w}\ \lbrace \ket{\phi^{+}}_{i}, \ket{\phi^{-}}_{i+x}\rbrace\ \ \rightarrow \lbrace \ket{\psi^{+}}_{i}, \ket{\psi^{-}}_{i+x}\rbrace$\\
\indent $(2)\ \sigma^{x}_{i+x,w}\sigma^{x}_{i+x+y,b}\ \lbrace \ket{\psi^{-}}_{i+x}, \ket{\phi^{+}}_{i+x+y}\rbrace \rightarrow \lbrace \ket{\phi^{-}}_{i+x}, \ket{\psi^{+}}_{i+x+y}\rbrace$
\\
\indent $(3)\ \sigma^{x}_{i+x+y,w}\sigma^{x}_{i+y,b}\ \lbrace \ket{\psi^{+}}_{i+x+y}, \ket{\phi^{-}}_{i+y}\rbrace\rightarrow \lbrace \ket{\phi^{+}}_{i+x+y}, \ket{\psi^{-}}_{i+y}\rbrace$
\\
\indent $(4)\ \sigma^{x}_{i,w}\sigma^{x}_{i+y,b}\ \lbrace \ket{\psi^{+}}_{i}, \ket{\psi^{-}}_{i+y}\rbrace \rightarrow \lbrace \ket{\phi^{+}}_{i}, \ket{\phi^{-}}_{i+y}\rbrace$
\\

%So, we see $\ket{\uparrow\downarrow}_{i}\rightarrow \ket{\downarrow\uparrow}_{i}$, $\ket{\downarrow\uparrow}_{i+x+y}\rightarrow \ket{\uparrow\downarrow}_{i+x+y}$ and the other two states at $i+x$ and $i+y$ remain same as initial ones.
The staggered field term ($\sim h_{z}$) causes transition from $\ket{\phi^{+}}_{i}$ to $\ket{\phi^{-}}_{i}$ (and vice-versa). Define the effective (bond) spins, $\tau^{x}_{i}=\ket{\uparrow\downarrow}\bra{\downarrow\uparrow}+\ket{\downarrow\uparrow}\bra{\uparrow\downarrow}$, and $\tau^{z}_{i}=\ket{\uparrow\downarrow}\bra{\uparrow\downarrow}-\ket{\downarrow\uparrow}\bra{\downarrow\uparrow}$, the fourth-order effective Hamiltonian obtains,
\begin{align}
H_{eff}=-J_{w}\sum_{i}\tau^{z}_{i}\tau^{z}_{i+x+y}-h_{z}\sum_{i}\tau^{x}_{i}\label{D6}
\end{align}
Here $J_{w}\sim O(J^{4}_{x}/J_{z}^{3})$. This is just a collection of decoupled TFIM chains along the diagonal of the dual square lattice. The domain wall excitations of these Ising chains corresponds to the anyons of the Wen's model \cite{Chen-Nussinov} which are dispersive due to finite $h_{z}$. Both \eqref{D5} and \eqref{D6} fall in the same $1d$ quantum Ising universality class.
%%%%%%%%%%%%%%%%%%%%%%%%%%%%%%%%%%%%%%%%%%%%%%%%%%%%%%%%%%%%%%%%%%%%%%%%%%%%%%%%%%%%%%%%%%
%%%%%%%%%%%%%%%%%%%%%%%%%%%%%%%%%%%%%%%%%%%%%%%%%%%%%%%%%%%%%%%%%%%%%%%%%%%%%%%%%%%%%%%%%%%%%%%%%%


\begin{thebibliography}{34}
\bibitem {Foot1}%
\BibitemOpen
To get the four robust corner modes, we require additional edge perturbations (that hybridizes the MZMs along the edges) to make the boundary edges fully gapped, leaving unpaired MZMs only at the high-symmetry corners of the lattice\BibitemShut {NoStop}%
\bibitem {Foot2}%
\BibitemOpen
Since the JW fermion model Eq.\eqref{eq6} explicitly contains a BCS Cooper-pairing term, the global charge $U(1)$ symmetry is reduced to $Z_{2}$, which corresponds to total fermion number parity conservation of the system.\BibitemShut {NoStop}%
\bibitem{Foot3}%
\BibitemOpen
Paramagnetic phase of Ising models can be described as off-diagonal long-range order of some appropriate disorder operators (observables), see Ref.\cite{Fisher}.\BibitemShut {NoStop}%
\end{thebibliography}
\end{document}